\documentclass[11pt]{elsarticle}
\usepackage[paper=a4paper,marginratio={1:1,5:6}]{geometry}
\makeatletter
\def\ps@pprintTitle{%
 \let\@oddhead\@empty
 \let\@evenhead\@empty
 \def\@oddfoot{\centerline{\thepage}}%
 \let\@evenfoot\@oddfoot}
\makeatother

\makeatletter
\let\@afterindenttrue\@afterindentfalse
\makeatother

\biboptions{sort&compress}
\usepackage[hyperref]{xcolor}
\definecolor{darkgreen}{rgb}{0.01, 0.75, 0.24}
\definecolor{darkblue}{HTML}{2B66D3}
\usepackage[colorlinks=true,
            linkcolor=darkblue,
            urlcolor=darkblue,
            citecolor=darkblue,linkcolor=darkblue]{hyperref}

\usepackage{amsmath}
\usepackage{amsfonts}
\usepackage{siunitx}
\usepackage{slashed}
\usepackage{accents}
\usepackage{amssymb}
\usepackage{mathrsfs}
\usepackage{mathtools}
\usepackage{comment}
\usepackage[utf8]{inputenc}
\usepackage[T1]{fontenc}
\let\oldbibliography\thebibliography
\renewcommand{\thebibliography}[1]{%
  \oldbibliography{#1}%
  \setlength{\itemsep}{1.4pt}%
}

\usepackage[cal=boondox,calscaled=1]{mathalfa}
\DeclareMathAlphabet{\bbvar}{U}{BOONDOX-ds}{m}{n}
\DeclareMathAlphabet{\mathsl}{\encodingdefault}{\familydefault}{m}{sl}

\usepackage[defaultsans,scale=0.9]{opensans}
\usepackage{psfrag}
\usepackage{pstool}
\usepackage{caption}
\usepackage{tabularx}
\usepackage{multirow}
\usepackage{pbox}
\usepackage[normalem]{ulem}

\usepackage{graphicx}
\usepackage{subcaption}
\usepackage{relsize}

\newcommand{\di}{\mathrm{d}}
\usepackage{tensor}
\newcommand{\ou}[3]{\tensor{#1}{^{#2}_{#3}}}

\newcommand{\I}{\mathrm{i}} 
\newcommand{\CC}{\mathrm{cc.}} 
\newcommand{\C}{\mathbb{C}}

\newcommand{\R}{\mathbb{R}}

\newcommand{\eref}[1]{(\ref{#1})}

\def\Ree{\operatorname{Re}}
\def\Imm{\operatorname{Im}}

\def\sl{\mathfrak{sl}}
\def\ii{{-1}}

\newcommand{\cA}{{\cal A}}
\newcommand{\cC}{{\cal C}}

\newcommand{\cE}{{\cal E}}
\newcommand{\cF}{{\cal F}}

\newcommand{\cB}{{\cal B}}

\newcommand{\cL}{{\cal L}}
\newcommand{\cH}{{\cal H}}

\newcommand{\mtext}[1]{\text{\it #1}}

\newcommand{\pif}{\rho}
\newcommand{\pis}{\pi}

\newcommand{\slc}{\sl(2,\C)}
\newcommand{\slcb}{\overline{\sl(2,\C)}}
\newcommand{\SLC}{\SL(2,\C)}
\newcommand{\SLCB}{\overline{\SL(2,\C)}}
\newcommand{\diff}{\operatorname{diff}}

\newcommand{{\Ap }}{{A'}}
\newcommand{{\Bp}}{{B'}}
\newcommand{{\tpsi}}{\tilde{\psi}}
\newcommand{{\tphi}}{\tilde{\phi}}
\newcommand{\tpif}{\tilde{\rho}}
\newcommand{\tpis}{\tilde{\pi}}
\newcommand{{\tH}}{\tilde{H}}
\newcommand{{\tA}}{\tilde{A}}
\newcommand{{\tq}}{\tilde{q}}
\newcommand{{\tgamma}}{\tilde{\gamma}}
\newcommand{\psiEE}{\mathcal{E}_\pi}
\newcommand{\phiEE}{\mathcal{E}_\rho}
\newcommand{\psiFF}{\mathcal{F}_\phi}
\newcommand{\phiFF}{\mathcal{F}_\psi}
\newcommand{\psiee}{E_\pi}
\newcommand{\phiee}{E_\rho}
\newcommand{\psiff}{F_\phi}
\newcommand{\phiff}{F_\psi}
\newcommand{\tpsiFF}{\tilde{\mathcal{F}}_{\tphi}}
\newcommand{\tphiFF}{\tilde{\mathcal{F}}_{\tpsi}}
\newcommand{\tpsiEE}{\tilde{\mathcal{E}}_{\tilde{\pi}}}
\newcommand{\tphiEE}{\tilde{\mathcal{E}}_{\tilde{\rho}}}
\newcommand{\cG}{\mathcal{G}}
\newcommand{\tcG}{\tilde{\mathcal{G}}}

\DeclareMathAlphabet{\bbgreek}{U}{bbold}{m}{n}

\usepackage{tikz-cd}
\usetikzlibrary{arrows.meta, automata,
                positioning,
                quotes}
\usepackage[low-sup]{subdepth}

\newcommand\vpm{\mathbin{\vcenter{\hbox{
  \oalign{\hfil$\scriptstyle+$\hfil\cr
          \noalign{\kern-.3ex}
          $\scriptscriptstyle({-})$\cr}}}}}
\DeclareMathAlphabet{\sfit}{OT1}{\sfdefault}{m}{it}
\DeclareMathAlphabet{\sfbfit}{OT1}{\sfdefault}{sb}{it}
\DeclareMathAlphabet{\mathsf}{OT1}{\sfdefault}{sb}{n}

\definecolor{darkgreen}{rgb}{0.01, 0.75, 0.24}
\definecolor{darkblue}{HTML}{2B66D3}

\usepackage[multiple, flushmargin]{footmisc}
\let\originalleft\left
\let\originalright\right
\renewcommand{\left}{\mathopen{}\mathclose\bgroup\originalleft}
\renewcommand{\right}{\aftergroup\egroup\originalright}

\usepackage{scalerel}

\newcommand{\dbarvar}{{\mathrm{d}\mkern-7.5mu\lower.18ex\hbox{$\textasciitilde$}\mkern-1.5mu}}

\renewcommand{\emph}[1]{{\it #1}}

\def\SU{\mathrm{SU}}
\def\SL{\mathrm{SL}}
\def\SO{\mathrm{SO}}
\newcommand{\frpi}{\tensor[^{\mathnormal{4}}]{\pi}{}}
\newcommand{\frrho}{\tensor[^{\mathnormal{4}}]{\rho}{}}
\newcommand{\frpsi}{\tensor[^{\mathnormal{4}}]{\psi}{}}
\newcommand{\frphi}{\tensor[^{\mathnormal{4}}]{\phi}{}}
\renewcommand{\d}{\mathrm{d}}
\newcommand{\downup}[3]{\tensor{#1}{_{#2}^{#3}}}
\newcommand{\updown}[3]{\tensor{#1}{^{#2}_{#3}}}
\newcommand{\frdownup}[3]{\tensor[^{\mathnormal{4}}]{#1}{_{#2}^{#3}}}
\newcommand{\frupdown}[3]{\tensor[^{\mathnormal{4}}]{#1}{^{#2}_{#3}}}
\newcommand{\pradownup}[4]{\tensor[^{\mathnormal{#1}}]{#2}{_{#3}^{#4}}}
\newcommand{\praupdown}[4]{\tensor[^{\mathnormal{#1}}]{#2}{^{#3}_{#4}}}

\usepackage{aligned-overset}

\begin{document}

\begin{abstract}
Four-dimensional gravity admits many equivalent formulations---metric, Einstein–Cartan, teleparallel, McDowell–Mansouri, among others---each offering distinct advantages, particularly, in view of quantization. We propose a new formulation based on Weyl spinor–val\-ued 1-forms, ultimately encoding the frame-field data. Starting from a topological field theory with a global  $\mathrm{SL}(2,\mathbb{C})$ symmetry, we show that promoting this symmetry to a local gauge symmetry leads to the emergence of gravity. We analyze  the covariant phase space of this theory, its symmetries and charge structure and  explore the role of admissible corner terms together with their impact on boundary charges and their algebra. We study several extensions of this framework, including the incorporation of a cosmological constant and a novel $ G \rightarrow 0 $ scaling limit obtained from this model. The presence of the frame field already at the topological level allows  point particles to be coupled uniformly in both the topological and gravitational theories. We perform a detailed Hamiltonian analysis of the theory and clarify the implementation of the reality conditions. We argue that this formulation provides structural features that make it particularly well suited for both discretization and quantization.

\end{abstract}

\title{
Topological field theory plus local Lorentz symmetry is gravity}

\author{Maïté Dupuis${}^1$, Florian Girelli${}^{2}$, Oleksandra Hrytseniak${}^{1,3}$, Wolfgang Wieland${}^4$\vspace{0.5em}}
\address{${}^1$Perimeter Institute for Theoretical Physics\\31 Caroline St North\\Waterloo N2L\,2Y5, Ontario, Canada}
\address{${}^2$Department of Applied Mathematics\\University of Waterloo\\200 University Avenue West\\Waterloo, Ontario, Canada\,N2L 3G1}
\address{${}^3$Department of Physics and Astronomy\\University of Waterloo\\200 University Avenue West\\Waterloo, Ontario, Canada\,N2L 3G1}
\address{${}^4$Institute for Quantum Gravity, Theoretical Physics III, Department of Physics\\Friedrich-Alexander-Universität Erlangen-Nürnberg\\Staudtstraße 7, 91052 Erlangen, Germany (EU)\\[1.5em]
{\normalfont \today}
}
\date{}
\maketitle
\hypersetup{
  linkcolor=black,
  urlcolor=black,
  citecolor=black
}
\newpage
{\tableofcontents}\vspace{-0.5em}
\hypersetup{
  linkcolor=darkblue,
  urlcolor=darkblue,
  citecolor=darkblue,
}
\begin{center}{\noindent\rule{\linewidth}{0.4pt}}\end{center}
\section*{Introduction}

In his book on gravitation \cite{Weinberg:1972kfs}, Steven Weinberg says that the only good reason for gravitational variables to appear in the matter Lagrangian is to guarantee local Lorentz invariance. This gauge principle reflects the freedom of choosing independent inertial frames in different spacetime regions. It is realized by upgrading partial derivatives to Lorentz-covariant derivatives. Here, we show that this principle not only applies to the matter sector, but can be extended to the gravitational action itself. 

We start from a topological field theory in four spacetime dimensions that admits a global Lorentz symmetry. We then promote this global Lorentz symmetry to a local gauge symmetry. This procedure breaks certain topological gauge symmetries---the so-called shift symmetry of BF-theory \cite{baez, Girelli_2008, Martins:2010ry, Zucchini:2021bnn, Borsten_2025}. The breaking of the topological symmetries makes otherwise unphysical gauge directions become physical, thus unlocking propagating degrees of freedom. The resulting field theory describes self-dual gravity, which has two complex physical degrees of freedom per point. Upon imposing reality conditions, we recover the dynamics of general relativity. In this sense,  gravity emerges from a gauge symmetry breaking of a topological field theory. A similar mechanism appears in Pleba{\'n}ski's formulation of gravity as a constrained topological field theory \cite{Plebanski:1977zz}.  The main conceptual difference is that in our case we break topological symmetries by adding local gauge  symmetries. 

As a result, we rediscover Robinson's formulation of complexified gravity in terms of Weyl spinor-valued $p$-forms \cite{Robinson_1996, Robinson:1998vf}. A similar formulation in terms of Dirac spinor--valued fields first appeared in \cite{Nester:1994zn}. A version in terms of Weyl spinor-valued $1$-forms was first introduced in Tung and Jacobson \cite{Tung:1995cj}, in which the expressions for quasi-local energy take a particularly simple form. Robinson  defined a first-order version of Tung and Jacobson's action
 and discussed  reality conditions to recover gravity \cite{Robinson_1996, Robinson:1998vf}.

Indeed, self-dual gravity can be understood as a complexification of general relativity. Thus, it has twice as many variables. To restore general relativity in terms of real-valued fields, one re-introduces an anti-self-dual part to the action. The two sectors are matched by reality conditions that select a real section of the  complex phase space. Upon imposing constraints and   quotienting  out gauge symmetries, we are left with two physical degrees of freedom, representing the two modes of gravitational radiation. 

Our paper provides a new perspective on some well-established results on gravitational symmetries and boundary charges. We study the charge algebra in the topological and gravitational theories using the covariant phase space formalism. In particular, we show how natural corner term ambiguities affect the values of the boundary charges and their algebra. Furthermore, we show that the reality conditions between the self-dual and anti-self-dual $\SL(2,\mathbb{C})$ sectors induce reality conditions on the corner charges.

\medskip 

Additionally, we consider several new developments of Robinson's theory. First, we discuss how to add  a cosmological constant term to the action. Second, by re-establishing  Newton's constant $G$ in the action, we show that there is  a new limit $G\rightarrow 0$ of gravity, which can be unlocked naturally in the Weyl  spinor variables. This limit can be considered at the level of the topological model as well, modifying the symmetry structure from the identity 2-group to a skeletal 2-group \cite{baez, Borsten_2025}. Finally, we discuss the coupling of point-like particle defects. Such particle defects can be coupled naturally to the topological theory. The symmetry-breaking mechanism that turns the topological field theory to (self-dual) gravity can be then also applied to the coupled system. The resulting coupled system consists of  point particles coupled to (self-dual) gravity. Such a procedure cannot be done in the Pleba\'{n}ski formulation, but it is similar to what is done in the McDowell-Mansouri approach \cite{Freidel:2006hv}. The key feature is the presence of the (co)frame-field data already at the topological level.

\smallskip

In the second half of the paper, we perform a detailed Hamiltonian analysis of the spinorial formulation of the gravitational action. The Poisson commutation relations among the fundamental first-class and second-class constraints take a very simple form: the Dirac matrix of constraints depends on the fundamental fields at most quadratically. We also recover the Hamiltonian originally introduced by Tung and Jacobson \cite{Tung:1995cj}, given by a sum of a bulk constraint and boundary charges. Going on shell, we obtain the familiar result that the gravitational Hamiltonian is a codimension-2 surface integral. We study the imposition of the reality conditions relative to a dynamical spin basis, which is built from the dynamical fields in a $3+1$ split of the theory. The construction allows to impose the reality conditions without breaking the internal $\SL(2,\C)$ gauge symmetry.

\subsection*{Paper organization} The paper is organized as follows. In \hyperref[sec:topo-id]{Section \ref{sec:topo-id}}, we introduce the Weyl spinor topological action and perform its symmetry analysis, from a covariant phase space perspective. We highlight, in particular, the global $\SLC$ symmetry, as well as a hidden $\SLCB$ local symmetry. In \hyperref[section:gravityCPS]{Section \ref{section:gravityCPS}}, we show how by including the Gauss constraint and the associated Lagrange multiplier, the connection, one can recover self-dual gravity. We perform the symmetry analysis of the gravitational action expressed in terms of  Weyl spinor-valued 1-forms and 2-forms  and  discuss how the cosmological constant is introduced in this context. In  \hyperref[sec:corner]{Section \ref{sec:corner}}, we discuss some of the corner terms we can add to change the charge structure of the gravity theory. We highlight how the reality conditions are realized in terms of the charges. 
In  \hyperref[app:g0]{Section \ref{app:g0}}, we discuss how this formalism allows to perform the limit $G\to 0$ in a new way. We review how the symmetries are affected by such a limit,  both in the topological and gravitational cases.  In \hyperref[sec-canonical]{Section \ref{sec-canonical}}, we perform the Hamiltonian analysis of the theory to identify the first- and second-class constraints and perform the counting of degrees of freedom. We also discuss how  reality conditions are stable under evolution.  In \hyperref[sec:part]{Section \ref{sec:part}}, we show that the (spinless) particle can be introduced in the same way in the topological and gravitational {theories}. Finally, in the \hyperref[sec:outlook]{Outlook}, we highlight the different strengths of the  spinorial formulation of gravity in view of discretization and quantization.

\subsection*{Notation} 

Following standard notation, \cite{penroserindler}, we use capital Latin letters to denote $\SLC$ indices $A,B,...$, which take values $0,1$, and are raised and lowered by the internal skew-symmetric $\epsilon$-tensors according to the pattern $\xi^A=\epsilon^{AB}\xi_B$, $\xi_A=\xi^B\epsilon_{BA}$. Primed indices $A',B',...$, correspond to the complex conjugate representation, $\SLCB$,  upon imposing reality conditions.  

\medskip 

In the following, $M$ is a four-dimensional manifold with a boundary $\partial M$. We denote with $\Omega^p(M)$  the space of $p$-forms on $M$. Let the boundary $\partial M$ be piecewise-smooth and composed of smooth three-dimensional hypersurfaces $\{\Sigma_i\}$ so that $\partial M=\bigcup_i\Sigma_i$. Each smooth boundary component $\Sigma_i$ is itself bounded by a co-dimension-2 submanifold $C_i=\partial\Sigma_i$, which will be called a \textit{corner}. 

In what follows,  $\psi_A,\phi_A\in\Omega^1(M)\otimes\C^2$ are Weyl spinor-valued 1-forms, and $\pi_A,\rho_A\in  \Omega^2(M)\otimes \C^2$  Weyl spinor-valued 2-forms. Spacetime indices are denoted by $\mu,\nu,\rho,\dots$ and $a,b,c,\dots$ denote three-dimensional indices on a hypersurface. Depending on the context, we will often suppress the spinor or spacetime indices, thereby denoting e.g.\ the self-dual connection 1-form as $A$, $\updown{A}{A}{B}$ or $\updown{A}{A}{Ba}$. The infinite-dimensional space of all these fields $\phi, \psi, A, \pi, \rho$ on $M$ is denoted as $\cF$. In what follows, $\bbvar{d}$  denotes the field-space exterior derivative, while $\bbvar{I}_\delta\Phi$ denotes the interior product of a field-space differential form $\Phi$ with a vector field $\delta\in T\mathcal{F}$ over the field space. By abuse of notation, we also denote by $\bbvar{I}_\xi\Phi$ the contraction of the field space $p$-form with a symmetry generator  $\delta_\xi\in T\mathcal{F}$, which generates a specific gauge transformation parametrized by a gauge parameter $\xi$.

We will note $\Omega^{p,q}(M,\cF)$ the space of spacetime $p$-forms and $q$-forms in field space. For example, $\psi_A\wedge \bbvar{d} \phi^A\in \Omega^{2,1}(M,\cF)$. 
In what follows, the wedge product $\wedge$ denotes the wedge product on spacetime, as well as on field space where necessary.

\medskip 

Consider a three-surface $\Sigma\in \{\Sigma_i\}$ with a boundary $C=\partial\Sigma$. 
The generators or momentum maps $H[\xi]$ associated to symmetry transformations $\delta_\xi$ that are parameterized by a gauge parameter $\xi$ are defined via the Hamiltonian flow equation. Given a symplectic potential $\Theta_\Sigma$ associated with a hypersurface $\Sigma$ and a corresponding symplectic form $\Omega_\Sigma=\bbvar{d}\Theta_\Sigma$, the Hamiltonian flow equation takes the form
\begin{equation}
    \bbvar{I}_{\xi}\Omega_\Sigma=-\bbvar{d} H_\Sigma[\xi].\label{HamFlowUnpr}
\end{equation}
The flow equations \eref{HamFlowUnpr} often pose stricter conditions on the transformation parameters than the direct application of the Noether theorem would suggest. 
\begin{align}
    H_\Sigma[\xi]&=\bbvar{I}_{\xi}\Theta_\Sigma-\!\!\int_\Sigma\! R[\xi],\label{Noether1}
\end{align}
where we denote
\begin{equation}
    \delta_\xi L=\d R[\xi], \qquad \delta_\xi S=\int_\Sigma R[\xi]
\end{equation}
for Lagrangian symmetries. Since most symmetries we are going to deal with are gauge symmetries, we expect that their generators vanish on shell in the interior of $\Sigma$, but not necessarily at its boundary $C$. The corner part of a generator $H[\xi]$ is called $q[\xi]$:
\begin{equation}
    H_\Sigma[\xi]\approx q_C[\xi].
\end{equation}
Here and in what follows, we denote with $\approx$ equality up to equations of motion. In what follows, we drop the subscripts $\Sigma$ and $C$ and assume that all generators are defined at the same surface.

\section{Abelian BF  theory with Weyl spinors} \label{sec:topo-id}

\subsection{Action, field equations and kinematical phase space}

Our starting point is the following first-order complex action:
\begin{equation}
    S[\psi,\phi,\pi,\rho]=\I\int_M \left(\pi_{ A}\wedge \d\psi^{A}+\rho_{ A}\wedge \d\phi^{A}+g\pi_{ A}\wedge\rho^{A}\right),\label{action-topo}
\end{equation}
where the symbol ``$\d$'' denotes the exterior derivative and $g$ is a coupling constant. This action defines a topological field theory, namely, a BF theory (or {2-BF} theory \cite{Girelli_2008, Martins:2010ry, Zucchini:2021bnn, Borsten_2025}). This can be seen by repackaging the fields to recover the usual form of a BF theory. We introduce
\begin{align}
    \mathcal{B}=\begin{pmatrix}
       \pi_A & \rho_A 
    \end{pmatrix}, \quad \mathcal{A}= \begin{pmatrix}
       \psi^A\\\phi^A 
    \end{pmatrix}, \quad \mathcal{F}= \d\cA, \quad t(\mathcal{B})=g\begin{pmatrix}
       \rho^A\\-\pi^A 
    \end{pmatrix}.\label{cal-BF}
\end{align}
The action $S[\psi,\phi,\pi,\rho]$ is then equivalent to
\begin{align}
    S[\mathcal{B}, \mathcal{A}]=\I \int_M\left(\langle\cB\wedge \cF \rangle +\frac{1}{2}\langle \cB\wedge t(\cB) \rangle\right).
\end{align}
The  equations of motion  are readily found to be
\begin{subequations}
\begin{align}
    &g\pis_A-\d\phi_A=0,\label{EOMab1}\\
    &g\pif_A+\d\psi_A=0, \label{EOMab2}\\ 
    &\d\pis_A =0,\label{EOMab3}\\
    &\d\pif_A=0, \label{EOMab4}
\end{align} 
\end{subequations}
Note that the latter two equations (\ref{EOMab3}, \ref{EOMab4}) are merely integrability conditions: they can be obtained for any $g\neq 0$ from the first pair of equations by acting with an exterior derivative on them. The special case of zero coupling will be described below. We will refer to the equations \eqref{EOMab1} and \eqref{EOMab2} as \emph{fake-flatness equations} and to the second pair of equations, i.e.\ \eqref{EOMab3} and \eqref{EOMab4}, as \emph{2-flatness equations} in correspondence with the higher gauge theory terminology sometimes used in the  topological field theory literature \cite{baez, Girelli_2008, Borsten_2025}.

In the variation of the action, a boundary term appears, which determines the symplectic current, from which we obtain the pre-symplectic potential $\Theta$, which is a 1-form on field space:
\begin{equation}
    \Theta_\Sigma=\I\int_{\partial M}(\pis_A\wedge\bbvar{d}\psi^A+  \pif_A\wedge\bbvar{d}\phi^A),\label{SymPot-topo}
\end{equation}
where $\bbvar{d}$ denotes the field-space exterior derivative and the implicit pullback to $\Sigma$ is understood. Namely, if $\varphi_\Sigma:\Sigma\hookrightarrow M$ is the canonical embedding of the hypersurface $\Sigma$ into $M$, integration over $\Sigma$ selects the pullback $\varphi^\ast_\Sigma\psi_A, \,\varphi^\ast_\Sigma\pi_A$, etc., of the phase-space fields. The pre-symplectic potential \eqref{SymPot-topo} immediately defines the pre-symplectic 2-form
\begin{equation}
    \Omega_\Sigma = \I\int_\Sigma  (\bbvar{d}\pis_A\wedge\bbvar{d}\psi^A+  \bbvar{d} \pif_A\wedge\bbvar{d}\phi^A).\label{SymForm-topo}
\end{equation} 

\subsection{Symmetries, generators and their algebra}

In this section, we analyse the symmetries of the topological theory, using the covariant phase space formalism. We will recover the standard topological symmetries of the BF-type of action \eref{action-topo}, specified by translations of the 1-connection $\mathcal{A}$ and exact shift symmetries of the $\mathcal{B}$-field, as introduced in \eref{cal-BF} above. Diffeomorphisms can be recovered from the latter as a combination of shift transformations with field-dependent gauge parameters. Unlike the usual BF-theories, we will identify some additional hidden symmetries. There is a global $\SL(2,\mathbb{C})$ symmetry, as well as a \emph{dual} $\SL(2,\mathbb{C})$ symmetry. To avoid confusion with the global symmetry, we denote it by $\SLCB$. This symmetry will be obtained by considering a certain class of field-dependent shift symmetries.

\paragraph{Shift symmetry}
Let $\eta_A, \zeta_A\in\C^2\otimes \Omega^1(M)$ be Weyl spinor--valued 1-forms which will parameterize the shift transformations. The action \eqref{action-topo} is invariant under two sets of shift transformations:
\begin{subequations}\label{shift-on-psi-topo}
\begin{alignat}{5}
\delta_{\eta}\psi_A&=g\,\eta_A, &\hspace{4em}&\delta_{\eta}\phi_A&=0,\\
    \delta_{\eta} \pif_A&=-\d\eta_A, &&\delta_{\eta} \pis_A&=0,  
\end{alignat}
\end{subequations}
and
\begin{subequations}
\label{shift-on-phi-topo}
\begin{alignat}{5}
\delta_\zeta\phi_A&=g\,\zeta_A,&\hspace{4em}&\delta_\zeta\psi_A&=0,\\
    \delta_\zeta \pis_A&=\d\zeta_A,&&\delta_\zeta \pif_A&=0.
\end{alignat}\end{subequations}
If the gauge parameters parameters $\eta$ and $\zeta$ are field-independent, the corresponding generators are integrable and given by
\begin{subequations}
\label{gen-topo-shift}
\begin{alignat}{5}
    \bbvar{I}_{\eta}\Omega&= -\bbvar{d} H_{\phi}[\eta],&\hspace{2.5em}&H_{\phi}[\eta] & =\psiff[\eta]+ q_{\phi}[\eta],\label{gen-topo-shift-psi}\\
    \bbvar{I}_{\zeta}\Omega&=-\bbvar{d} H_{\psi}[\zeta],&&H_{\psi}[\zeta] & =\phiff[\zeta]+ q_{\psi}[\zeta],\label{gen-topo-shift-phi}
\end{alignat}
\end{subequations}
where the momentum maps correspond to the equations of motion \eqref{EOMab1}--\eqref{EOMab2},
\begin{subequations}
\begin{align}
    &\psiff[\eta]=\I\int_\Sigma (g\pis_A-\d\phi_A)\wedge\eta^A,\label{mommapP1}\\
    &\phiff[\zeta]=\I\int_\Sigma (g\pif_A+\d\psi_A)\wedge\zeta^A\label{mommapP2}
\end{align}
\end{subequations}
(mnemonic: $F$ stands for \emph{fake-flatness}), and corner charges are given by
\begin{subequations}
\label{shift-charges-topo}
\begin{align}
    &q_{\phi}[\eta]=\I\int_{C}\eta_A\wedge\phi^A,\label{shift-charge-topo-1}\\
    &q_{\psi}[\zeta]=\I\int_{C}\zeta^A\wedge\psi_A.\label{shift-charge-topo-2}
\end{align}
\end{subequations}

\paragraph{Exact shift symmetries}

Let us focus on the subspace of exact shift symmetries. It is easy to notice that $\d$-exact shift transformations leave the momenta invariant.

Let $\lambda_A,\mu_A\in\C^2\otimes \Omega^0(M)$ be Weyl spinor fields. The two sets of the $\d$-exact shifts  have the following action:
\begin{subequations}\label{exact-shift-topo}
\begin{alignat}{5}
    \delta_{\lambda}\psi_A&=\d\lambda_A, &\hspace{4em}&\delta_{\lambda}\phi_A=\delta_{\lambda} \pis_A=\delta_{\lambda} \pif_A=0,\\
    \delta_\mu\phi_A&=\d\mu_A,&\hspace{4em}&\delta_\mu\psi_A=\delta_\mu\pis_A=\delta_\mu\pif_A=0.
\end{alignat}
\end{subequations}
The Hamiltonian flow equation now takes the form:
\begin{subequations}\label{gen-topo-exshift}
\begin{alignat}{5}
    \bbvar{I}_{\lambda}\Omega&\equiv -\bbvar{d} H_\pis[\lambda], &\hspace{2.5em}&H_\pis[\lambda]&=\psiee[\lambda]+q_\pis[\lambda],\label{gen-topo-exshift-psi}\\
    \bbvar{I}_{\mu}\Omega&\equiv -\bbvar{d} H_\pif[\mu],&&H_\pif[\mu]&=\phiee[\mu]+ q_\pif[\mu],\label{gen-topo-exshift-phi}
\end{alignat}\end{subequations}
where the momentum maps correspond to the remaining equations of motion \eqref{EOMab3}--\eqref{EOMab4},
\begin{subequations}
\begin{align}
&\psiee[\lambda]=\I\int_\Sigma \d\pis^A\lambda_A,\\
    &\phiee[\mu]=\I\int_\Sigma \d\pif^A\mu_A.
\end{align}
\end{subequations}
(mnemonic: $E$ stands for \emph{Einstein equations} in correspondence with their role in the gravitational sector), and corner charges are, respectively:
\begin{subequations}
\begin{align}
    &q_{\pi}[\lambda]=\I\int_{C} \pis_A\lambda^A,\\
    &q_{\phi}[\mu]=\I\int_{C} \pif_A\mu^A.
\end{align}
\end{subequations}
We note that the on-shell charges for the shift and the exact shifts are  related:
\begin{subequations}
\label{ex-shift-charges-topo}
\begin{align}
    &q_\pis[\lambda]=\I\int_{C} \pis_A\lambda^A\overset{\partial C=\emptyset}{\approx} \frac{\I}{g}\int_{C} \d\lambda_A\wedge \phi^A=q_{\phi}[g^{-1}\d\lambda],\\
    &q_\pif[\mu]=\I\int_{C} \pif_A\mu^A\overset{\partial C=\emptyset}{\approx} \frac{\I}{g}\int_{C} \d\mu^A\wedge \psi_A=q_{\psi}[g^{-1}\d\mu].
\end{align}
\end{subequations}
The fact that $\d$-exact shifts generated by the 2-flatness constraint are a subset of shifts generated by the fake-flatness conditions is consistent with the redundancy of the system of constraints: in the abelian theory, the 2-flatness equations \eqref{EOMab3}--\eqref{EOMab4} follow from fake-flatness \eqref{EOMab1}--\eqref{EOMab2}. 

Higher-gauge theories \cite{baez, Borsten_2025} are normally symmetric under two tiers of gauge transformations satisfying certain coherence relations. Exact shifts \eqref{exact-shift-topo} and shifts \eqref{shift-on-psi-topo}--\eqref{shift-on-phi-topo} represent this structure in disguise. The division between the 1-gauge (exact shifts) and 2-gauge (full shifts) has been washed away due to 
the gauge algebra being abelian, but will be restored in the $g\rightarrow0$ (skeletal) limit, see \hyperref[app:g0]{Section \ref{app:g0}}.

\paragraph{Diffeomorphisms}
The next gauge symmetry to be considered here is the four-di\-men\-sional diffeomorphism symmetry. They are generated on phase-space by the Lie derivative, $\delta^{di\!f\!\!f}_\xi= L_\xi\equiv \d \iota_\xi+\iota_\xi \d$. This is where the difference between the definitions of the generator in terms of the Hamiltonian flow equation  \eqref{HamFlowUnpr} or from Noether theorem \eqref{Noether1} is manifest. The Noether charge can be expressed according to \eqref{Noether1} for an arbitrary vector field $\xi\in \Gamma(TM)$ in terms of the momentum maps introduced above and a corner charge as
\begin{subequations}
\begin{align}
    H^{di\!f\!\!f}[\xi]&\overset{\eqref{Noether1}}{=}\phiff[\iota_\xi \pi]-\psiff[\iota_\xi \pif]+\psiee[\iota_\xi \psi]+\phiee[\iota_\xi \phi]+q^{di\!f\!\!f}[\xi],\label{dif-gen-top}\\
    q^{di\!f\!\!f}[\xi]&=\I\int_{C}(\iota_\xi \psi^A\pis_A+\iota_\xi \phi^A\pif_A ).
\end{align}\end{subequations}
Hence we have recovered here a standard result of BF-theory: On shell, the diffeomorphism charges can be built from the shift symmetries with field dependent parameters.

Note that in the special case where the vector field $\xi\in \Gamma(TM)$ is tangential to the corner, i.e.\ $\xi|_{C}\in T{C}$, we obtain further that
\begin{subequations}
\begin{align}
    q^{di\!f\!\!f}[\xi]&\equiv q_\pis[\iota_\xi\psi]+q_\pif[\iota_\xi\phi]\overset{\xi|_{C}\in T{C}}{=} -q_{\phi}[\iota_\xi\pif]+q_{\psi}[\iota_\xi\pis]\equiv \label{q-diff-thru-shifts}\\
    &\equiv \frac12\left(-q_{\phi}[\iota_\xi\pif]+q_{\psi}[\iota_\xi\pis]+q_\pis[\iota_\xi\psi]+q_\pif[\iota_\xi\phi]\right).
\end{align}\end{subequations}
This in turn implies that the corner contribution of shift generators corresponds to one-half of the bulk generators. The full field-dependent shift generators will be recovered upon addition of the anti-self-dual sector, which we will discuss in \href{sec:corner}{Section \ref{sec:corner}}. In fact, the identity \eqref{q-diff-thru-shifts} holds due to the following two observations:
\begin{subequations}
    \begin{align}
        q_{\phi}[\iota_\xi\pif]+q_\pif[\iota_\xi\phi]&=\I\int_{C} \iota_\xi(\pif_A\wedge\phi^A)\overset{\xi|_{C}\in T{C}}{=}0,\\
        -q_{\psi}[\iota_\xi\pis]+q_\pis[\iota_\xi\psi]&=\I\int_{C} \iota_\xi(\pis_A\wedge\psi^A)\overset{\xi|_{C}\in T{C}}{=}0.
    \end{align}
\end{subequations}
Hamilton's flow equation for diffeomorphisms is, however, integrable on $\Sigma$ only if $\xi|_\Sigma\in T\Sigma$ and $\xi|_{C}\in T{C}$. It then takes the form
\begin{equation}
    \bbvar{I}_{\xi}\Omega=-\bbvar{d} H^{di\!f\!\!f}[\xi].\label{xi-intgrblty-cond}
\end{equation}
On-shell, the tangency of $\xi$ to $\Sigma$ is relaxed and the Hamilton's flow equation is integrable provided the gauge parameter is tangent to the corner, i.e.\ $\xi|_C\in TC$.

\paragraph{$\SLCB$ local symmetry} 
There are yet another kind of field-dependent shift transformations which turn out to be integrable and play a crucial role in the gravitational theory. These transformations are parametrized by three complex fields $\{a,b,c\}\equiv\beta$ on spacetime. They act on the phase space as rotations in the spinor plane, as
\begin{subequations}
\label{slcb-top}
\begin{align}
    &\delta_\beta\psi_A= a\psi_A+b\phi_A,\label{dualsl2conpsi-top}\\
    &\delta_\beta\phi_A=c\psi_A-a\phi_A,\label{dualsl2conphi-top}\\
    &\delta_\beta\pis_A= -c\pif_A-a\pis_A+\frac1g\d c\wedge\psi_A-\frac1g\d a\wedge\phi_A,\label{dualsl2conpipsi-top}\\
    &\delta_\beta\pif_A= a\pif_A-b\pis_A-\frac1g\d a\wedge\psi_A-\frac1g\d b\wedge\phi_A.\label{dualsl2conpiphi-top}
\end{align}
\end{subequations}
The corresponding off-shell generator is
\begin{equation}
    H^{\overline{\sl}}[\beta]=\frac1g\psiff[a\psi+b\phi]+\frac1g\phiff[c\psi-a\phi]+q^{\overline{\sl}}[\beta],\label{slcb-gen-topo}
\end{equation}
where the corner charge is given by
\begin{align}
    q^{\overline{\sl}}_{}[\beta]&=\frac{\I}{g}\int_{C}\left(\frac{c}{2} \,\psi^A\wedge\psi_A-\frac{b}{2}\,\phi^A\wedge\phi_A-a\,\psi^A\wedge\phi_A\right)=\nonumber\\
    &\equiv\frac{\I}{2g}\int_{C}\left(\psi^A\wedge\delta_\beta\phi_A-\delta_\beta\psi^A\wedge\phi_A\right)=\nonumber\\
    &=\frac{1}{2g}\left(q_{\psi}[\delta_\beta\phi]+q_{\phi}[\delta_\beta\psi]\right).\label{slcb-charge-as-shift-topo}
\end{align}
While not obvious from the action \eqref{action-topo}, this symmetry can be  guessed from the connection of this model to gravity. It becomes manifest that these transformations generate a dual $\SL(2,\C)$ symmetry, when considering its Poisson algebra below in \hyperref[section:gravityCPS]{Section \ref{section:gravityCPS}}.

\paragraph{$\SLC$ global symmetry}
The action \eqref{action-topo} is also invariant under a \textit{global} $\SLC$ symmetry. If $\alpha\in\slc$, the infinitesimal $\SLC$ action is given by
\begin{subequations}
\label{slc-on-top}
\begin{align}
    &\delta_\alpha\psi_A=\alpha_A{}^B\psi_B,\\
    &\delta_\alpha\phi_A=\alpha_A{}^B\phi_B,\\
    &\delta_\alpha\pis_A=\alpha_A{}^B\pis_B,\\
    &\delta_\alpha\pif_A=\alpha_A{}^B\pif_B.
\end{align}
\end{subequations}
Global symmetries are not typically associated with corner charges. Here, although the symmetry is global, its generator  is localized  at the corner on shell:
\begin{equation}
    H^{\sl}[\alpha]=\I\alpha^{AB}\int_\Sigma(\pis_A\wedge\psi_B+\pif_A\wedge\phi_B)\approx- \I g^{-1}\alpha^{AB}\int_{C} \psi_{(A}\wedge\phi_{B)}\equiv q^\sl[\alpha].\label{SLCgen-top}
\end{equation}
We will also show that the Poisson algebra of such charges is isomorphic to $\slc$.

\paragraph{Poisson algebra} In what follows, we determine the Poisson algebra of the different momentum maps using the identity
\begin{equation}
    \{H_{\xi_1},H_{\xi_2}\}=\bbvar{I}_{\xi_2}\bbvar{I}_{\xi_1}\Omega.\label{genPBfromOm}
\end{equation}
 The generators associated to the shift and the exact shifts (\ref{gen-topo-shift}, \ref{gen-topo-exshift})  Poisson commute off-shell up to central extensions
 \begin{subequations}
\begin{align}
    &\{H_{\phi}[\eta],H_\pif[\mu]\}=\I\int_{C}\mu_A \d\eta^A,\\
    &\{H_{\psi}[\zeta],H_\pis[\lambda]\}=\I\int_{C}\lambda^A \d\zeta_A,\\
    &\{H_{\phi}[\eta],H_{\psi}[\zeta]\}=\I \,g\int_{C}\eta^A\wedge\zeta_A.
\end{align}\end{subequations}
Considering now the Poisson algebra associated with the diffeomorphisms generated by  $\xi,\chi|_\Sigma\in T\Sigma$ and $\xi,\chi|_{C}\in T{C}$, we recover the expected algebra, 
\begin{equation}
    \{H^{di\!f\!\!f}[\chi],H^{di\!f\!\!f}[\xi]\}=H^{di\!f\!\!f}\big[[\xi,\chi]\big],
\end{equation}
where $[\cdot,\cdot]$ is the Lie bracket of vector fields. We can also determine the Poisson algebra mixing the different sectors, namely the shift and d-exact shift symmetries and diffeomorphisms
\begin{subequations}
\begin{align}
    \{H_{\phi}[\eta],H^{di\!f\!\!f}[\xi]\}&=H_{\phi}[L_\xi\eta],&&\{H_\pis[\lambda],H^{di\!f\!\!f}[\xi]\}=H_\pis[L_\xi\lambda],\\
    \{H_{\psi}[\zeta],H^{di\!f\!\!f}[\xi]\}&=H_{\psi}[L_\xi\zeta]&&\{H_\pif[\mu],H^{di\!f\!\!f}[\xi]\}=H_\pif[L_\xi\mu].
\end{align}\end{subequations}
We have recovered, in this way, the standard Poisson algebra of the topological theory.

\medskip

A direct calculation shows that the Poisson algebra of the  field-dependent shift generators \eqref{slcb-gen-topo} associated with what we called $\SLCB$,  is indeed  the $\slc$ algebra:
\begin{equation}\label{Hslcb}
    \{H^{\overline{\sl}}[\beta_1],H^{\overline{\sl}}[\beta_2]\}=-H^{\overline{\sl}}\big[[\beta_1,\beta_2]\big]
\end{equation}
with its parameters identified with traceless matrices as
\begin{equation}
    \{\beta_{A'}{}^{B'}\}=\begin{pmatrix}
        a&b\\
        c&-a
    \end{pmatrix},\label{beta-slcb}
\end{equation}
and
\begin{equation}
    \downup{[\beta_1,\beta_2]}{A'}{B'}=\downup{\beta_1{}}{A'}{C'}\downup{\beta_2{}}{C'}{B'}-\downup{\beta_2{}}{A'}{C'}\downup{\beta_1{}}{C'}{B'}
\end{equation}
is the $\slc$ Lie bracket.

The global $\SLC$ generators \eqref{SLCgen-top} form a closed Poisson algebra too,
\begin{equation}\label{Hslc}
    \{H^{\sl}[\alpha_1],H^{\sl}[\alpha_2]\}=-H^{\sl}\big[[\alpha_1,\alpha_2]\big].
\end{equation}

Let $(\alpha\eta)_A\equiv\alpha_A{}^B\eta_B$ denote an action of an $\slc$ algebra element $\alpha$ on a spinor $\eta$. If $\xi|_{\Sigma}\in T\Sigma$ and $\xi|_{C}\in T{C}$,
\begin{subequations}
\begin{align}
    &\{H_{\phi}[\eta],H^{\sl}[\alpha]\}=H_{\phi}[\alpha\eta],\qquad\{H_\pis[\lambda],H^{\sl}[\alpha]\}=H_\pis[\alpha\lambda],\\
    &\{H_{\psi}[\zeta],H^{\sl}[\alpha]\}=H_{\psi}[\alpha\zeta],\qquad\{H_\pif[\mu],H^{\sl}[\alpha]\}=H_\pif[\alpha\mu],\\
    &\{H^{di\!f\!\!f}[\xi],H^{\sl}[\alpha]\}=0.
\end{align}\end{subequations}

Thus, while the $\SLC$ symmetry of the action \eqref{action-topo} is global, it behaves like a  generator of a gauge transformation. We are going to use this insight to promote the global symmetry into a local gauge symmetry, thereby transforming the topological theory into self-dual gravity. Throughout this section, we assumed $g\neq 0$. For the study of the $g=0$ case, see \hyperref[app:g0-topo]{Section \ref{app:g0-topo}} below.

\section{Gravity and self-dual gravity from the topological theory}\label{section:gravityCPS}

\subsection{Action}\label{section:grav-action}

In this section, we are going to show why the topological theory \eqref{action-topo} is our starting point for studying gravity. Namely, we will argue that gravity can be formulated as this topological theory with a local $\SLC$  symmetry imposed. To do so, we demand that the $\SLC$  bulk generator $\eqref{SLCgen-top}$ vanishes on-shell by adding a Lagrange-multiplier term to the action \eqref{action-topo}. The Lagrange multiplier 1-form $A_{AB}$ can be interpreted as a self-dual $\SLC$ connection, which renders the action locally $\SLC$-invariant via the introduction of covariant derivatives:
\begin{subequations}
\begin{align}
    S&=\I\int_M \left(\pis_{ A}\wedge \d\psi^{A}+\pif_{ A}\wedge \d\phi^{A}+g\pis_{ A}\wedge\pif^{A}+A^{AB}\wedge(\pis_{ A}\wedge \psi_{B}+\pif_{ A}\wedge \phi_{B})\right)\equiv\\
    &\equiv \I\int_M \left(\pis_{ A}\wedge \nabla\psi^{A}+\pif_{ A}\wedge \nabla\phi^{A}+g\pis_{ A}\wedge\pif^{A}\right),\label{action-Robinson}
\end{align}\end{subequations}
where $\nabla=\d+A$ acts on spinor-valued $p$-forms as
\begin{equation}
    \nabla\psi_A=\d\psi_A+A_A^{\,\,\,\,B}\wedge\psi_B.
\end{equation}
The action \eqref{action-Robinson} is the same as the one proposed by Robinson in \cite{Robinson_1996} up to the relative coupling $g$ in the last term, which we will explain to be related to Newton's constant.

It is also useful to define the curvature of such a connection\footnote{Note that we are using the convention of \cite{Capovilla:1991qb,Ashtekar:1987qx}.}
\begin{equation}
    F_{AB}=\d A_{AB}+A_A{}^C\wedge A_{CB}.
\end{equation}
Note that while both $A_A{}^B$ and $F_A{}^B$ are $\slc$-valued, their all-upper- and all-lower-index counterparts are symmetric matrices obtained through a contraction with the $\epsilon$-tensor:
\begin{equation}
    A_{AB}=A_A{}^C\epsilon_{CB}=A_{(AB)},\qquad A^{AB}=\epsilon^{AC}A_C{}^B=A^{(AB)},\quad \text{etc.}
\end{equation}
This property is extensively used in the calculations below.

\medskip

Let us now step back for a moment and show the relationship between this action and gravity. Let the capital Latin letters from the second half of the alphabet $I,J,...$ label internal Lorentz four-vector components and take values in $\{0,...,3\}$. Following the conventions from \cite{Krasnov:2020lku}, we use an anti-hermitian basis\footnote{Here, $\sigma^i_{AA'}$ corresponds to an $i$-th Pauli matrix and $\sigma^0_{AA'}=\mathbf{1}_{AA'}$ is the identity. Objects with a different placement of indices can be obtained from these via multiplication by an $\epsilon$-tensor and contraction with internal Minkowski metric $\eta_{IJ}$. The choice of an anti-hermitian basis corresponds to the $(-,+,+,+)$ metric signature.} $\{\I\sigma^I_{AA'}\}$ to convert internal four-vectors into $2\times2$ complex matrices via 
\begin{equation}
    x_{AA'}=\frac{\I}{\sqrt{2}}\sigma_{AA'}{}^Ix_I.\label{spinor-from-lor-vector}
\end{equation}
An image of such a map is an anti-hermitian matrix \textit{if and only if} the vector $x^I$ is real. The inverse map is, similarly,
\begin{equation}
    x^I=\frac{\I}{\sqrt{2}}\sigma_{AA'}{}^Ix^{AA'}.\label{lor-vector-from-spinor}
\end{equation}
Using this correspondence, we replace a tetrad 1-form $\updown{e}{I}{\mu}$ with a \textit{soldering form} $\updown{e}{AA'}{\mu}$. 

The same map can be used to relate the connection $A_A{}^B$ to the Lorentz connection 1-form $\omega_I{}^J$. We first define
\begin{equation}
    \omega_{AA'BB'}=-\frac12\downup{\sigma}{AA'}{I}\downup{\sigma}{BB'}{J}\omega_{IJ}.
\end{equation}
This spinor-valued 1-form can be split into its self-dual and anti-self-dual parts:
\begin{equation}
\omega_{AA'BB'}=A_{AB}\epsilon_{A'B'}+\tilde{A}_{A'B'}\epsilon_{AB}.
\end{equation}
The partial traces $A_{AB}=\frac{1}{2}\omega_{AA'BB'}\epsilon^{A'B'}$ and $\tilde{A}_{A'B'}=\frac{1}{2}\omega_{AA'BB'}\epsilon^{AB}$ are then called a self-dual and an anti-self-dual $\slc$ connection respectively\footnote{Our naming convention matches those of e.g.\  \cite{Capovilla:1991qb}}. An analogous relation holds between the curvature $F^{IJ}=\d \omega^{IJ}+\omega^{I}{}_K\wedge\omega^{KJ}$ and the pair $F_A{}^B$ and $\tilde{F}_{A'}{}^{B'}$ with
\begin{subequations}
\begin{equation}
F_{A}{}^{B}=\d A_{A}{}^{B}+A_{A}{}^{C}\wedge A_{C}{}^{B}, \qquad
    \tilde{F}_{A'}{}^{B'}=\d\tilde{A}_{A'}{}^{B'}+\tilde{A}_{A'}{}^{C'}\wedge\tilde{A}_{C'}{}^{B'},
\end{equation}
\begin{equation}
    F_{AA'BB'}=F_{AB}\epsilon_{A'B'}+\tilde{F}_{A'B'}\epsilon_{AB}.
\end{equation}\end{subequations}
The Einstein--Cartan action can be rewritten as an $\SLC$ gauge theory with complex variables:
\begin{equation}\label{EC action}
    S=\frac{1}{16\pi G}\int_M \epsilon_{IJKL}e^I\wedge e^J\wedge F^{KL}_\omega=-\frac{\I}{16\pi G}\int_M e_{AA'}\wedge e_{BB'}\wedge (F^{AB}\epsilon^{A'B'}-\Tilde{F}^{A'B'}\epsilon^{AB}).
\end{equation}
The two terms in the right-hand side correspond to the self-dual and the anti-self-dual sectors of gravity. In this formulation, the primed spinors transform in a complex conjugate representation to that of the unprimed ones.

Equation \eqref{spinor-from-lor-vector} implies that the soldering forms must be anti-hermitian matrices for the metric tensor to be real. In the same way, the reality of the connection 1-form $\updown{\omega}{I}{J}$ implies that the self-dual and anti-self-dual connections $A$ and $\tA$ must be complex conjugate of each other \cite{Plebanski:1977zz}. These conditions guarantee that the two terms in the r.h.s.\ of \eqref{EC action} are the complex conjugate of each other and that the action is real. Thus, we start with a complex phase space consisting of two disjoint sectors and restore the real gravitational degrees of freedom by imposing the reality conditions---the anti-hermicity of the soldering form---later.

\medskip

Let us introduce a basis $o_A$, $\iota_A$ in the spinor space, which satisfies
\begin{align}
    &o_A\iota^A=1, &&o_A o^A=0=\iota_A\iota^A, \label{dyad}
\end{align}
and their Hermitian conjugates $\bar{o}_{A'}$, $\bar{\iota}_{A'}$. The totally anti-symmetric tensor can be rewritten as
\begin{equation}
    \epsilon_{AB}=o_A\iota_B-\iota_A o_B.
\end{equation}
This suggests the identification of columns and rows of the soldering form in the action \eqref{EC action} as separate unprimed and prime spinor-valued 1-forms:
\begin{subequations}
\label{spinors-to-tetrads}
\begin{alignat}{5}
    \psi_A&=e_{AA'}\bar{\iota}^{A'},&\hspace{4em}&&\Tilde{\psi}_{A'}&=e_{AA'}\iota^{A},\\
    \phi_A&=-e_{AA'}\bar{o}^{A'},&&&\Tilde{\phi}_{A'} &=-e_{AA'}o^{A}.
\end{alignat}
\end{subequations}
This brings the action to the form
\begin{subequations}
\begin{align}
    S&=-\I g^\ii\int (\psi_A\wedge \phi_B\wedge F^{AB}-\Tilde{\psi}_{A'}\wedge \Tilde{\phi}_{B'}\wedge \Tilde{F}^{A'B'})\label{SelfDualActionInSpinors}\\
    &=\I g^\ii\int (\psi^A\wedge \nabla^2\phi_A-\Tilde{\psi}^{A'}\wedge \tilde{\nabla}^2\Tilde{\phi}_{A'}),
\end{align}\end{subequations}
where we  define $g$ in terms of Newton's constant as
\begin{equation}
    g=8\pi G.
\end{equation}
Integration by part and addition of the boundary action
\begin{equation}
    S_\partial=\I g^\ii\int_{\partial M}(\psi^A\wedge \nabla\phi_A-\Tilde{\psi}^{A'}\wedge \Tilde{\nabla}\Tilde{\phi}_{A'})\label{action-bdry-term}
\end{equation}
result in the action proposed by Tung and Jacobson \cite{Tung:1995cj},
\begin{equation}
    S=ig^\ii\int _M(\nabla\psi^A\wedge \nabla\phi_A-\Tilde{\nabla}\Tilde{\psi}^{A'}\wedge \Tilde{\nabla}\Tilde{\phi}_{A'}).\label{DiracLikeAction}
\end{equation}
The first-order action for such a theory can be split into an anti-self-dual and self-dual parts:
\begin{equation}
    S=S_{SD}+S_{ASD}.\label{action-full-dual-anti-dual}
\end{equation}
Let $\tpis,\tpif \in\C^2\otimes\Omega^2(M)$, similarly to the unprimed sector. The two parts of the action take the form
\begin{subequations}
\begin{align}
    S_{\text{SD}}&=\I\,g^\ii\int_M (\pis_A\wedge \nabla\psi^{A}+\pif_A\wedge \nabla\phi^{A}+\pis_A\wedge\pif^{A}),\label{action-full-SD}\\
    S_{\text{ASD}}&=-\I\, g^\ii\int_M(\Tilde{\pi}_{A'}\wedge \Tilde{\nabla}\Tilde{\psi}^{A'}+\Tilde{\rho}_{A'}\wedge \Tilde{\nabla}\Tilde{\phi}^{A'}+\Tilde{\pi}_{A'}\wedge\Tilde{\rho}^{A'}).\label{action-full-ASD}
\end{align}\end{subequations}
To obtain the action \eqref{action-Robinson}, we simply rescale the variables as $\psi\rightarrow\psi$, $\phi\rightarrow\phi$, $\pi\rightarrow g\pi$, $\rho\rightarrow g\rho$. For the dual sector, the analogous rescaling gives:
\begin{align}
    S_{\text{ASD}}&=-\I\int_M(\Tilde{\pi}_{A'}\wedge \Tilde{\nabla}\Tilde{\psi}^{A'}+\Tilde{\rho}_{A'}\wedge \Tilde{\nabla}\Tilde{\phi}^{A'}+g\,\Tilde{\pi}_{A'}\wedge\Tilde{\rho}^{A'}).\label{action-full-anti-dual}
\end{align}

\subsection{Equations of motion and  phase space}

We begin by studying the self-dual sector \eqref{action-Robinson}, from which Robinson's action can be obtained as a particular case in which  $g=1$. In what follows, we assume $g\neq 0$ and dedicate Section \ref{app:g0-topo} to the special case of $g=0$. From the variation of action, we get the following equations of motion:
\begin{subequations}
\begin{align}
    \psiEE{}_A&=\I\,\nabla \pis_A\approx0,\label{2-flatness-pi-psi}\\
    \phiEE{}_A&=\I\,\nabla \pif_A\approx0,\label{2-flatness-pi-phi}\\
    \psiFF{}_A&=\I\,(g\pis_A-\nabla\phi_A)\approx0,\label{fake-flatness-pi-psi}\\
    \phiFF{}_A&=\I\,(g\pif_A+\nabla\psi_A)\approx0,\label{fake-flatness-pi-phi}\\
    \cG_{AB}&=\I\,(\pis_{(A}\wedge\psi_{B)}+\pif_{(A}\wedge\phi_{B)})\approx0.\label{GaussConstr}
\end{align}\end{subequations}
The constraints \eqref{2-flatness-pi-psi}--\eqref{2-flatness-pi-phi} and \eqref{fake-flatness-pi-psi}--\eqref{fake-flatness-pi-phi} constitute $\SLC$-covariant versions of the 2-flatness and fake-flatness conditions, respectively, while the latter \eqref{GaussConstr} will be called Gauss constraint following common terminology of theories with internal Lorentz gauge symmetries. The 2-flatness constraints are also called Einstein equations in this section since this is what they correspond to in the Einstein--Cartan theory on shell of fake-flatness and Gauss constraints:
\begin{subequations}
\begin{align}
    &\psiEE{}_A\approx \I\,g^\ii\nabla^2\phi_A=\I\, g^\ii F_A{}^{B}\wedge\phi_B,\label{EE1}\\
    &\phiEE{}_A\approx -\I\,g^\ii\nabla^2\psi_A=-\I \,g^\ii F_A{}^{B}\wedge\psi_B.\label{EE2}
\end{align}
\end{subequations}
We emphasize that the Einstein equations have now their own potentials: they are  expressed as total exterior derivatives of the momenta $\pi_A$ and $\rho_A$. \smallskip

Next, we consider the covariant phase space formalism. The symplectic potential
\begin{equation}
    \Theta_\Sigma=\I\,\int_{\Sigma}\left(\pis_A\wedge \bbvar{d}\psi^A+\pif_A\wedge \bbvar{d}\phi^A\right),\label{SymPotSD}
\end{equation}
implies that the gravitational theory has the same kinematical phase space as the topological theory:
\begin{equation}
    \Omega_\Sigma=\I\int_{\Sigma}\left(\bbvar{d}\pis_A\wedge \bbvar{d}\psi^A+\bbvar{d}\pif_A\wedge \bbvar{d}\phi^A\right).\label{SymFormSD}
\end{equation}
A similar phase space in the context of gravity has been derived in \cite{Freidel:2019ofr,Wieland:2017zkf}.

The original pre-symplectic 2-form of the self-dual-gravity phase space  can be recovered up to a corner term from \eqref{SymFormSD} by going on-shell of the fake-flatness constraint. Indeed,
\begin{equation}
    \Omega\approx-\frac{\I}{g}\int_{\Sigma}\bbvar{d} A^{AB}\wedge\bbvar{d}(\phi_A\wedge\psi_B)+\frac{\I}{g}\int_{C}\bbvar{d}\psi^A\wedge\bbvar{d}\phi_A.\label{SDphasespace-relation}
\end{equation}
At this stage, the causal nature of $\Sigma$ is unspecified. It might be either spacelike, timelike or null. In \hyperref[sec-canonical]{Section \ref{sec-canonical}}, we will restrict ourselves to the spacelike case. The case in which $\Sigma$ is null implies further conditions on $\phi_A$ and $\psi_A$ at the boundary. If we align the spin dyad $(o^A,\iota^A)$ to the geometry of the null boundary, these conditions simplify. They  impose that the pull-back of either $\phi_Ao^A$ or $\psi_A\iota^{A}$ to the boundary vanishes, depending on whether the null normal to the hypersurface is given by $\I\,o_A\bar{o}_{A'}$ or $\I\,\iota_A\bar{\iota}_{A'}$. The resulting bulk plus boundary phase space can be then matched to earlier results developed in \cite{Wieland:2017cmf,Wieland:2021vef,Wieland:2017zkf}.  

Note that \eqref{SDphasespace-relation} is not the only possible parametrization of the on-shell phase space. The fake-flatness equations (\ref{fake-flatness-pi-psi}, \ref{fake-flatness-pi-phi}) can be equally well solved for connection $A$ in terms of the phase-space variables:
\begin{equation}
    A^{AB}_\alpha=\psi^C_\alpha\phi^{\mu}_C \psi^{\nu (A}(\pi_{\mu\nu}^{B)}-\partial_\mu\phi_\nu^{B)}+\partial_\nu\phi_\mu^{B)})-\phi^C_\alpha\psi^{\mu}_C \phi^{\nu (A}(\rho_{\mu\nu}^{B)}+\partial_\mu\psi_\nu^{B)}-\partial_\nu\psi_\mu^{B)}),\label{Asoln}
\end{equation}
where we introduced the contravariant components of spinors $\psi^\mu$ and $\phi^\mu$. They correspond to the frame field through the same relations as \eqref{spinors-to-tetrads} and satisfy the following orthogonality relations:
\begin{subequations}
\begin{align}
    &\psi^A_\mu\psi^{\mu B}=0, &&2\psi^A_{(\mu}\psi_{\nu) A}=0,\\
    &\phi^A_\mu\phi^{\mu B}=0, &&2\phi^A_{(\mu}\phi_{\nu) A}=0,\\
    &\psi^A_\mu\phi^{\mu B}=\epsilon^{BA}, &&2\psi^A_{(\mu}\phi_{\nu) A}=g_{\mu\nu},\label{metricinspinors}
\end{align}\end{subequations}
which, in particular, imply that the Weyl spinor--valued 4-vectors $\psi_A^\mu$, $\phi_A^\mu$ are null.

Thus, the connection $A_{AB}$ and area variables $\Sigma_{AB}=\phi_{(A}\wedge\psi_{B)}$ of the self-dual formulation do not have to be treated as on-shell fundamental degrees of freedom. Instead, an on-shell theory can be written in terms of the line-element data, represented by $\psi_A$, $\phi_A$, and torsion-like data $\pi_A$, $\rho_A$. This choice would make our on-shell theory more akin to the teleparallel gravitation \cite{Aldrovandi:2013wha,Tung:1998dw}. 

\subsection{Symmetries, charges and generator algebra}

Let $\eta,\zeta\in\C^2\otimes\Omega^1(M)$, $\lambda,\mu\in\C^2\otimes\Omega^0(M)$ and $\alpha\in\sl(2,\C)\otimes \Omega^0(M)$. We introduce notation for the equations of motion of the self-dual sector of gravity \eqref{2-flatness-pi-psi}--\eqref{GaussConstr} smeared with corresponding test functions in full analogy with the topological case:
\begin{subequations}
\label{mommaps-GR}
\begin{align}
    \psiFF{}_\Sigma[\eta]&=\int_\Sigma \psiFF{}_A\wedge\eta^A\equiv \I\int_\Sigma (g\pis_A-\nabla\phi_A)\wedge\eta^A,\label{mommapFFpsi}\\
    \phiFF{}_\Sigma[\zeta]&=\int_\Sigma \phiFF{}_A\wedge\zeta^A\equiv \I\int_\Sigma (g\pif_A+\nabla\psi_A)\wedge\zeta^A,\label{mommapFFphi}\\
    \psiEE{}_\Sigma[\lambda]&=\int_\Sigma \psiEE{}^A\lambda_A\equiv \I\int_\Sigma \nabla\pis^A\lambda_A,\\
    \phiEE{}_\Sigma[\mu]&=\int_\Sigma \phiEE{}^A\mu_A\equiv \I\int_\Sigma \nabla\pif^A\mu_A,\label{mommapEEphi}\\
    \cG_\Sigma[\alpha]&=\int_\Sigma\cG_{AB}\alpha^{AB}\equiv \I\int_\Sigma(\pis_A\wedge\psi_B+\pif_A\wedge\phi_B)\alpha^{AB}.\label{SLCgen}
\end{align}
\end{subequations}
The gravitational action \eqref{action-Robinson} possesses three types of gauge symmetries: internal $\SLC$ and $\SLCB$ transformations and four-dimensional diffeomorphisms. We will perform the covariant phase space analysis for the bulk phase space, equipped with the pre-symplectic 2-form \eqref{SymFormSD} first. The corner phase space contribution will be discussed separately in \hyperref[sec:corner]{Section \ref{sec:corner}}.

\paragraph{$\SL(2,\C)$ gauge symmetry}
The $\SL(2,\C)$ gauge transformations act on the spinor-valued fields infinitesimally as
\begin{subequations}
\begin{align}
    &\delta^\sl_\alpha\psi_A=\alpha_A^{\,\,\,\,B}\psi_B,\label{sl2conPsi}\\
    &\delta^\sl_\alpha\phi_A=\alpha_A^{\,\,\,\,B}\phi_B,\\
    &\delta^\sl_\alpha\pis_A=\alpha_A^{\,\,\,\,B}\pis_B,\\
    &\delta^\sl_\alpha\pif_A=\alpha_A^{\,\,\,\,B}\pif_B,\\
    &\delta^\sl_\alpha A_B^{\,\,\,\,C}=-\nabla\alpha_B^{\,\,\,\,C}.\label{sl2conA}
\end{align}
\end{subequations}
Such transformations leave the action \eqref{action-Robinson} invariant. Its generator is given by the Gauss constraint \eqref{SLCgen} and becomes trivial on-shell:
\begin{align}
    &\bbvar{I}_\alpha\Omega=-\bbvar{d} H^{\sl}[\alpha], && H^{\sl}[\alpha]=\cG[\alpha].\label{slc-gen-gr}
\end{align}

\paragraph{$\SLCB$ gauge symmetry}
The  $\SLCB$ gauge symmetry is parametrized  by three complex-valued scalar fields $a,b,c$ on $M$, denoted collectively by $\beta$. The infinitesimal action of this symmetry on the field space is given by
\begin{subequations}
\begin{align}
    &\delta_\beta\psi_A= a\psi_A+b\phi_A,\label{dualsl2conpsi}\\
    &\delta_\beta\phi_A=c\psi_A-a\phi_A,\label{dualsl2conphi}\\
    &\delta_\beta\pi_A^\psi= -c\pif_A-a\pis_A+\frac1g \d c\wedge\psi_A-\frac1g \d a\wedge\phi_A,\label{dualsl2conpipsi}\\
    &\delta_\beta\pi_A^\phi= a\pif_A-b\pis_A-\frac1g \d a\wedge\psi_A-\frac1g \d b\wedge\phi_A,\label{dualsl2conpiphi}\\
    &\delta_\beta A_A{}^B=0.
\end{align}
\end{subequations}
The infinitesimal generator is integrable and given by a linear combination of fake-flatness constraints. The situation mirrors the topological case considered in \eqref{slcb-gen-topo}. We obtain,
\begin{subequations}
\begin{align}
    \bbvar{I}_\beta\Omega&=-\bbvar{d} H^{\overline{\sl}}[\beta],\\
    H^{\overline{\sl}}[\beta]&=\frac1g\psiFF{}[\delta_\beta\psi]+\frac1g\phiFF{}[\delta_\beta\phi]+q^{\overline{\sl}}[\beta],\label{slcb-gen-gr}\\
    q^{\overline{\sl}}[\beta]&=\frac{\I}{g}\int_{C}\left(\frac{c}{2}\,\psi^A\wedge\psi_A-\frac{b}{2}\,\phi^A\wedge\phi_A-a\,\psi^A\wedge\phi_A\right).\label{cornercurrent1}
\end{align}
\end{subequations}
The emergence of the $\SLCB$ gauge transformations is easy to justify. Recall that the spinors $\psi_A$ and $\phi_A$ are defined as components of $e_{AA'}$, which transform infinitesimally under $\SLCB$ as
\begin{equation}
    e_{AA'}\rightarrow \beta_{A'}{}^{B'}e_{AB'},\qquad \beta\in\slcb.
\end{equation}
Denoting $\beta_{0'}{}^{0'}=a=-\beta_{1'}{}^{1'}$, $\beta_{0'}{}^{1'}=b$, $\beta_{1'}{}^{0'}=c$, one arrives at the expressions \eqref{dualsl2conpsi}--\eqref{dualsl2conphi}. Equations \eqref{dualsl2conpipsi} and \eqref{dualsl2conpiphi} are {then consistent} with  the equations of motion \eqref{fake-flatness-pi-psi}--\eqref{fake-flatness-pi-phi}. The reader might wonder why we do not use the straightforward transformation $g\,\delta_\beta\pis=\nabla(\delta_\beta\phi)$ and $g\,\delta_\beta\pif=-\nabla(\delta_\beta\psi)$ instead. The reason is, however, that the Hamiltonian flow equation for this transformation is not integrable off-shell. 

\paragraph{Diffeomorphisms}
The diffeomorphisms act on fields by Lie derivation $\delta^{di\!f\!\!f}_\xi\equiv L_\xi$ and leave the action invariant up to the boundary term,
\begin{align}
    R[\xi]&=\iota_\xi L.
\end{align}
The generator computed via the Noether theorem for an arbitrary vector field $\xi\in \Gamma(TM)$ is given by
\begin{equation}
    H^{di\!f\!\!f}[\xi]\overset{\eqref{Noether1}}{=}\diff[\xi]+q^{di\!f\!\!f}[\xi],\label{diff-gen-gr}
\end{equation}
where the off-shell diffeomorphism generator includes each of the functionals \eqref{mommapFFpsi}--\eqref{SLCgen} with field-dependent parameters, to compare with the topological case \eqref{dif-gen-top}:
\begin{equation}
    \diff[\xi]=\psiEE{}[\iota_\xi\psi]+\phiEE{}[\iota_\xi\phi]-\psiFF{}[\iota_\xi\pif]+\phiFF{}[\iota_\xi\pis]-\cG[\iota_\xi A].\label{dif-gen-gr}
\end{equation}
The corner charge is the same as that in the topological theory:
\begin{equation}
    q^{di\!f\!\!f}[\xi]=\I\int_{C}(\iota_\xi \psi^A\pis_A +\iota_\xi \phi^A\pif_A )\approx \frac{\I}{g}\int_{C}\psi^A\wedge\cL_\xi\phi_A,\label{dif-charge-gr}
\end{equation}
where $\cL_\xi=\nabla \iota_\xi+\iota_\xi\nabla$ is the \textit{covariant} Lie derivative.  However, similarly to the topological case, the Hamiltonian flow equation is integrable only if $\xi|_\Sigma\in T\Sigma$ and $\xi|_{C}\in T{C}$. It then takes the form
\begin{equation}
    \bbvar{I}_\xi\Omega=-\bbvar{d} H^{di\!f\!\!f}[\xi],
\end{equation}
where the generator $H^{di\!f\!\!f}[\xi]$ is defined in \eqref{diff-gen-gr}. On-shell, this equation is satisfied if $\xi|_{C}\in T{C}$, and the generator simplifies to $q^{di\!f\!\!f}[\xi]$. 

\paragraph{Poisson algebra}
So far, we identified different momentum maps. The resulting Poisson algebra  is calculated using \eqref{genPBfromOm} as usual. Off-shell momentum maps are algebra anti-homomorphisms:
\begin{subequations}
\begin{align}
    &\{H^\sl[\alpha_1],H^\sl[\alpha_2]\}=-H^\sl[[\alpha_1,\alpha_2]],&&\{H^\sl[\alpha],H^{\overline{\sl}}[\beta]\}=0,\\
    &\{H^{\overline{\sl}}[\beta_1],H^{\overline{\sl}}[\beta_2]\}=-H^{\overline{\sl}}[[\beta_1,\beta_2]],&& \{H^{di\!f\!\!f}[\xi],H^\sl[\alpha]\}=-H^\sl[L_\xi\alpha],\\
    &\{H^{di\!f\!\!f}[\xi],H^{di\!f\!\!f}[\chi]\}=-H^{di\!f\!\!f}[[\xi,\chi]],&& \{H^{di\!f\!\!f}[\xi],H^{\overline{\sl}}[\beta]\}=-H^{\overline{\sl}}[L_\xi\beta].
\end{align}
\label{1st-class-off-shell-algebra}
\end{subequations}
Closure conditions are identical to those for integrability: $\xi,\chi|_\Sigma\in T\Sigma$ and $\xi,\chi|_{C}\in T{C}$. 

\medskip

\paragraph{Shift generators} 

Recall that in the topological theory we were able to decompose the $\SLCB$ and diffeomorphism corner charges in terms of the shift charges (see (\ref{slcb-charge-as-shift-topo}, \ref{q-diff-thru-shifts})). Moreover, the same held true for off-shell generators (\ref{dif-gen-top}, \ref{slcb-gen-topo}).  That diffeomorphisms are on shell equivalent to a combination of field-dependent shift and Lorentz transformations has been long known in the context of three-dimensional gravity \cite{Achucarro:1986uwr,Witten:1988hc}. More recent developments extended this statement to gravity in four dimensions \cite{Montesinos:2017epa,Freidel:2019ofr,Langenscheidt:2025lha}. The interest in this is driven, in particular, by an observation that shifts are more convenient for discretization and quantization \cite{ Freidel_2003, Dittrich:2008pw, Baratin_2011}. This viewpoint is especially natural in formulations of gravity as deformed topological theories, of which our theory is an example. Then, diffeomorphisms represent the subset of shifts left unbroken by the deformation. In our formulation, this is also the case with the $\SLCB$ gauge.

While the meaning of the decompositions (\ref{dif-gen-top}, \ref{slcb-gen-topo}) is obvious in the topological theory, the meaning of individual terms in the expressions (\ref{slcb-gen-gr}, \ref{dif-gen-gr}) of $\SLCB$ and diffeomorphism generators in self-dual gravity has not been yet discussed. As one might infer by analogy, the smeared equations of motion on $\Sigma$ \eqref{mommapFFpsi}--\eqref{mommapEEphi} generate shift transformations of phase-space variables, with the only difference being that $\d$-exact shifts become $\nabla$-exact. However, shifts are not a symmetry of the gravitational action (with the exception of diffeomorphisms and $\SLCB$ transformations), and their generators require a phase-space extension   to be integrable. In the remainder of this section, we provide such an extension, which then allows us to introduce shift generators \eqref{mommapFFpsi}--\eqref{mommapFFphi} as phase-space symmetry generators in the interior of $\Sigma$ and justify an interpretation of $\SLCB$ and diffeomorphism transformations as field-dependent shifts from the intrinsically gravitational point of view. At the level of the corner charges, this interpretation translates into the decompositions (\ref{q-diff-thru-shifts}, \ref{slcb-charge-as-shift-topo}).

Let $\Pi_{A}{}^B$ be an $\slc$-valued 2-form. Consider an extended symplectic form
\begin{equation}
    \Omega^{ext}_\Sigma=\I\int_{\Sigma}\left(\bbvar{d}\pis_A\wedge \bbvar{d}\psi^A+\bbvar{d}\pif_A\wedge \bbvar{d}\phi^A+\bbvar{d}\Pi_{AB}\wedge\bbvar{d} A^{AB}\right).\label{SymFormSDext}
\end{equation}
$\Pi_{AB}$ plays the role of the conjugate momentum to the connection $A^{AB}$. In order not to affect the physical phase space, we require that $\Pi_{AB}$ vanishes on shell. This phase space corresponds to the action \eqref{action-Robinson} extended by a term 
\begin{equation}
    \I\int_M\Pi_{AB} \wedge(F^{AB}-\gamma^{AB}). 
\end{equation}
A similar extension will be discussed in the context of canonical analysis in \hyperref[sec-canonical]{Section \ref{sec-canonical}}. The extra equations of motion are given by
\begin{align}
    \cG_{AB}+\nabla\Pi_{AB}=0, \qquad F_{AB}=\gamma_{AB}, \qquad \Pi_{AB}=0.
\end{align}
We now move on to the discussion of symmetries. Let $\eta_A, \zeta_A\in\C^2\otimes \Omega^1(M)$ and $\lambda_A,\mu_A\in\C^2\otimes \Omega^0(M)$. Consider the shift transformation
 \begin{subequations}
\begin{alignat}{5}
    \delta_{\eta}\psi_A&=g\,\eta_A, &\hspace{4em}\delta_{\eta} \pis_A&=0,\\
    \delta_{\eta}\phi_A&=0, &\delta_{\eta} \pif_A&=-\nabla\eta_A,\\
    \delta_\eta A_{AB}&=0, &\delta_\eta\Pi_{AB}&=-\phi_{(A}\wedge\eta_{B)}.
\end{alignat}
 \end{subequations}
The Hamiltonian flow equation for such a transformation is integrable off-shell:
\begin{alignat}{5}
    \bbvar{I}_{\eta}\Omega&= -\bbvar{d} \cH_{\phi}[\eta],&\hspace{2.5em}&\cH_{\phi}[\eta]&=\psiFF[\eta]+ q_{\phi}[\eta],\label{gen-shift-psi}
\end{alignat}
where $\psiFF[\eta]$ is defined in \eqref{mommapFFpsi}, and $q_{\phi}[\eta]$ is the same as in the topological case, \eqref{shift-charge-topo-1}. Analogously, the shift transformation with non-trivial components
\begin{subequations}
\begin{alignat}{5}
    \delta_{\zeta}\phi_A&=g\,\zeta_A,\\
    \delta_{\zeta} \pis_A&=\nabla\zeta_A,\\
    \delta_\eta\Pi_{AB}&=\psi_{(A}\wedge\zeta_{B)}
\end{alignat}
\end{subequations}
is generated by 
\begin{equation}
    \cH_{\psi}[\zeta]=\phiFF[\zeta]+ q_{\psi}[\zeta].\label{gen-shift-phi}
\end{equation}
The covariantly-exact shifts with non-trivial components
\begin{alignat}{5}
\delta_{\lambda}\psi_A&=\nabla\lambda_A,&\hspace{2.5em}&\text{and}&\hspace{2.5em} \delta_{\mu}\phi_A&=\nabla\mu_A,\nonumber\\
    \delta_\lambda\Pi_{AB}&=-\lambda_{(A}\pis_{B)}&&\text{and}& \delta_\mu\Pi_{AB}&=-\mu_{(A}\pif_{B)},
\end{alignat}
are generated, respectively, by
\begin{subequations}\label{gen-gr-shift}
\begin{align}
&\cH_\pis[\lambda]=\psiEE[\lambda]+q_\pis[\lambda],\label{gen-exshift-psi}\\
    &\cH_\pif[\mu]=\phiEE[\mu]+ q_\pif[\mu].\label{gen-exshift-phi}
\end{align}
\end{subequations}We note that corner shift generators on the gravitational phase space coincide with the topological charges (\ref{shift-charges-topo}, \ref{ex-shift-charges-topo}). Thus, while $q_\phi$ and $q_\psi$ are not corner charges, they still appear in the theory of the self-dual gravity with an extended phase space as corner shift generators. In the next section, we will use this decomposition to show that the corner shift generators can be switched off with a certain corner phase-space extension.

\subsection{Cosmological constant}

There are three different ways to rewrite the cosmological constant contribution in the spinor variables,
\begin{subequations}
\begin{align}
    S_\Lambda&=-\frac{1}{16 \pi G}\int \frac\lambda{6}\epsilon_{IJKL}e^I\wedge e^J\wedge e^{K}\wedge e^L=\nonumber\\
    &=-\frac\lambda{12 g}\int \epsilon^{AA'BB'CC'DD'}e_{AA'}\wedge e_{BB'}\wedge e_{CC'}\wedge e_{DD'}=\nonumber\\
    &=-\frac{\I\Lambda}{2g}\int\psi_A\wedge\psi^A\wedge\phi_B\wedge\phi^B\label{cosmconst-allunrimed}\\
    &=-\frac{\I\Lambda}{4g}\int(\psi_A\wedge\psi^A\wedge\phi_B\wedge\phi^B-\tpsi_\Ap \wedge\tpsi^\Ap \wedge\tphi_\Bp\wedge\tphi^\Bp)\label{cosmconst-allmixed}\\
    &= \frac{\I\Lambda}{2g}\int\tpsi_\Ap \wedge\tpsi^\Ap \wedge\tphi_\Bp\wedge\tphi^\Bp.\label{cosmconst-primd}
\end{align}
\end{subequations}
It is interesting that the cosmological constant contribution can be distributed symmetrically between the self-dual and anti-self-dual sectors \eqref{cosmconst-allmixed}, or concentrated fully in one of them, see \eref{cosmconst-allunrimed} or \eref{cosmconst-primd}. All of these options will match upon imposing the reality conditions, studied in \href{sec:corner}{Section \ref{sec:corner}}. 

Assume the choice \eqref{cosmconst-allunrimed}. While fake-flatness and Gauss equations of motion are not modified, the Einstein equation acquires a cosmological-constant term,
\begin{subequations}\begin{align}
    \nabla\pis_A&+\frac{\Lambda}{g}\psi_{A}\wedge\phi_{B}\wedge\phi^B=0,\\
    \nabla\pif_A&+\frac{\Lambda}{g}\phi_{A}\wedge\psi_{B}\wedge\psi^B=0.
\end{align}\end{subequations}
This means that the following momentum maps are modified in the case $\Lambda\neq 0$:
\begin{subequations}\begin{align}
 \psiEE{}_\Lambda[\lambda]&=\I\int_\Sigma (\nabla\pis^A+\frac{\Lambda}{g}\psi^A\wedge\phi_B\wedge\phi^B)\lambda_A,\\
    \phiEE{}_\Lambda[\mu]&= \I\int_\Sigma (\nabla\pif^A+\frac{\Lambda}{g}\phi^A\wedge\psi_B\wedge\psi^B)\mu_A.
\end{align}\end{subequations}
The generator of diffeomorphisms, which is the only  generator containing $\cE$-momentum maps, is integrable on-shell if $\xi|_C\in TC$, and off-shell if also $\xi|_\Sigma\in T\Sigma$. If $\xi_\Sigma\in T\Sigma$, 
\begin{subequations}
\begin{align}
    H^{di\!f\!\!f}_{\Lambda}[\xi]&=\diff_\Lambda[\xi]+q^{di\!f\!\!f}[\xi],\\
    \diff_\Lambda[\xi]&=\psiEE{}_\Lambda[\iota_\xi\psi]+\phiEE{}_\Lambda[\iota_\xi\phi]-\psiFF[\iota_\xi\pif]+\phiFF[\iota_\xi\pis]-\cG[\iota_\xi A].
\end{align}
\end{subequations}
However, if $\xi$ has a time-like component, the on-shell generator does not vanish in the interior of $\Sigma$ but contains a volume-form contribution:
\begin{equation}
    H^{di\!f\!\!f}_{\Lambda}[\xi]\approx -\frac{\I\Lambda}{2g}\int_\Sigma\iota_\xi(\psi_A\wedge\psi^A\wedge\phi_B\wedge\phi^B)+q^{di\!f\!\!f}[\xi].
\end{equation}
If $\xi|_\Sigma\in T\Sigma$ and $\xi|_C\in TC$, the Poisson algebra of diffeomorphism generators is analogous to the $\Lambda=0$ case \eqref{1st-class-off-shell-algebra}.

\section{The corner phase space ambiguity}\label{sec:corner} 

In field theories, there are two major ambiguities in the definition of a symplectic potential. First, we can always add a total field-space variation, thus shifting $\Theta$ into $\Theta-\bbvar{d}F$, where $F$ is a generating functional, corresponding to a canonical transformation of the canonical phase space coordinates. At the classical level, such canonical transformations do not change the physics. At the quantum level they may, because not every canonical transformation is also represented as a unitary map on the Hilbert space \cite{Haag:1963dh}. In other words, different choices of canonical variables can correspond to unitarily inequivalent quantum theories. Second, the variation of the action fixes the pre-symplectic potential only up to a corner term---the symplectic current \eqref{SDphasespace-relation} is defined only up to a $\d$-exact 3-form. The addition of such $\d$-exact 3-forms adds a corner term to the symplectic potential. What are the possible such corner terms? If we only allow terms that are polynomial in the fundamental fields, the two non-trivial gauge-covariant $(2,1)$-forms, i.e.\ $(2,1)$-forms that are invariant under the internal $\SLC$ action and under diffeomorphisms whose generators are tangent to the corner, are then given by
\begin{equation}
    \psi_A\wedge\bbvar{d}\phi^A,\qquad A_{AB}\wedge\bbvar{d} A^{AB}.
\end{equation}
This defines the most general polynomial ambiguity of the corner symplectic potential. The resulting  symplectic form consists, therefore, of two terms parametrized by two \textit{complex-valued} coupling constants $\gamma,\nu\in\C$:
\begin{subequations}
\begin{align}
    &\Theta^{C}_{\gamma}=\I(\gamma g)^\ii\int_{C}\psi_A\wedge\bbvar{d}\phi^A,&&\Omega^{C}_{\gamma}=\I(\gamma g)^\ii\int_{C}\bbvar{d}\psi_A\wedge\bbvar{d}\phi^A,\label{SymFormChГ}\\
    &\Theta^{C}_\nu=-\I\nu\int_{C} A_{AB}\wedge\bbvar{d} A^{AB},&&\Omega^{C}_\nu=-\I\nu\int_{C} \bbvar{d} A_{AB}\wedge\bbvar{d} A^{AB},\label{SymFormChNu}
\end{align}\end{subequations}
with $\gamma\rightarrow\infty,\,\nu\rightarrow 0$ corresponding to the phase space \eqref{SymFormSD}, and $\gamma=1,\,\nu=0$ to the self-dual gravity phase space as can be easily seen from \eqref{SDphasespace-relation}. Such an extension matches exactly the phase space contribution found in the literature due to the Holst and Chern class topological terms if $\gamma$ is related to the Immirzi parameter $\gamma_{\text{Im}}$ as $\gamma_{\text{Im}}=\I\gamma$ (see \cite{Freidel:2020xyx,Freidel:2015gpa,Geiller:2020edh}). The corner phase space of the kind (\ref{SymFormChГ}, \ref{SymFormChNu}) was obtained also in \cite{Durka:2021ftc} from the MacDowell-Mansouri action. In our case, however, the constants $\gamma$ and $\nu$ are treated independently. Implications of such a modification on the corner charge algebra are discussed in this section.

\subsection{Holst term}
First, we focus on the Holst term ambiguity parameterized by $\gamma$. 
As a reminder, the following on-shell corner charges, corresponding to the symmetries of the action \eqref{action-Robinson}, were obtained in \eqref{slc-gen-gr}, \eqref{cornercurrent1}, and \eqref{dif-charge-gr}:
\begin{subequations}
\begin{align}
    q^{\sl}[\alpha]&=0,\\
    q^{\overline{\sl}}[\beta]&=\I\,g^\ii\!\!\int_{C}\left(\frac{c}{2} \,\psi^A\wedge\psi_A-\frac{b}{2}\,\phi^A\wedge\phi_A-a\,\psi^A\wedge\phi_A\right),\\
    q^{di\!f\!\!f}[\xi]&=\I\,\int_{C}(\iota_\xi \psi^A\pis_A +\iota_\xi \phi^A\pif_A ).
\end{align}\end{subequations}
The on-shell corner-charge algebra follows the same pattern as the off-shell generator algebra \eqref{1st-class-off-shell-algebra}. The vanishing $\SLC$ charge makes some of the Poisson brackets trivially zero on shell. For any diffeomorphism parameter $\xi$, such that $\xi|_{C}\in T{C}$,
\begin{subequations}
\begin{align}
    &\{q^\sl[\alpha_1],q^\sl[\alpha_2]\}\equiv 0,&&\{q^\sl[\alpha],q^{\overline{\sl}}[\beta]\}\equiv 0,\\
    &\{q^{\overline{\sl}}[\beta_1],q^{\overline{\sl}}[\beta_2]\}=-q^{\overline{\sl}}[[\beta_1,\beta_2]],&& \{q^{di\!f\!\!f}[\xi],q^\sl[\alpha]\}\equiv 0,\\
    &\{q^{di\!f\!\!f}[\xi_1],q^{di\!f\!\!f}[\xi_2]\}=-q^{di\!f\!\!f}[[\xi_1,\xi_2]],&& \{q^{di\!f\!\!f}[\xi],q^{\overline{\sl}}[\beta]\}=-q^{\overline{\sl}}[\partial_\xi\beta].
\end{align}
\end{subequations}
Such a set of corner charges implies that the corresponding quantum theory will not carry a representation of $\SLC$. However, let us remember that we can make a different choice of the corner phase space. The corner charges for a general pre-symplectic 2-form $\Omega_\gamma$ \eqref{SymFormChГ} modify the charges. They now read
\begin{subequations}\begin{align}
    q^\sl_\gamma[\alpha]&=\I(\gamma g)^\ii\!\!\int_{C} \alpha^{AB}\psi_A\wedge\phi_B=\\
    &= \gamma^\ii q^\sl_1[\alpha],\\
    q^{\overline{\sl}}_\gamma[\beta]&=\I(1-\gamma^\ii)g^\ii\!\int_{C}\left(\frac{c}{2} \,\psi^A\wedge\psi_A-\frac{b}{2}\,\phi^A\wedge\phi_A-a\,\psi^A\wedge\phi_A\right)=\\
    & =(1-\gamma^\ii)q^{\overline{\sl}}[\beta],\\
    q^{di\!f\!\!f}_\gamma[\xi]&=\I(1-\gamma^\ii)\!\int_{C}(\iota_\xi \psi^A\pis_A+\iota_\xi \phi^A\pif_A)-\I(\gamma g)^\ii\!\! \int_{C} \iota_\xi A^{AB}\,\psi_A\wedge\phi_B,\label{DiffCornerChargeGamma}
\end{align}\end{subequations}
where the notation $q_1$ stands for $q_{\gamma=1}$ and is mentioned in more detail below, while $q\equiv q_\infty$ refers to the corner charges with a trivial corner phase space ($\gamma\rightarrow\infty$) derived in the previous section. The $\gamma$-modified corner charge algebra $\forall \xi:\xi|_{C}\in T{C}$ becomes
\begin{subequations}
    \begin{align}
    &\{q^\sl_\gamma[\alpha_1],q^\sl_\gamma[\alpha_2]\}=-q^\sl_\gamma[[\alpha_1,\alpha_2]],&&\{q^\sl_\gamma[\alpha],q^{\overline{\sl}}_\gamma[\beta]\}= 0,\\
    &\{q^{\overline{\sl}}_\gamma[\beta_1],q^{\overline{\sl}}_\gamma[\beta_2]\}=-q^{\overline{\sl}}_\gamma[[\beta_1,\beta_2]],&& \{q^{di\!f\!\!f}_\gamma[\xi],q^\sl_\gamma[\alpha]\}=-q^\sl_\gamma[\partial_\xi\alpha],\\
    &\{q^{di\!f\!\!f}_\gamma[\xi_1],q^{di\!f\!\!f}_\gamma[\xi_2]\}=-q^{di\!f\!\!f}_\gamma[[\xi_1,\xi_2]],&& \{q^{di\!f\!\!f}_\gamma[\xi],q^{\overline{\sl}}_\gamma[\beta]\}=-q^{\overline{\sl}}_\gamma[\partial_\xi\beta],
\end{align}\label{CornerChargeAlgebraGamma}
\end{subequations}
which provides a possibility for a richer representation theory. 

The choice of $\gamma=1$ ($\gamma_{\text{Im}}=\I$)  restores the self-dual gravity phase space. It leads to a swap of a non-trivial charge from $\SLCB$ to $\SLC$, as well as a change in the diffeomorphism charge:
\begin{subequations}\begin{align}
    q^\sl_1[\alpha]&=\I\,g^\ii \int_{C}\alpha^{AB}\psi_A\wedge\phi_B,\\
    q^{\overline{\sl}}_1[\beta]&=0,\\
    q^{di\!f\!\!f}_1[\xi]&=-\I\,g^\ii\int_{C} \iota_\xi A^{AB}\,\psi_A\wedge\phi_B.\label{diffcharge1}
\end{align}\end{subequations}
However, to understand this at a deeper level, let us rewrite \eqref{DiffCornerChargeGamma} using the charge  decompositions (\ref{q-diff-thru-shifts}, \ref{slcb-charge-as-shift-topo}) to single out Lorentz and shift components of corner generators: 
\begin{subequations}
\begin{align}
    q^\sl_\gamma[\alpha]&=\gamma^\ii q^\sl_1[\alpha],\\
    q^{\overline{\sl}}_\gamma[\beta]&= \frac{1-\gamma^\ii}{2g}\left(q_{\phi}[\delta_\beta\psi^A]+q_{\psi}[\delta_\beta\phi_A]\right)\\
    q^{di\!f\!\!f}_\gamma[\xi]&=\frac{1-\gamma^\ii}{2g}\left( -q_{\phi}[\iota_\xi\pif]+q_{\psi}[\iota_\xi\pis]\right)-\gamma^\ii q^\sl_1[\iota_\xi A].
\end{align}\end{subequations}
This suggests that $\gamma$ is a switch between \textit{Lorentz} and \textit{shift} generator components. Namely, $\gamma=\infty$ corresponds to a non-zero shift generator contribution and a trivial $\SLC$ contribution, whereas $\gamma=1$ enables $\SLC$ charges and makes shift components across all charges vanish.

Note that the phase-space extension \eqref{SymFormChГ} can be added to the topological phase space \eqref{SymForm-topo} as well. Since the connection is absent from the theory, such corner term would simply cancel the corner charges whenever $\gamma=1$.

For completeness, we mention that off-shell diffeomorphisms generators  $H_\gamma$ obtained from $H$ via the deformation $q\rightarrow q_\gamma$ of their corner components satisfy the same algebra \eqref{1st-class-off-shell-algebra}. Off-shell, the corner part of the diffeomorphism charge, instead of \eqref{DiffCornerChargeGamma}, becomes more involved:
\begin{equation}
    q^{di\!f\!\!f}_\gamma[\xi]=\I\int_{C}(\iota_\xi \psi^A(\pis_A-(\gamma g)^\ii\nabla\phi_A)+\iota_\xi \phi^A(\pif_A+(\gamma g)^\ii \nabla\psi_A) )-i(\gamma g)^\ii\!\! \int_{C} \iota_\xi A^{AB}\,\psi_A\wedge\phi_B.\label{DiffCornerChargeGammaOffshell}
\end{equation}
It can be simplified to \eqref{DiffCornerChargeGamma}, in particular, by imposing the boundary conditions $\psiFF{}^A|_{C}=0$ and $\phiFF{}^A|_{C}=0$ (see \eqref{fake-flatness-pi-psi}, \eqref{fake-flatness-pi-phi}) at ${C}$. Note that the addition of the corner phase space \eqref{SymFormChГ} in the $g=0$ case is impossible since it would diverge in the limit.

\subsection{Chern--Simons corner charges}

The Chern-Simons corner charge contribution corresponding to the \textit{symmetries of the bulk action} is easily integrable. First let us consider the gauge transformations, with $\alpha\in\slc$ a field-independent parameter: 
\begin{equation}
    \bbvar{I}_\alpha\Omega^{C}_\nu=-\bbvar{d} q^\sl_\nu[\alpha],\qquad q^\sl_\nu[\alpha]=\I\nu\int_{C}\alpha^{AB}F_{AB}.
\end{equation}
Similarly, charges of diffeomorphisms generated by a field-independent vector field $\xi$ are integrable if $\xi|_{C}\in T{C}$:
\begin{equation}
    \bbvar{I}_\xi\Omega^{C}_\nu=-\bbvar{d} q^\sl_\nu[\alpha],\qquad q^{di\!f\!\!f}_\nu[\alpha]=-\I\nu\int_{C} \iota_\xi A^{AB}dA_{AB}.
\end{equation}
Note that via the addition of 
\begin{equation}
    -\I\nu\int_{C} \iota_\xi A^{AB}A_A{}^C\wedge A_{CB}=-\frac{\I\nu}{3}\int_{C} \iota_\xi (A^{AB}\wedge A_A{}^C\wedge A_{CB})\overset{\xi|_{C}\in T{C}}{=}0
\end{equation}
the charge can be brought into a nice form:
\begin{equation}
    q^{di\!f\!\!f}_\nu[\alpha]=-\I\nu\int_{C} \iota_\xi A^{AB}F_{AB}=-q^\sl_\nu[\iota_\xi A],
\end{equation}
which matches the charge relation in the rest of the analysis. The Chern--Simons charges satisfy the already familiar algebra:
\begin{subequations}\label{CornerChargeAlgebraNu}
    \begin{align}
    &\{q^\sl_\nu[\alpha_1],q^\sl_\nu[\alpha_2]\}=-q^\sl_\nu[[\alpha_1,\alpha_2]],\\
    & \{q^{di\!f\!\!f}_\nu[\xi],q^\sl_\nu[\alpha]\}=-q^\sl_\nu[\partial_\xi\alpha],\\
    &\{q^{di\!f\!\!f}_\nu[\xi_1],q^{di\!f\!\!f}_\nu[\xi_2]\}=-q^{di\!f\!\!f}_\nu[[\xi_1,\xi_2]].
\end{align}
\end{subequations}
The Chern--Simons charges do not affect the other corner generators like Holst contributions do; hence, we end their discussion here. Related developments have been reported in \cite{Geiller:2017xad,Durka:2021ftc,Langenscheidt:2026fii}.

\subsection{The dual sector}

Due to the similarity of the actions \eqref{action-Robinson} and \eqref{action-full-anti-dual}, the analysis above can be straightforwardly extended to the dual sector. We proceed as follows. 
Let $\tilde{f}[\alpha]$ denote a functional that is equal to $f[\alpha]$ up to the substitution $\psi\rightarrow\tpsi$, $\phi\rightarrow \tphi$, $\pi\rightarrow\tpis$, $\rho\rightarrow\tpif$ and an overall minus sign. Then,
\begin{equation}
    \tilde{\Omega}_\Sigma=-\I\int_{\Sigma}\left(\bbvar{d}\tpis_{\Ap }\wedge \bbvar{d}\tpsi^{\Ap }+\bbvar{d}\tpif_{\Ap }\wedge \bbvar{d}\tphi^{\Ap }\right),\label{SymFormASD}
\end{equation}
and the considerations above follow through without any change according to the following formulas:
\begin{equation}
    \bbvar{I}_\xi\tilde{\Omega}_\Sigma=-\bbvar{d} \tilde{H}_\Sigma[\xi],\label{HamFlowPri}
\end{equation}
\begin{subequations}\begin{align}
    \tilde{H}^\sl[\alpha]&=\tcG[\alpha],\\
    \tH^{\overline{\sl}}[\beta]&=\frac1g\tpsiFF[\delta_\beta\tpsi]+\frac1g\tphiFF[\delta_\beta\tphi]+\tq^{\overline{\sl}}[\beta],\\
    \tilde{H}^{di\!f\!\!f}[\xi]&=\tpsiEE[\iota_\xi\tpsi]+\tphiEE[\iota_\xi\tphi]-\tpsiFF[\iota_\xi\tpif]+\tphiFF[\iota_\xi\tpis]-\tcG[\iota_\xi \tA]+\tilde{q}^{di\!f\!\!f}[\xi].
\end{align}\end{subequations}
The tilde-sector charges satisfy the same algebra \eqref{1st-class-off-shell-algebra} and \eqref{CornerChargeAlgebraGamma}.  Note that if the current variables are to be treated as fundamental, the corner phase space of the dual sector can \emph{a priori} have a different Immirzi parameter\footnote{The choice of the opposite sign in this relation in contrast to the unprimed sector is due to the Holst terms of the two sectors having equal signs: \\
$S=-(g\gamma_{\text{Im}})^\ii\int_M e_I\wedge e_J\wedge F^{IJ}_\omega=-2(g\gamma_{\text{Im}})^\ii\int_M (\psi_{A}\wedge \phi_{B}\wedge F^{AB}+\tpsi_{A}\wedge \tphi_{B}\wedge\Tilde{F}^{A'B'})$.} $\tgamma_{\text{Im}}=-\I\tgamma$,
\begin{equation}
    \tilde{\Omega}_\Sigma+\tilde{\Omega}^{C}_{\tilde{\gamma}}=-\I\int_\Sigma\left(\delta\tpis_{\Ap }\wedge \delta\tpsi^{\Ap }+\delta\tpif_{\Ap }\wedge \delta\tphi^{\Ap }\right)-\I(\tilde{\gamma} g)^\ii\int_{C}\delta\tpsi_{\Ap }\wedge\delta\tphi^{\Ap }.
\end{equation}
The Chern--Simons extension can also be added to the dual sector. However, we will not discuss it in what follows.

\subsection{Corner charges under reality conditions}

Let us  now study the impact of the reality conditions on the charges in the presence of both the self-dual and anti-self-dual sectors of gravity. Owing to the decoupling of the two sectors in \eqref{action-full-dual-anti-dual} and linearity of the vector operations applied in the covariant phase space analysis, the total gravitational symplectic form, symmetry generators, and charges are simply a sum of contributions from the two sectors. This, however, no longer holds, if we go on-shell of the reality conditions, which relate the variables in the two sectors.

We recall that the soldering form can be decomposed into either unprimed or primed spinors:
\begin{subequations}\begin{equation}\label{tetradasunprimed}
    e_{AA'}=\psi_A \bar{o}_{A'}+\phi_A \bar{\iota}_{A'},
\end{equation}
\begin{equation}\label{tetradasprimed}
    e_{AA'}=\tpsi_{A'} o_{A}+\tphi_{A'} \iota_{A}.
\end{equation}\end{subequations}
A pseudo-Riemannian tetrad with a signature $(-,+,+,+)$ corresponds to an anti-hermitian soldering form (see Section \ref{section:grav-action}):
\begin{equation}\label{reality}
    e_{AA'}^\dagger=-e_{AA'}.
\end{equation}
We can obtain two kinds of reality conditions on the phase space variables from this equation. The first set of conditions can be obtained via the substitution of \eqref{tetradasunprimed} into \eqref{reality}. These guarantee reality of a tetrad defined by the soldering form \eqref{tetradasunprimed} \cite{Robinson:1998vf}:
\begin{subequations}
    \begin{align}
    (\psi_0)^*&=-\psi_0,\\
    (\phi_1)^*&=-\phi_1,\\
    (\psi_1)^*&=-\phi_0,
\end{align}\label{reality-unprimed}
\end{subequations}
where $(a)^\ast$ is the complex conjugate of $a\in\C$ and the projections of fields along the basis directions are defined as
\begin{subequations}
\begin{align}
    \psi_0&=\psi_A\iota^A,\\
    \psi_1&=-\psi_A o^A.
\end{align}
\end{subequations}
Analogously, substitution of \eqref{tetradasprimed} into \eqref{reality} leads to
\begin{subequations}\label{reality-primed}
    \begin{align}
    (\tpsi_0)^*&=-\tpsi_0,\\
    (\tphi_1)^*&=-\tphi_1,\\
    (\tpsi_1)^*&=-\tphi_0.
\end{align}
\end{subequations}The second kind of conditions is obtained upon substitution of \eqref{tetradasunprimed} and \eqref{tetradasprimed} into the right- and left-hand sides of \eqref{reality} respectively. This yields consistency of the definitions \eqref{tetradasunprimed} and \eqref{tetradasprimed} and compares fields from opposite sectors:
\begin{subequations}\label{consistency}
\begin{align}
    \tpsi_{A'}&=-\psi^\dagger_{A'},\\
    \tphi_{A'}&=-\phi^\dagger_{A'}.
\end{align}
\end{subequations}
The Hermitian conjugate of a two-component spinor is defined as an anti-linear map ${}^\dagger:\C^2\rightarrow\overline{\C^2}$ and $\overline{\C^2}\rightarrow \C^2$, which preserves the anti-symmetric tensor, $(\epsilon_{AB})^\dagger=\epsilon_{A'B'}$ and satisfies
\begin{subequations}
\begin{align}
    \lambda^\dagger_A&:=(\lambda_{A'})^*,\\
    \mu^\dagger_{A'}&:=(\mu_{A})^*.
\end{align}
\end{subequations}
The reality conditions are not preserved by either $\SLC$ or $\SLCB$ transformations. Requiring that a joint $\SLC$ and $\SLCB$ gauge transformation preserves \eqref{reality-unprimed} (or, equivalently, \eqref{reality}),
\begin{equation}
    (\delta_\alpha+\delta_\beta)\big(\psi_0+(\psi_0)^*\big)=0, \quad \text{etc.,}
\end{equation}
leads to a relationship between the two gauge algebra parameters:
\begin{equation}
    \beta=\alpha^*.
\end{equation}
This restores the conjugacy relationship not only at the level of phase space fields, but also at the level of the gauge transformations.

An explicit computation shows that the following relations\footnote{We recall the notation $q\equiv q_\infty$.} can be established between the corner charges on shell of the conditions \eqref{reality-unprimed}--\eqref{consistency}: 
\begin{subequations}
\begin{align}
    q^{\overline{\sl}}[\alpha^*]&\overset{\cdot}{\approx}(q^\sl_1[\alpha])^*,\\
    \tq^{\overline{\sl}}[\alpha]&\overset{\cdot}{\approx}(\tq^\sl_1[\alpha^*])^*,\\
    \tilde{q}^\sl_1[\alpha^*]&\overset{\cdot}{\approx}(q^\sl_1[\alpha])^*,\\
    \tq^{\overline{\sl}}[\alpha]&\overset{\cdot}{\approx}(q^{\overline{\sl}}[\alpha^*])^*.
\end{align}
\end{subequations}
Here, and in the rest of this section, the symbol $\overset{\cdot}{\approx}$ is reserved to denote application of the reality conditions. We recall that we considered the possibility that a different factor $\tilde\gamma$ could encode the corner ambiguity for the tilde variables sector. Putting the two sectors together with the reality conditions, we obtain different relations for the corner charges: 
\begin{subequations}
\begin{align}
    &q^\sl_\gamma[\alpha]+q^{\overline{\sl}}_\gamma[\alpha^*]\overset{\cdot}{\approx} q^{\overline{\sl}}[\alpha^*]+\frac{2\I}{\gamma}\Imm(q^\sl_1[\alpha]),\\
    &\tq^\sl_{\tilde{\gamma}}[\alpha^*]+\tq^{\overline{\sl}}_{\tilde{\gamma}}[\alpha]\overset{\cdot}{\approx} \tq^{\overline{\sl}}[\alpha]+\frac{2\I}{\tilde{\gamma}}\Imm( \tq^\sl_1[\alpha^*])\overset{\cdot}{\approx}(q^{\overline{\sl}}[\alpha^*])^*-\frac{2\I}{\tilde{\gamma}}\Imm(q^\sl_{1}[\alpha]),\\
    &q^\sl_\gamma[\alpha]+q^{\overline{\sl}}_\gamma[\alpha^*]+\tq^\sl_{\tilde{\gamma}}[\alpha^*]+\tq^{\overline{\sl}}_{\tilde{\gamma}}[\alpha]\overset{\cdot}{\approx} 2\Ree(q_1^\sl[\alpha])+2\I\left(\frac{1}{\gamma}-\frac{1}{\tilde{\gamma}}\right)\Imm(q^\sl_1[\alpha])=\\
    &\qquad \overset{\tilde{\gamma}=\gamma}{=}2\Ree(q^\sl_1[\alpha])\overset{\cdot}{\approx} 2\Ree(q^{\overline{\sl}}[\alpha^*]).
\end{align}
\end{subequations}
Thus, the extended corner phase space \eqref{SymFormChГ} does not affect the total $\SLC$ charge of the real on-shell theory if the Immirzi parameters $\gamma, \,\tilde\gamma$ of the two sectors match. If only the real parts of the inverse Immirzi parameters are equal,
\begin{equation}
    \Ree\frac1\gamma=\Ree\frac{1}{\tilde{\gamma}},\label{inv-Immirzi-real-eq}
\end{equation}
the total corner charge is real and contains an Immirzi-dependent contribution. If 
\begin{equation}
    \Ree\frac1\gamma\neq\Ree\frac{1}{\tilde{\gamma}},
\end{equation}
the extension controls the imaginary contribution of the $\SLC$ on-shell charge. Thus, matching real parts of the inverse Immirzi parameters are necessary for the reality of the corner charges. In the case $\gamma,\tilde{\gamma}\in\R$, the requirement that the corner charges be real simplifies to $\gamma=\tilde{\gamma}$.

Similarly, diffeomorphism corner charges add up to a real corner charge on shell of the reality conditions. It is easy to see that momenta $\pi$ and $\rho$ follow the same reality conditions as the spinors $\psi$ and $\phi$:
\begin{subequations}
\begin{align}
    \tpis^\dagger_{A}&=-\pis_A,\\
    \tpif^\dagger_{A}&=-\pif_A.
\end{align}\label{consistency-momenta}
\end{subequations}
Taking this into account, we obtain
\begin{multline}
    q^{di\!f\!\!f}_\gamma[\xi]+\tq^{di\!f\!\!f}_{\tilde{\gamma}}[\xi]=q^{di\!f\!\!f}[\xi]+(q^{di\!f\!\!f}[\xi])^*+\frac{1}{\gamma}q^{di\!f\!\!f}_1[\xi]+\frac{1}{\tilde{\gamma}}(q^{di\!f\!\!f}_1[\xi])^*=\\=2\Ree q^{di\!f\!\!f}[\xi]+\left(\frac{1}{\gamma}+\frac{1}{\tilde{\gamma}}\right)\Ree(q^{di\!f\!\!f}_1[\xi])+\I\left(\frac{1}{\gamma}-\frac{1}{\tilde{\gamma}}\right)\Imm(q^{di\!f\!\!f}_1[\xi])\overset{\tilde{\gamma}=\gamma}{=} 2\Ree q^{di\!f\!\!f}_\gamma[\xi].
\end{multline}
The reality of the diffeomorphism charge, in addition to \eqref{inv-Immirzi-real-eq}, also requires
\begin{equation}
    \Imm\frac1\gamma=-\Imm\frac{1}{\tilde{\gamma}}.\label{inv-Immirzi-im-eq}
\end{equation}
Hence, the reality of corner charges implies
\begin{equation}
    \tilde{\gamma}=\gamma^*\qquad \iff\qquad \tilde{\gamma}_{\text{Im}}=\gamma^*_{\text{Im}}.
\end{equation}
At last, on shell of the reality conditions, full shift charges are restored in \eqref{slcb-charge-as-shift-topo} and \eqref{q-diff-thru-shifts}:
\begin{subequations}    \begin{align}
        &q^\sl[\alpha]+q^{\overline{\sl}}[\alpha^*]+\tq^\sl[\alpha^*]+\tq^{\overline{\sl}}[\alpha]\overset{\cdot}{\approx} \frac1g\Ree\left(q_{\phi}[\delta^{\overline{\sl}}_{\alpha^*}\psi]+q_{\psi}[\delta^{\overline{\sl}}_{\alpha^*}\phi]\right),\\
        &q^{di\!f\!\!f}[\xi]+\tq^{di\!f\!\!f}[\xi]\overset{\cdot}{\approx} -q_{\phi}[\iota_\xi\pif]+q_{\psi}[\iota_\xi\pis]+q_\pis[\iota_\xi\psi]+q_\pif[\iota_\xi\phi].
    \end{align}
    \end{subequations}
This permits a concise off-shell expression
\begin{subequations}\begin{align}
        &H^{\overline{\sl}}[\beta]+\tH^{\overline{\sl}}[\beta^*]\overset{\cdot}{\approx}\frac1g\Ree\left(H_{\phi}[\delta^{\overline{\sl}}_{\alpha^*}\psi]+H_{\psi}[\delta^{\overline{\sl}}_{\alpha^*}\phi]\right),\\
        &H^{di\!f\!\!f}[\xi]+\tH^{di\!f\!\!f}[\xi]\overset{\cdot}{\approx} \\
        &\overset{\cdot}{\approx}\Ree\left(-H_{\phi}[\iota_\xi\pif]+H_{\psi}[\iota_\xi\pis]+H_\pis[\iota_\xi\psi]+H_\pif[\iota_\xi\phi]-H^\sl[\iota_\xi A]\right),
    \end{align}
\end{subequations}
where shift generators in the topological case are defined as in (\ref{gen-topo-shift}, \ref{gen-topo-exshift}), while in the gravity case they are given by (\ref{gen-shift-psi}, \ref{gen-shift-phi}, \ref{gen-gr-shift}) and defined on an extended phase space.

\section{Weak-coupling limit}\label{app:g0}
Thanks to the new first-order formulation in terms of the spinor variables, we can perform in a new manner the limit $G\propto g\rightarrow0$ by rescaling the momentum variables.  
At the topological level, this  corresponds to dealing with \textit{skeletal} 2-groups \cite{baez, Borsten_2025} as symmetries. At the gravitational level, we obtain a new theory, physical aspects of which still need to be understood. 
 
\subsection{The $g\rightarrow0$ limit of the topological theory}\label{app:g0-topo}

Let us revisit the  analysis performed in \hyperref[sec:topo-id]{Section \ref{sec:topo-id}} but with $g=0$ now. It has certain distinct features despite an algebraic smoothness of the limit. First, the momenta become dynamically independent of the configuration variables. Second, the exact shifts become an independent symmetry transformation.

The action has in this case one  term less. It now reads:
\begin{equation}
    S_0=\I\int_M \left(\pis_{ A}\wedge \d\psi^{A}+\pif_{ A}\wedge \d\phi^{A}\right),\label{action-topo-g0}
\end{equation}
but the off-shell phase space remains the same as before.

Let us count the dynamical degrees of freedom. In the case $g=0$, the two pairs of equations \eqref{EOMab1}--\eqref{EOMab2} and \eqref{EOMab3}--\eqref{EOMab4} disentangle,
\begin{subequations}
\label{EOMab-0}
\begin{align}\label{EOMab01}
    &\d\psi_A=0, \\ 
    &\d\phi_A=0\label{EOMab02}\\
    &\d\pif_A=0,\label{EOMab03} \\
    &\d\pis_A =0.\label{EOMab04}
\end{align}
\end{subequations}
In contrast to the topological theory \eqref{action-topo} with $g\neq0$, equations (\ref{EOMab03}, \ref{EOMab04}) are independent of equations (\ref{EOMab01}, \ref{EOMab02}) and constrain two degrees of freedom each. At the same time, since $\d(\d\psi_A)\equiv0\equiv \d(\d\phi_A)$, each of the constraints (\ref{EOMab01}, \ref{EOMab02}) implies two trivial equations. Thus, the theory still has 12 non-trivial independent first-class constraints and has no local dynamical degrees of freedom.

\paragraph{Shift symmetry}
Similarly to the constraints \eqref{EOMab-0},  the corresponding gauge transformations disentangle in the $g\rightarrow0$ limit and exhibit a  2-gauge structure. The remaining non-trivial shifts affect only the canonical moments: 
\begin{equation}
    \delta_{\eta} \pif_A=-\d\eta_A\qquad\text{and}\qquad  \delta_{\zeta} \pis_A=\d\zeta_A.
\end{equation}
Their corresponding generators simplify to 
\begin{subequations}\begin{align}
    \bbvar{I}_\eta\Omega&=-\bbvar{d} H'_{\psi}[\eta],&H'_{\psi}[\eta]&=\psiff'{}{}[\eta]+q_{\phi}[\eta],\\
    \bbvar{I}_\zeta\Omega&=-\bbvar{d} H'_{\phi}[\eta],&H'_{\phi}[\eta]&=\phiff'[\zeta]+q_{\psi}[\zeta],
\end{align}\end{subequations}
where the new momentum maps correspond to the equations of motion in the $g\rightarrow 0$ limit,
\begin{subequations}
\label{mommapsF-top-0}
\begin{align}
    &\psiff'[\eta]=\I\int_\Sigma \d\phi^A\wedge\eta_A,\label{P1-top-0}\\
    &\phiff'[\zeta]=\I\int_\Sigma \d\psi_A\wedge\zeta^A.\label{P2-top-0}
\end{align}
\end{subequations}
These momentum maps can be directly obtained from the limit $g\rightarrow0$.  

\paragraph{Gauge symmetry} The exact shifts become an independent gauge transformation generated by $H_\pis[\lambda]$ and $H_\pif[\mu]$ introduced in \eqref{gen-topo-exshift}, which are not affected by the limit. 

\paragraph{Global $\SLC$ symmetry}
The global $\SLC$ transformation \eqref{slc-on-top} is generated by a physical conserved bulk charge
\begin{equation}
    H^{\sl}[\alpha]=\I\alpha^{AB}\int_\Sigma(\pis_A\wedge\psi_B+\pif_A\wedge\phi_B),
\end{equation}
since the equations of motion no longer relate spinors and their momenta. Considering the Poisson algebra of such generators gives the $\slc$ algebra as in \eqref{Hslc}.  

\paragraph{$\SLCB$ local symmetry}
The new realization of the  $\SLCB$ local symmetry  is given by
\begin{subequations}
\label{slcb-g0-topo}
\begin{align}
    &\delta'^{\overline{\sl}}_{\beta'}\psi_A= 0,\\
    &\delta'^{\overline{\sl}}_{\beta'}\phi_A=0,\\
    &\delta'^{\overline{\sl}}_{\beta'}\pi_A^\psi= \d c'\wedge\psi_A-\d a'\wedge\phi_A,\\
    &\delta'^{\overline{\sl}}_{\beta'}\pi_A^\phi=-\d a'\wedge\psi_A-\d b'\wedge\phi_A.
    \end{align}
\end{subequations}
Its Hamiltonian flow equation takes the form
\begin{subequations}
\begin{align}
    \bbvar{I}_{\beta'}\Omega&=-\bbvar{d} H'^{\overline{\sl}}[\beta'],\\
    H'^{\overline{\sl}}[\beta']&=\psiff{}{}'[a\psi+b\phi]+\phiff{}{}'[c\psi-a\phi]+q'^{\overline{\sl}}[\beta'],\\
    q'^{\overline{\sl}}[\beta']&=\I\int_{C}\left(\frac{c'}{2} \,\psi^A\wedge\psi_A-\frac{b'}{2}\,\phi^A\wedge\phi_A-a'\,\psi^A\wedge\phi_A\right)=\label{q-slcb-g0}\\
    &=\frac{1}{2}\left(q_{\phi}[a'\psi+b'\phi]+q_{\psi}[c'\psi-a'\phi]\right).
\end{align}
\end{subequations}
The notation $\overline{\sl}$ is due to the fact that this transformation can be obtained from $\slcb$ \eqref{slcb-top} by performing a rescaling 
\begin{equation}
    \beta'=\beta/g\label{beta-rescaling}
\end{equation}
in \eqref{slcb-top}--\eqref{slcb-charge-as-shift-topo} and setting $g=0$. Considering the Poisson algebra of such generators gives the $\slc$ algebra as in \eqref{Hslcb}.

\paragraph{Diffeomorphism symmetry}
The diffeomorphism generator is integrable on shell only if $\xi|_C\in TC$ and off shell if, additionally, $\xi|_\Sigma\in  T\Sigma$. It can be expressed through the momentum maps \eqref{mommapsF-top-0}:
\begin{equation}
    H'^{di\!f\!\!f}[\xi]=-\psiff'{}{}[\iota_\xi \pif]+\phiff'{}{}[\iota_\xi \pis]+\psiee{}[\iota_\xi \psi]+\phiee{}[\iota_\xi \phi]+q^{di\!f\!\!f}[\xi].
\end{equation}

\paragraph{Dual diffeomorphisms}
 This theory is additionally symmetric under dual diffeomorphisms defined as \cite{Geiller:2020edh, Geiller_2023} 
\begin{subequations}
\label{dualdiffeo}    
\begin{align}
    \delta^{\overline{di\!f\!\!f}}_\chi\psi&=L_\chi\phi, &\delta^{\overline{di\!f\!\!f}}_\chi\pis&=L_\chi\pif\\
    \delta^{\overline{di\!f\!\!f}}_\chi\phi&=L_\chi\psi &\delta^{\overline{di\!f\!\!f}}_\chi\pif&=L_\chi\pis.
\end{align}\end{subequations}
The corresponding generator is given by
\begin{equation}
    H^{\overline{di\!f\!\!f}}[\chi]=-\psiff'{}{}[\iota_\chi\pis]+\phiff'[\iota_\chi\pif]+\psiee{}[\iota_\chi\phi]+\phiee{}[\iota_\chi\psi]+q^{\overline{di\!f\!\!f}}[\chi],
\end{equation}
where the on-shell charge,
\begin{equation}
    q^{\overline{di\!f\!\!f}}[\chi]=\I\int_{C} (\iota_\chi\psi^A\pif_A+\iota_\chi\phi^A\pis_A),
\end{equation}
resembles the regular diffeomorphism corner charge up to a swap in momenta with respect to the corresponding coordinates. One can notice that
\begin{equation}
    q^{\overline{di\!f\!\!f}}[\chi]\equiv -q_{\phi}[\iota_\xi\pis]+q_{\psi}[\iota_\xi\pif]=q_\pis[\iota_\xi\phi]+q_\pif[\iota_\xi\psi],
\end{equation}
which is due to
\begin{subequations}
    \begin{align}
        &q_{\phi}[\iota_\xi\pis]+q_\pis[\iota_\xi\phi]=\I\int_{C} \iota_\xi(\pis_A\wedge\phi^A)\overset{\xi|_{C}\in T{C}}{=}0,\\
        -&q_{\psi}[\iota_\xi\pif]+q_\pis[\iota_\xi\psi]=\I\int_{C} \iota_\xi(\pif_A\wedge\psi^A)\overset{\xi|_{C}\in T{C}}{=}0.
    \end{align}
\end{subequations}
The flow equation is integrable in the same conditions as that for the regular diffeomorpshisms.

\paragraph{Poisson algebra}
We observe that all corner charges corresponding to the gauge transformations are independent of $g$, and the only difference in the algebra of generators in the $g= 0$ case is the absence of the central extension in the $\{H_{\phi}[\eta],H_{\psi}[\zeta]\}$ commutator.

The dual diffeomorphisms form a closed algebra with the rest of the constraints and act by Lie derivation of their parameter and swapping some of the momentum maps:
\begin{subequations}\begin{align}
    &\{H^{\overline{di\!f\!\!f}}[\chi],H^{\overline{di\!f\!\!f}}[\xi]\}=H^{di\!f\!\!f}[[\xi,\chi]],&&\{H^{\overline{di\!f\!\!f}}[\chi],H^{di\!f\!\!f}[\xi]\}=H^{\overline{di\!f\!\!f}}[[\xi,\chi]],\\
    &\{H_{\phi}[\eta],H^{\overline{di\!f\!\!f}}[\chi]\}=H_{\psi}[L_\chi\eta],&
    &\{H_\pis[\lambda],H^{\overline{di\!f\!\!f}}[\chi]\}=H_\pif[L_\chi\lambda],\\
    &\{H_{\psi}[\zeta],H^{\overline{di\!f\!\!f}}[\chi]\}=H_{\phi}[L_\chi\zeta],&&\{H_\pif[\mu],H^{\overline{di\!f\!\!f}}[\chi]\}=H_\pis[L_\chi\mu].
\end{align}\end{subequations}
Let us now carry over to the study of gravity in the vanishing-coupling case.

\subsection{The $g\rightarrow0$ limit of gravity}\label{g0-GR}

In the limit $g\rightarrow 0$, the action \eqref{action-Robinson} simplifies to
\begin{equation}\label{g=0-gr}
    S'_{SD}=\I\int_M (\pis_A\wedge \nabla\psi^{A}+\pif_A\wedge \nabla\phi^{A}).
\end{equation}
Instead of the $\SLCB$ symmetry, the action is invariant under the transformation \eqref{slcb-g0-topo}. Such a transformation is generated exactly by
\begin{equation}
    H'^{\overline{\sl}}[\beta']=\psiFF{}{}'[a\psi+b\phi]+\phiFF{}{}'[c\psi-a\phi]+q'^{\overline{\sl}}[\beta'],
\end{equation}
where, similarly to the topological case, the momentum maps
\begin{subequations}
\label{mommapFF-g0}
\begin{align}
    \psiFF{}'[\eta]&=\int_\Sigma \psiFF{}_A|_{g=0}\wedge\eta^A\equiv i\int_\Sigma \nabla\phi^A\wedge\eta_A,\label{mommapFFpsi-0}\\
    \phiFF{}'[\zeta]&=\int_\Sigma \phiFF{}_A|_{g=0}\wedge\zeta^A\equiv i\int_\Sigma \nabla\psi_A\wedge\zeta^A\label{mommapFFphi-0}
\end{align}
\end{subequations}
correspond to the simplified fake-flatness equations of motion, and $q'^{\overline{\sl}}[\beta']$ is defined in \eqref{q-slcb-g0}.

Similarly, the generator of diffeomorphisms contains simplified momentum maps \eqref{mommapFF-g0}:
\begin{equation}
    H'^{di\!f\!\!f}[\xi]=\psiEE[\iota_\xi\psi]+\phiEE{}[\iota_\xi\phi]-\psiFF{}'[\iota_\xi\pif]+\phiFF{}'[\iota_\xi\pis]-\cG[\iota_\xi A]+q^{di\!f\!\!f}[\xi].
\end{equation}

The results in this \hyperref[app:g0]{Section \ref{app:g0}}  can be obtained equally well by performing an $\slcb$ rescaling \eqref{beta-rescaling} and setting $g=0$ at any point of the calculation. 

\medskip 

Note that the limit { $g\rightarrow\infty$} can be performed  analogously. One starts with the action \eqref{action-full-SD} and performs the rescaling $\psi\rightarrow g\psi$, $\phi\rightarrow g\phi$, leaving unchanged $\rho$ and $\pi$. This rescaling  results in 
\begin{equation}
    S_{SD}\rightarrow\I\int_M (\pis_A\wedge \nabla\psi^{A}+\pif_A\wedge \nabla\phi^{A}+g^\ii\pis_A\wedge\pif^{A}),
\end{equation}
and in the limit $g\rightarrow\infty$ we recover the same action  in \eqref{g=0-gr}. It is quite remarkable that the different regimes, $G\to 0$ and $G\to\infty$, are governed by the same action. We will explore the physical meaning of this duality elsewhere.

\section{Canonical analysis of the Weyl spinor gravity formulation}\label{sec-canonical}

In what follows, we present the Hamiltonian analysis in the continuum. Note a slight geometric notation change in this section. The starting point is to introduce a foliation, thereby assuming
\begin{equation}
M\simeq[0,1]\times \varSigma.
\end{equation}
Thus there is a natural embedding
\begin{equation}
\varphi_t:\varSigma\rightarrow M,\quad \varphi_t(\varSigma)=:\varSigma_t\simeq\{t\}\times\varSigma.\label{embddng}
\end{equation}
In this section, we add a prefix to denote fields living on the four-dimensional manifold $M$ as ${}^4X$ and distinguish them from their pullback to $\Sigma$, denoted as $X$. We introduce a foliation by choosing a clock $t:M\rightarrow\Sigma$, which increases monotonically along every future-directed curve. Consider then a dual time-flow vector field $t\in\Gamma(TM)$ such that
\begin{equation}
t^\mu\nabla_\mu t=1.
\end{equation}

Given the embedding \eref{embddng} together with a choice of a time-flow vector field $t^\mu$, we can perform the $3+1$-split of the action of the self-dual gravity action \eqref{action-Robinson}. To compute the full constraint algebra, we extend the phase space \eqref{SymFormSD} by a connection term. The fundamental configuration variables now consist of:
\begin{subequations}
\begin{align}
\updown{A}{A}{Ba}(t)&=[\varphi_t^\ast\frupdown{A}{A}{B}]_{a},\\
\updown{\psi}{A}{a}(t)&=[\varphi_t^\ast\frpsi^A]_a,\\
\updown{\phi}{A}{a}(t)&=[\varphi_t^\ast\frphi^A]_a.
\end{align}\end{subequations}
Notice then that all fields depend parametrically on the time variable, which we shall suppress for simplicity. Time derivatives can be then defined as
\begin{subequations}\begin{align}
\updown{\dot{A}}{A}{Ba}&=\frac{\d}{\d t}\varphi_t^\ast[\frupdown{A}{A}{B}]_{a}=\varphi_t^\ast[L_t\frupdown{A}{A}{B}]_{a},\label{Adot}\\
\updown{\dot{\psi}}{A}{a}&=\frac{\d}{\d t}\varphi_t^\ast[\frpsi^A]_a=\varphi_t^\ast[L_t\frpsi^A]_a,\label{phidot}\\
\updown{\dot{\phi}}{A}{a}&=\frac{\d}{\d t}\varphi_t^\ast[\frphi^A]_a=\varphi_t^\ast[L_t\frphi^A]_a,\label{psidot}
\end{align}\end{subequations}
where $L_t(\cdot)=\d(\iota_t\cdot)+(\iota_t\di\cdot)$ is the Lie derivative along the time-flow vector field $t$, which commutes with the pullback. Next, we introduce the momentum variables
\begin{subequations}\begin{align}
\pi_{Aab}&=\varphi_t^\ast[\frpi_A]_{ab},\\
\rho_{Aab}&=\varphi_t^\ast[\frrho_A]_{ab},
\end{align}\end{subequations}
and the Lagrange multipliers
\begin{subequations}\begin{align}
\updown{\alpha}{A}{B}&=\varphi^\ast_t[\iota_t\frupdown{A}{A}{B}],\label{lambd-def}\\
\lambda^A&=\varphi^\ast_t[\iota_t\frupdown{\psi}{A}{}], \label{lambd-def1}\\
\mu^A&=\varphi^\ast_t[\iota_t\frupdown{\phi}{A}{}],\label{lambd-def2}\\
\updown{\eta}{A}{a}&=-\varphi^\ast_t[\iota_t\frupdown{\rho}{A}{}]_a,\\
\updown{\zeta}{A}{a}&=\varphi^\ast_t[\iota_t\frupdown{\pi}{A}{}]_a.
\end{align}\end{subequations}
Next, consider the pullback of the exterior derivatives. The time-space components satisfy
\begin{subequations}\begin{align}
\varphi_t^\ast[\iota_t\frupdown{F}{A}{B}]_{a}&=\updown{\dot{A}}{A}{Ba}-D_a\updown{\alpha}{A}{B},\\
\varphi_t^\ast[\iota_t D\frpsi^A]_a&=\updown{\dot{\psi}}{A}{a}-D_a\lambda^A+\updown{\alpha}{AB}{}\updown{\psi}{}{Ba},\\
\varphi_t^\ast[\iota_t D\frphi^A]_a&=\updown{\dot{\phi}}{A}{a}-D_a\mu^A+\alpha^{AB}\updown{\phi}{}{Ba},
\end{align}\end{subequations}
Since exterior derivatives commute with the pullback, the spatial components are simply
\begin{subequations}\begin{align}
\varphi_t^\ast[\frupdown{F}{}{}^{AB}]_{ab}&=\updown{F}{AB}{ab}=2\partial_{[a}\updown{A}{AB}{b]}+2\updown{A}{AC}{[a}\downup{A}{C}{B}{}_{b]},\\
\varphi_t^\ast[D\frpsi^A]_{ab}&= 2D_{[a}\updown{\psi}{A}{b]},\\
\varphi_t^\ast[D\frphi^A]_{ab}&=2D_{[a}\updown{\phi}{B}{b]}.
\end{align}\end{subequations}
where $D_a$ is the pull-back of the $\SL(2,\C)$ derivative $D_a\xi^A=\partial_a\xi^A + \ou{A}{AB}{a}\xi_B$ on $\varSigma$, such that e.g.\  
\begin{equation}
D\psi^A=\varphi^\ast_t(\nabla\praupdown{4}{\psi}{A}{}).
\end{equation}

We can now perform the $3+1$ decomposition of the action \eqref{action-Robinson}. It assumes a standard Hamiltonian form consisting of a symplectic potential term and a Hamiltonian:
\begin{align}
S&=\I\int_0^1\di t\left[\int_{\varSigma_t}\left(\pi_A\wedge\dot{\psi}^A+\rho_A\wedge\dot{\phi}^A+\Pi_{AB}\wedge\dot{A}^{AB}\right)-H_o\right].
\end{align}
Here, we introduced an auxiliary $\mathfrak{sl}(2,\C)$-valued 2-form $\updown{\Pi}{A}{Bab}$, which is a conjugate momentum to the connection $\updown{A}{A}{Ba}$. Since the action \eqref{action-Robinson} contains no derivatives of the connection, the momentum variables $\updown{\Pi}{A}{Bab}$ vanish as constraints.\smallskip

The Hamiltonian, on the other hand, is a sum of a boundary Hamiltonian $H_{\partial\varSigma}$ and constraints \eqref{mommaps-GR}
\begin{align}
H_o=\psiFF[\eta]+\phiFF[\zeta]+\psiEE[\lambda]+\phiEE[\mu]+\cG[\alpha]+\cC[\gamma]+H_{\partial\varSigma},\label{H0-def}
\end{align}
This Hamiltonian drives the evolution equations, which follow from the fundamental Poisson brackets on the kinematical phase space. The fundamental Poisson brackets among the canonical variables are
\begin{subequations}\begin{align}
    \big\{\downup{\hat{\pi}}{A}{a}(x),\updown{\psi}{B}{b}(y)\big\}&=-\I\delta^a_b\delta_A^B\hat{\delta}(x,y),\label{rho-phi-Poiss}\\
    \big\{\downup{\hat{\rho}}{A}{a}(x),\updown{\phi}{B}{b}(y)\big\}&=-\I\delta^a_b\delta_A^B\hat{\delta}(x,y),\label{pi-psi-Poiss}\\
    \big\{\downup{\hat{\Pi}}{AB}{a}(x),\updown{A}{CD}{b}(y)\big\}&=-\I\delta^a_b\delta^{(C}_A\delta^{D)}_B\hat{\delta}(x,y),\label{Pi-A-Poiss}
\end{align}\label{Poiss}\end{subequations}
where we introduced the following spinor-valued vector densities
\begin{subequations}\begin{align}
    \downup{\hat{\pi}}{A}{a}&=\frac{1}{2}\hat{\varepsilon}^{abc}\pi_{Abc},\\
    \downup{\hat{\rho}}{A}{a}&=\frac{1}{2}\hat{\varepsilon}^{abc}\rho_{Abc},\\
    \downup{\hat{\Pi}}{AB}{a}&=\frac{1}{2}\hat{\varepsilon}^{abc}\Pi_{ABbc},
\end{align}\end{subequations}
with $\hat{\varepsilon}^{abc}$ denoting the three-dimensional Levi-Civita tensor density. All other brackets among the canonical variables vanish on the kinematical phase space.

The constraints at the level of the canonical variables, agree with what we introduced earlier at the level of the covariant phase space. First of all, we obtain, in agreement with \eqref{mommaps-GR} that 
\begin{subequations}\begin{align}
    \psiFF[\eta]&=\I\int_\varSigma\left(g\downup{\hat{\pi}}{A}{a}-\hat{\varepsilon}^{abc}D_b\phi_{Ac}\right)\updown{\eta}{A}{a},\label{R-cons}\\
    \phiFF[\zeta]&=\I\int_\varSigma\left(g\downup{\hat{\rho}}{A}{a}+\hat{\varepsilon}^{abc}D_b\psi_{Ac}\right)\updown{\zeta}{A}{a},\label{P-cons}\\
    \psiEE[\lambda]&=-\I\int_\varSigma D_a\downup{\hat{\pi}}{A}{a}\lambda^A,\label{E-cons}\\
    \phiEE[\mu]&=-\I\int_\varSigma D_a\downup{\hat{\rho}}{A}{a}\mu^A.\label{H-cons}
\end{align}\end{subequations}
In addition, we also have the Gauss constraint, which obtains an extension necessary to preserve its role as a generator of an $\SLC$ gauge transformation,
\begin{subequations}\begin{align}
\nonumber\cG[\alpha]&=\I\int_\varSigma{\alpha}^{AB}\left(\pi_{(A}\wedge\psi_{B)}+\rho_{(A}\wedge\phi_{B)}+D\Pi_{AB}\right)=\\
&\equiv\I\int_\varSigma{\alpha}^{AB}\left(\downup{\hat{\pi}}{(A}{a}\psi_{B)a}+\downup{\hat{\rho}}{(A}{a}\phi_{B)a}+D_a\downup{\hat{\Pi}}{AB}{a}\right).
\end{align}\label{G-cons}\end{subequations}
Finally, we have a constraint on the auxiliary momentum $\hat \Pi$, smeared with a $\mathfrak{sl}(2,\C)$-valued Lagrange multiplier:
\begin{align}
    \cC[\gamma]=\I\int_\varSigma\downup{\hat{\Pi}}{AB}{a}\updown{\gamma}{AB}{a}\equiv\I\int_\varSigma\Pi_{AB}\wedge\gamma^{AB}=0.\label{C-cons}
\end{align}
We call this constraint the \textit{void momentum constraint}. 
It ensures that the momentum $\Pi_{AB}$ vanishes on-shell and, thus, the physical phase space is not affected by the extension. In particular, one can notice that the extended Gauss constraint \eqref{G-cons} differs from the smeared Gauss constraint given in \eqref{SLCgen}. However, on shell of the void-momentum constraint, they coincide. If, in addition, we impose all constraints simultaneously, the  Hamiltonian is a mere boundary term: 
\begin{equation}
H_o\approx H_{\partial\varSigma}=\I\oint_{\partial \varSigma}\left(\pi_A \lambda^A+\rho_A \mu^A\right).
\end{equation}
We have recovered the result of Tung and Jacobson \cite{Tung:1995cj}. 

\subsection{Algebraic solution of the 2-curvature constraints}
Before considering the constraint analysis, let us discuss here the algebraic structure of the 2-curvature constraints, which will be useful for identifying the first-class and second-class constraints of the system. In what follows, we restrict ourselves to those configurations in which the 3-forms $\psi^A\wedge\phi^B\wedge\phi^C$ and $\phi^A\wedge\psi^B\wedge\psi^C$ are linearly independent on $\Sigma$, i.e.\footnote{Upon imposing the reality conditions and restricting ourselves to a spacelike or timelike hypersurface $\varSigma$, this condition is always satisfied.} 
\begin{equation}
\forall a,b\in\C: a\psi^A\wedge\phi^B\wedge\phi^C+b\phi^A\wedge\psi^B\wedge\psi^C=0\Leftrightarrow a=b=0.\label{phipsi-cond}
\end{equation}
Let us expand on this condition further. First of all, we note that $\psi^A\wedge\phi^B\wedge\phi^C=\tfrac{1}{2}\epsilon^{BC}\psi^A\wedge\phi_D\wedge \phi^D$ and $\phi^A\wedge\psi^B\wedge\psi^C=\tfrac{1}{2}\epsilon^{BC}\phi^A\wedge\psi_D\wedge\psi^D$ are top-forms on $\varSigma$. Upon choosing some reference volume 3-form $d^3v$ on $\varSigma$, we can thus find spinor fields $u^A$ and $v^A$ such that\footnote{At this point $d^3v$ is complex, upon imposing the reality conditions, $d^3v$ winds up to be the canonical three-volume $d^3v=e^1\wedge e^2\wedge e^3$, which is real-valued.}
\begin{subequations}\begin{align}
\psi^A\wedge\phi^B\wedge\phi^C&=-\sqrt{2}u^A\epsilon^{BC}d^3v,\label{three-vol1}\\
\phi^A\wedge\psi^B\wedge\psi^C&=-\sqrt{2}v^A\epsilon^{BC}d^3v.\label{three-vol2}
\end{align}\end{subequations}
The relative normalisation between $u^A$ and $v^A$ can always be reabsorbed into a redefinition of $d^3v$. Without loss of generality, we can then also always assume
\begin{equation}
u_Av^A=1.
\end{equation}
Notice also that the spinors $(\updown{\psi}{A}{a},\updown{\phi}{A}{a})$ naturally define a map from a $\C^4$ bundle over $\varSigma$ into complexified cotangent space $T ^\ast\varSigma$. It is easy to check that the  Weyl spinors $u_A$ and $v_A$ introduced in \eref{three-vol1} and \eref{three-vol2} determine the kernel of this map, i.e.\
\begin{equation}
u_A\updown{\psi}{A}{a}+v_A\updown{\phi}{A}{a}=0.\label{uv-def}
\end{equation}
If the 3-forms $\psi^A\wedge\phi^B\wedge\phi^C$ and $\phi^A\wedge\psi^B\wedge\psi^C$ are linearly independent, i.e.\ if \eref{phipsi-cond} is satisfied, we can use the self-dual 2-forms 
\begin{equation}
    \Sigma^{AB}=\phi^{(A}\wedge\psi^{B)},\label{Sgma-phipsi-3d}
\end{equation}
as a basis of 2-forms in $\bigwedge^2 T^\ast_\C\varSigma$. We can then split the $\SL(2,\C)$ curvature 2-form $\updown{F}{A}{B}$ on $\varSigma$ into components $F_{ABCD}$ according to 
\begin{equation}
\updown{F}{A}{B}=\frac{1}{2}\updown{F}{A}{BCD}\Sigma^{CD}.
\end{equation}
The rank-four curvature spinor $F_{ABCD}=F_{(AB)(CD)}$ can be now naturally split into irreducible spin-0, spin-1 and spin-2 parts
\begin{equation}
\updown{F}{AB}{CD}=\frac{1}{3}S\delta^{(A}_C\delta^{B)}_D-\delta^A_C\updown{T}{B}{D}-\delta^B_D\updown{T}{A}{C}+\updown{\Psi}{AB}{CD},
\end{equation}
where $S$ is a scalar, $T_{AB}=T_{BA}$ is the spin-1 contribution and $\Psi_{ABCD}=\Psi_{(ABCD)}$ is the spin-2 component (Weyl spinor). 

Combining \eref{three-vol1} and \eref{three-vol2} {with the equations of motion} \eqref{EE1} and \eqref{EE2}, it is then easy to show that all but the Weyl component vanishes, 
\begin{equation}
S\approx 0,\quad T_{AB}\approx 0.
\end{equation}
The curvature tensor can be thus written solely in terms of the spin-2 Weyl spinor: 
\begin{equation}
\updown{F}{A}{B}\approx\updown{\Psi}{A}{BCD}\phi^C\wedge\psi^D.\label{F-Weyl-spinr}
\end{equation}
It is important to note that this equation allows us to infer \emph{all} components of the Weyl tensor from data solely intrinsic to $\varSigma$. The left-hand side is the curvature 2-form of the self-dual connection on $\varSigma$. The Weyl spinor on right-hand side, on the other hand, is sufficient to infer all components of the Weyl tensor via
\begin{equation}
    C_{A A'B B'C C'D D'}=\bar{\epsilon}_{\Ap \Bp}\bar{\epsilon}_{C'D'}\Psi_{ABCD}+\CC.
\end{equation}

\subsection{Constraint algebra}\label{sec3.4}
In this section, we list the Poisson brackets between the constraints. First of all, we recall that the Gauss constraint is the generator of local $\SL(2,\C)$ transformations 
\begin{subequations}\begin{align}
\big\{\cG[\alpha],\updown{\psi}{A}{a}\big\}&=-\updown{\alpha}{A}{B}\updown{\psi}{B}{a},\\
\big\{\cG[\alpha],\downup{\hat{\pi}}{A}{a}\big\}&=\updown{\alpha}{B}{A}\downup{\hat{\pi}}{B}{a},\\
\big\{\cG[\alpha],\updown{\phi}{A}{a}\big\}&=-\updown{\alpha}{A}{B}\updown{\phi}{B}{a},\\
\big\{\cG[\alpha],\downup{\hat{\rho}}{A}{a}\big\}&=\updown{\alpha}{B}{A}\downup{\hat{\rho}}{B}{a},\\
\big\{\cG[\alpha],\updown{A}{A}{Ba}\big\}&={-}D_a\updown{\alpha}{A}{B},\\
\big\{\cG[\alpha],\downup{\hat{\Pi}}{AB}{a}\big\}&=2\updown{\alpha}{C}{(A}\downup{\hat{\Pi}}{B)C}{a}.
\end{align}\end{subequations}
These equations immediately imply,
\begin{subequations}\begin{align}
\big\{\cG[\alpha],\psiFF[\eta]\big\}&=-\psiFF[\alpha\eta],\label{Gauss-1}\\
\big\{\cG[\alpha],\phiFF[\zeta]\big\}&=-\phiFF[\alpha\zeta],\label{Gauss-2}\\
\big\{\cG[\alpha],\psiEE[\lambda]\big\}&=-\psiEE[\alpha\lambda],\label{Gauss-3}\\
\big\{\cG[\alpha],\phiEE[\mu]\big\}&=-\phiEE[\alpha\mu],\label{Gauss-4}\\
\big\{\cG[\alpha],\cC[\gamma]\big\}&=-\cC[\alpha\gamma],\label{Gauss-5}\\
\big\{\cG[\alpha_1],\cG[\alpha_2]\big\}&={-}\cG\big[[\alpha_1,\alpha_2]\big],\label{Gauss-6}
\end{align}\label{Gauss-1-6}\end{subequations}
where $\updown{(\alpha\gamma)}{AB}{a}=2\alpha^{C(A}\gamma^{B)}{}_{Ca}$. 
The remaining non-trivial Poisson brackets between the primary constraints are given by
\begin{subequations}\begin{align}
\big\{\phiFF[\zeta],\cC[\gamma]\big\}&=\I\int_\varSigma \zeta_A\wedge \updown{\gamma}{A}{B}\wedge\psi^B,\label{ConstrntAlgbr-1}\\
\big\{\psiFF[\eta],\cC[\gamma]\big\}&=-\I\int_\varSigma \eta_A\wedge \updown{\gamma}{A}{B}\wedge\phi^B,\label{ConstrntAlgbr-2}\\
\big\{\phiFF[\zeta],\psiEE[\lambda]\big\}&=\I\int_\varSigma \zeta_A\wedge\updown{F}{A}{B}\lambda^B,\label{ConstrntAlgbr-3}\\
\big\{\psiFF[\eta],\phiEE[\mu]\big\}&=-\I\int_\varSigma \eta_A\wedge\updown{F}{A}{B}\mu^B,\label{ConstrntAlgbr-4}\\
\big\{\cC[\gamma],\psiEE[\lambda]\big\}&=\I\int_\varSigma \lambda^A\updown{\gamma}{B}{A}\wedge\pi_B,\label{ConstrntAlgbr-5}\\
\big\{\cC[\gamma],\phiEE[\mu]\big\}&=\I\int_\varSigma\mu^A\updown{\gamma}{B}{A}\wedge\rho_B.\label{ConstrntAlgbr-6}
\end{align}\label{ConstrntAlgbr-1-6}\end{subequations}
All other Poisson brackets among the constraints vanish. It is instructive at this point to compare our results with the situation for gravity in terms of ADM variables \cite{adm}. In the ADM formalism, all constraints are first-class, whereas there will be both first-class and second-class constraints in our formalism. This may sound slightly disadvantageous, but the case is subtle. The advantage is that the right hand side of \eref{ConstrntAlgbr-1-6} depends on the fundamental phase space variables at most quadratically (the curvature tensor depends on the connection quadratically). For the ADM (Arnowitt--Deser-Misner) hypersurface deformation algebra the situation is different. The ADM algebra contains structure functions that depend on the inverse metric. 
\subsection{Stability of constraints}\label{sec3.5}
In this section, we identify the conditions that are necessary for the stability of the primary constraints under the Hamiltonian evolution equations. These conditions imply no further secondary constraints. Instead, we only need to impose conditions on the Lagrange multipliers for the constraints to be preserved in time. We comment more about that in \hyperref[sec3.6]{Section \ref{sec3.6}} below, where we will show that the conditions on the Lagrange multipliers coincide with the time-space components of the equations of motion. 

\paragraph{Gauss constraint}
First of all, we note that the Poisson brackets (\ref{Gauss-1-6}) imply that the Hamiltonian \eref{H0-def} preserves the Gauss constraint \eref{G-cons}, i.e. for all $\updown{\alpha}{\prime A}{B}$
\begin{equation}
    \big\{H_o,\cG[\alpha^{\prime}]\big\}\approx 0.
\end{equation}

\paragraph{Fake-flatness constraints}  
For the fake-flatness constraints, we obtain, on the other hand:
\begin{subequations}
\begin{align}
    \big\{H_o,\psiFF[\updown{\eta}{\prime }{}]\big\}&\approx \I\int_\varSigma 
    \eta^\prime_A\wedge\big(\updown{\gamma}{A}{B}\wedge\phi^B+\updown{F}{A}{B}\mu^B\big),\label{H0-R}\\
    \big\{H_o,\phiFF[\zeta^{\prime }]\big\}&\approx -\I\int_\varSigma \zeta^\prime_A\wedge\left(\updown{\gamma}{A}{B}\wedge\psi^B+\updown{F}{A}{B}\lambda^B\right)\label{H0-P}.
\end{align}\end{subequations}
For these constraints to be conserved in time, both \eqref{H0-R} and \eqref{H0-P} must vanish for all choices of $\eta^{\prime A}$ and $\zeta^{\prime A}$. This is possible by fixing the Lagrange multiplier $\updown{\gamma}{A}{Ba}$ in terms of $\lambda^A$ and $\mu^A$. If we set
\begin{equation}
    \updown{\gamma}{A}{B}\stackrel{!}{=} \updown{\gamma}{A}{B}[\lambda,\mu]:=\updown{\Psi}{A}{BCD}\left(\psi^C\mu^D-\phi^C\lambda^D\right),\label{mu-cons}
\end{equation}
 the constraints will be conserved, i.e.\ for all smearing functions $\eta^{\prime A}$ and $\zeta^{\prime A}$,
\begin{subequations}\begin{align}
\big\{H_o,\psiFF[\eta']\big\}&\approx 0,\label{H0-R-var}\\
\big\{H_o,\phiFF[\zeta^{\prime}]\big\}&\approx 0.\label{H0-P-var}
\end{align}\end{subequations}
This is a direct consequence of the decomposition of the curvature tensor in terms of the Weyl spinor \eref{F-Weyl-spinr}, which immediately implies
\begin{subequations}\label{Weyl-spin-with-antisym}
\begin{align}
    &\Psi_{ABCD}\psi^C\wedge\psi^D=\frac{1}{2}\Psi_{ABCD}\epsilon^{CD}\psi_E\wedge\psi^E=0,\\&\Psi_{ABCD}\phi^C\wedge\phi^D=\frac{1}{2}\Psi_{ABCD}\epsilon^{CD}\phi_E\wedge\phi^E=0.
\end{align}
\end{subequations}
It is easy to check that there is no other solution for $\updown{\gamma}{A}{B}$, see \ref{app:LM-constr}.

Next, we consider the stability of the void-momentum constraint \eref{C-cons}. The Poisson brackets (\ref{ConstrntAlgbr-1}--\ref{ConstrntAlgbr-6}) imply
\begin{equation}
\big\{H_o,\cC[\gamma']\big\}\approx -\I\int_\varSigma\left(g(\eta_{(A}\wedge\phi_{B)}-\zeta_{(A}\wedge\psi_{B)})-\lambda_{(A} D\phi_{B)}+\mu_{(A} D\psi_{B)}\right)\wedge\gamma^{\prime AB}.\label{C-stblty}
\end{equation}
For the constraint $\cC[\gamma']\stackrel{!}{=}0$ to be conserved in time, the right-hand side of \eref{C-stblty} must vanish for all $\mathfrak{sl}(2,\C)$-valued 1-forms $\updown{\gamma}{\prime A}{Ba}$.  We thus obtain the condition
\begin{equation}
g(\eta_{(A}\wedge\phi_{B)}-\zeta_{(A}\wedge\psi_{B)})-\lambda_{(A} D\phi_{B)}+\mu_{(A} D\psi_{B)}\stackrel{!}{=}0.\label{uv-cond}
\end{equation}
For any given configuration of $\lambda^A$, $\mu^A$ and $(\psi_{Aa},\phi_{Aa})$, this is a system of $3\times 3=9$ complex-valued linear equations for $\eta_{Aa}$ and $\zeta_{Aa}$, which are $2\times 2\times 3=12$ variables. If $(\psi_{Aa},\phi_{Aa})$ are regular, i.e.\ the three-volume (\ref{three-vol1}, \ref{three-vol2}) is non-degenerate, these equations will be linearly independent such that the solutions to the homogenous equations
\begin{equation}
{}^{\mtext{hom}}\eta_{(A}\wedge\phi_{B)}-{}^{\mtext{hom}}\zeta_{(A}\wedge\psi_{B)}\stackrel{!}{=}0
\end{equation}
span a three-dimensional complex subspace of Lagrange multipliers that can be always added to any given solution to the inhomogenous equations \eref{uv-cond} for $\eta_{Aa}$ and $\zeta_{Aa}$. It is easy to find this subspace, generic elements can be parametrized as
\begin{subequations}\begin{align}
g \,\praupdown{\mtext{hom}}{\eta}{A}{a}=a\updown{\psi}{A}{a}+b\updown{\phi}{A}{a},\label{u-hom}\\
g \, \praupdown{\mtext{hom}}{\zeta}{A}{a}=c\updown{\psi}{A}{a}-a\updown{\phi}{A}{a},\label{v-hom}
\end{align}\end{subequations}
where $a,b,c\in\C$ can be naturally arranged into $\beta\in\slcb$ (see \eqref{beta-slcb}).

In \hyperref[appdxA2]{Appendix \ref{appdxA2}}, we show how to construct a solution to the inhomogeneous equations. 
The inhomogenous solution for the Lagrange multipliers $\updown{\eta}{A}{a}$ and $\updown{\zeta}{A}{a}$ depends parametrically on  $\lambda^A$ and $\mu^A$. Schematically,
\begin{subequations}\begin{align}
\updown{\eta}{A}{a}&=\updown{\eta}{A}{a}[\lambda,\mu],\label{eta-cons}\\
\updown{\zeta}{A}{a}&=\updown{\zeta}{A}{a}[\lambda,\mu].\label{zeta-cons}
\end{align}\end{subequations}

Finally, we turn to the 2-curvature constraints \eref{E-cons} and \eref{H-cons}. If the 2-curvature constraint $\psiEE[\lambda^{\prime A}]=0$ is conserved for all spinor-valued test functions $\lambda^{\prime A}$, we must have that
\begin{align}\nonumber
\big\{H_o,\psiEE[\lambda^{\prime A}]\big\}\approx \I\int_\varSigma\left(g^{-1}D\phi_A\wedge\updown{\gamma}{A}{B}\lambda^{\prime B}+D \zeta_A\wedge D\lambda^{\prime A}\right)
\\
\approx\I \int_\varSigma\left ( g^{-1}D\phi_B\wedge\updown{\gamma}{B}{A}-F_{AB}\wedge \zeta^B\right)\lambda^{\prime A}\stackrel{!}{=}0.
\end{align}Given the constraints and conditions on Lagrange multipliers, we can now show that this equation will be always satisfied.
Taking into account the on-shell decomposition of the curvature tensor into irreducible components, see \eref{F-Weyl-spinr}, as well as the Gauss constraint $D(\psi_{(A}\wedge\phi_{B)})\approx 0$ and the condition on the Lagrange multipliers $\updown{\eta}{A}{a}$, $\updown{\zeta}{A}{a}$ and $\updown{\gamma}{A}{Ba}$, see \eref{mu-cons} and \eref{uv-cond}, we obtain
\begin{subequations}\begin{align}
    F_{AB}\wedge \zeta^B &\overset{\eqref{F-Weyl-spinr}}{\approx}  -\Psi_{ABCD}\psi^C\wedge\phi^D\wedge \zeta^B\approx\\
    &\overset{\eqref{Weyl-spin-with-antisym}}{\approx}  -\Psi_{ABCD}\phi^B\wedge\left(\zeta^C\wedge\psi^D-\eta^C\wedge\phi^D\right)\approx
    \\
    &\overset{\eqref{uv-cond}}{\approx}  g^{-1} \Psi_{ABCD}\phi^B\wedge\left(\lambda^C D\phi^D-\mu^C D\psi^D\right)\approx
    \\
    & \overset{\text{\tiny{Gauss}}}{\approx}  g^{-1}\Psi_{ABCD}D\phi^B\wedge\left(\lambda^C \phi^D-\mu^C \psi^D\right)\approx
    \\
    &\overset{\eqref{mu-cons}}{\approx} g^{-1} \gamma_{AB}\wedge D\phi^B.
\end{align}\end{subequations}
This implies that the 2-curvature constraint $\psiEE[\lambda]=0,\forall\lambda^A$ is conserved on shell provided the Lagrange multipliers $\eta^A$, $\zeta^A$ and $\updown{\gamma}{A}{B}$ satisfy \eref{mu-cons} and \eref{uv-cond}, in other words,
\begin{equation}
\big\{H_o,\psiEE[\lambda^{\prime A}]\big\}\approx 0.
\end{equation}
The calculation for the second 2-curvature constraint proceeds analogously. The evolution of $\phiEE[\mu']=0$ under the Hamiltonian flow for a spinor-valued test function $\mu'^A$ gives
\begin{align}\nonumber
\big\{H_o,\phiEE[\mu^{\prime}]\big\}\approx  -\I\int_\varSigma\left(g^{-1}D\psi_A\wedge\updown{\gamma}{A}{B}\mu^{\prime B}+D \eta_A\wedge D\mu^{\prime A}\right)=\\
\approx  -\I\int_\varSigma\left(g^{-1}D\psi_B\wedge\updown{\gamma}{B}{A}-F_{AB}\wedge \eta^B\right)\mu^{\prime A}.
\end{align}
Taking into account the constraints \eref{R-cons}--\eref{G-cons} and conditions for the Lagrange multipliers \eref{uv-cond} and \eref{mu-cons}, we now have:
\begin{subequations}\begin{align}
    F_{AB}\wedge \eta^B&\approx -\Psi_{ABCD}\psi^C\wedge\phi^D\wedge \eta^B=\\
    &\approx -\Psi_{ABCD}\psi^C\wedge\left(\phi^D\wedge \eta^B-\psi^D\wedge \zeta^B\right)=\\
    &\approx -g^{-1}\Psi_{ABCD}\psi^C\wedge\left(\lambda^DD \phi^B-\mu^D D\psi^B\right)=\\
    &\approx -g^{-1}\Psi_{ABCD}D\psi^C\wedge\left(\lambda^D\phi^B-\mu^D \psi^B\right)=\\
    &\approx g^{-1}\gamma_{AC}\wedge D\psi^C=0.
\end{align}\end{subequations}
This implies that there are no secondary constraints required for the stability of $\phiEE[\mu]=0$ for all spinor-valued test functions $\mu^A$. In other words,
\begin{equation}
\big\{H_o,\phiEE[\mu^{\prime }]\big\}\approx 0,
\end{equation}
where $\approx $ means equality up to terms constrained to vanish through  \eref{R-cons}--\eref{G-cons} and conditions for the Lagrange multipliers \eref{uv-cond} and \eref{mu-cons}.

\bigskip

In summary, there is no constraint on $\lambda^A$ and $\mu^A$. The Lagrange multipliers $\updown{\eta}{A}{a}$, $\updown{\zeta}{A}{a}$, and $\updown{\gamma}{A}{B}$ can be solved for in terms of $\lambda^A$ and $\mu^A$ by imposing the stability conditions for the primary constraints, see \eref{mu-cons}, \eref{eta-cons} and \eref{zeta-cons}. For each choice of $\lambda^A$ and $\mu^A$, there is a corresponding first-class constraint consisting of a linear combination of the 2-curvature, fake-flatness, and void-momentum constraints. Schematically,
\begin{multline}
\mathcal{E}^{\mtext{first-class}}_\pi[\lambda]+\mathcal{E}^{\mtext{first-class}}_\rho[\mu]\\:=\psiEE[\lambda]+\mathcal{E}_\rho[\mu]+\mathcal{F}_\phi\big[\eta[\lambda,\mu]\big]+\mathcal{F}_\psi\big[\zeta[\lambda,\mu]\big]+\mathcal{C}\big[\gamma[\lambda,\mu]\big].
\end{multline}
The Gauss constraint \eref{G-cons} and the 2-curvature constraints (\ref{E-cons}, \ref{H-cons}) weakly commute with the primary Hamiltonian \eref{H0-def}, which is a sum of constraints plus a boundary term. Thus, the Lagrange multipliers for the Gauss constraint \eref{G-cons} and the 2-curvature constraints (\ref{E-cons}, \ref{H-cons}) are independent gauge parameters. These correspond to $3+2+2=7$ complex-valued first-class constraints. Next, we have the $3\times 3=9$ complex-valued void-momentum constraints \eref{C-cons}. The corresponding Lagrange multiplier $\updown{\gamma}{A}{Ba}$ is completely determined by \eref{mu-cons}. This in turn implies that the nine void-momentum constraints \eref{C-cons} are all second-class.  The fake-flatness constraints \eref{R-cons} and \eref{P-cons}, on the other hand, are $2\times 2\times 3=12$ constraints. The corresponding Lagrange multipliers are constrained to satisfy \eref{uv-cond}. This equation has a kernel, which is three-complex-dimensional, see (\ref{u-hom}, \ref{v-hom}). The kernel parametrizes three (complex) first-class constraints, generating an internal $\slcb$ algebra, with the corresponding gauge element given in \eqref{beta-slcb}. The list of first-class constraints consists, therefore,  of four complex-valued constraints corresponding to Lagrange multipliers $\lambda^A$ and $\mu^A$, three complex-valued Gauss constraints for $\mathfrak{sl}(2,\C)$-valued Lagrange multipliers $\updown{\alpha}{A}{B}$, and three additional complex-valued Gauss constraints for $\overline{\mathfrak{sl}(2,\C)}$ given by $\mathcal{F}_\phi[\praupdown{\mtext{hom}}{\eta}{}{}]+\mathcal{F}_\psi[\praupdown{\mtext{hom}}{\zeta}{}{}]$, see \eref{u-hom} and \eref{v-hom}. The four complex degrees of freedom of $\lambda^A$ and $\mu^A$ correspond to the four-dimensional complexified diffeomorphism parameter given in \eqref{diff-gen-gr}.

In total, there are $2\times 2\times 2\times 3+2\times 3\times 3=42$ complex phase-space dimensions. Counting complex-valued constraints, there are in total $2\times 2\times 3=12$ fake-flatness constraints (\ref{R-cons}, \ref{P-cons}), $2\times 2=4$ are 2-curvature  constraints (\ref{E-cons}, \ref{H-cons}), $3\times 3=9$ are  void-momentum constraints \eref{C-cons}, and there are three Gauss constraints \eref{G-cons}. These are 28 constraints of which there are $3+2\times 2+3=10$ first-class and 18 second-class constraints. The reduced physical phase space has $42-2\times 10-18=2\times 2$ dimensions, representing the two complex physical degrees of freedom of self-dual gravity.

\subsection{Interpretation of the Lagrange multiplier constraints as evolution equations}\label{sec3.6}
In this section, we discuss how the constraints that we obtained on the Lagrange multipliers are equivalent to some of the components of the equations of motion. Going back to the definition of the Hamiltonian \eref{H0-def}, and taking into account the fundamental Poisson brackets \eref{Poiss}, we determine the  evolution equations for the configuration variables,
\begin{subequations}\begin{align}
\updown{\dot{\psi}}{A}{a}&=\big\{H_o,\updown{{\psi}}{A}{a}\big\}=g\updown{\eta}{A}{a}+\nabla_a\lambda^A-\updown{\alpha}{A}{B}\updown{\psi}{B}{a},\\
\updown{\dot{\phi}}{A}{a}&=\big\{H_o,\updown{{\phi}}{A}{a}\big\}=g\updown{\zeta}{A}{a}+\nabla_a\mu^A-\updown{\alpha}{A}{B}\updown{\phi}{B}{a},\\
\updown{\dot{A}}{A}{Ba}&=\big\{H_o,\updown{{A}}{A}{Ba}\big\}=\updown{\gamma}{A}{Ba}+\nabla_a\updown{\alpha}{A}{B}.
\end{align}\end{subequations}
The rightmost terms in each line are infinitesimal $\mathfrak{sl}(2,\C)$ frame rotations of the connection 1-forms $\updown{{A}}{A}{B}$, $\updown{{\psi}}{A}{a}$ and $\updown{{\phi}}{A}{a}$. If we bring these terms to the left-hand side, the partial time derivatives combine with the infinitesimal $\mathfrak{sl}(2,\C)$ frame rotations to form the $\SL(2,\C)$-gauged covariant derivative. This, in turn, allows us to immediately infer the covariant meaning of the Lagrange multipliers:
\begin{subequations}\begin{align}
\updown{\eta}{A}{Ba}&=\varphi^\ast_t[\iota_t(\nabla\praupdown{4}{\psi}{A}{})]_a,\label{u-geom}\\
\updown{\zeta}{A}{Ba}&=\varphi^\ast_t[\iota_t(\nabla\praupdown{4}{\phi}{A}{})]_a,\label{v-geom}\\
\updown{\gamma}{A}{Ba}&=\varphi^\ast_t[\iota_t\praupdown{4}{F}{A}{B}]_a,\label{mu-geom}.
\end{align}\end{subequations}
Let us introduce a basis of self-dual 2-forms on the manifold $M$:
\begin{equation}
\frdownup{\Sigma}{AB}{}=\frphi_{(A}\wedge\frpsi_{B)}.\label{Sgma-phipsi}
\end{equation}
If we now compare equation \eref{mu-geom} with the condition on the Lagrange multiplier $\updown{\gamma}{A}{Ba}$ found in \eref{mu-cons} and the constraint \eref{F-Weyl-spinr} on the pullback of the curvature tensor, we see that both equations, namely \eref{F-Weyl-spinr} and \eref{mu-geom}, can be summarized as
\begin{equation}
\praupdown{4}{F}{A}{B}=\updown{\Psi}{A}{BCD}\praupdown{4}{\Sigma}{CD}{},\label{4d-Weyl-spnir}
\end{equation}
such that the condition \eref{mu-cons} and the constraint \eref{F-Weyl-spinr}  on respectively $\gamma$ and curvature tensor $F$,   are merely the temporal and purely spatial components of equation \eref{4d-Weyl-spnir}. 

\smallskip

In the same way, we proceed for the Gauss constraint and the conditions on the Lagrange multipliers $\updown{\eta}{A}{a}$ and $\updown{\zeta}{A}{a}$. If the void-momentum constraint and the fake-flatness constraints are satisfied, the Gauss constraint \eref{G-cons} is merely the statement that
\begin{equation}
D(\psi_{(A}\wedge\phi_{B)})=0.\label{gauss-on-sigma1}
\end{equation}
The exterior covariant derivative $D$ on $\varSigma$ is simply the pull-back of $\nabla$ to $\varSigma$. It acts on both primed and unprimed indices, e.g. 
\begin{equation}
D e_{AA'}=\d e_{A A'}+\updown{A}{B}{A}\wedge e_{B\Ap }+\updown{\Bar{A}}{\Bp}{\Ap }\wedge e_{A B'}.
\end{equation}
Since $\phi_{(A}\wedge\psi_{B)}=\varphi^\ast_t\pradownup{4}{\Sigma}{AB}{}$, see \eref{Sgma-phipsi}, we can write \eref{gauss-on-sigma1} as
\begin{equation}
\varphi^\ast_t(\nabla\pradownup{4}{\Sigma}{AB}{})=0.\label{gauss-on-sigma2}
\end{equation}
The conditions on the Lagrange multipliers $\updown{\eta}{A}{a}$ and $\updown{\zeta}{A}{a}$, \eref{u-geom} and \eref{v-geom}, are given in \eref{uv-cond} above. This condition also involves the Lagrange multipliers $\lambda^A$ and $\mu^A$, which are the temporal components of $\praupdown{4}{\psi}{A}{a}$ and $\praupdown{4}{\phi}{A}{a}$. We thus have
\begin{multline}
\varphi^\ast_t[\iota_t\nabla\pradownup{4}{\Sigma}{AB}{}]=\varphi^\ast_t[\iota_t\nabla(\pradownup{4}{\phi}{(A}{}\wedge \pradownup{4}{\psi}{B)}{})]=\varphi^\ast_t[\iota_t(\nabla\pradownup{4}{\phi}{(A}{}\wedge \pradownup{4}{\psi}{B)}{}-\pradownup{4}{\phi}{(A}{}\wedge \nabla\pradownup{4}{\psi}{B)}{})]=\\
=g(\zeta_{(A}\wedge\psi_{B)}-\eta_{(A}\wedge\phi_{B)})+D\phi_{(A}\lambda_{B)}-D\psi_{(A}\mu_{B)},\label{gauss-on-sigma3}
\end{multline}
which is precisely the condition \eqref{uv-cond}.

\subsection{Secondary constraint from reality conditions}

So far, we have considered  the Hamiltonian formulation of self-dual gravity in terms of the spinor-valued configuration variables $(\updown{\psi}{A}{a},\updown{\phi}{A}{a})$, their conjugate momenta $(\downup{\hat{\pi}}{A}{a},\downup{\hat{\rho}}{A}{a})$, and an auxiliary canonical pair consisting of the self-dual connection $\updown{A}{A}{Ba}$ and its momentum $\downup{\hat{\Pi}}{A}{Ba}$, see \eref{Pi-A-Poiss}. The partial observables \cite{Dittrich:2004cb,Rovelli:2001bz} for this theory are analytic functionals on a complex kinematical phase space. Hence, there are twice as many local degrees of freedom as in the real-world gravity. Let us now discuss here the reality conditions that would take us back to general relativity in terms of real-valued variables from the canonical split perspective. We recover the two real degrees of freedom of general relativity by imposing reality conditions on phase space. 

\smallskip

\paragraph{Torsionless condition and reality conditions} We saw in the previous section that the purely self-dual theory has two complex degrees of freedom per point on $\Sigma$. 
In what follows, we explain how the addition of the reality conditions \eref{realcond-1} alters our system of constraints and reduces the number of physical degrees of freedom further down to two (real) physical degrees of freedom per point on $\Sigma$. 
To achieve this, it will be enough to check how the torsionless condition is compatible with the reality conditions. This can be understood as follows. 

First of all, we note that the Gauss constraint \eref{G-cons} and the Lagrange multiplier conditions \eref{uv-cond} for $\updown{\eta}{A}{a}$ and $\updown{\zeta}{A}{a}$, are merely the spatial and temporal components of the four-dimensional closure constraint on the initial hypersurface:
\begin{equation}
\nabla \pradownup{4}{\Sigma}{AB}{}\approx 0.\label{4d-gauss-on-sigma}
\end{equation}
If the tetrad \eref{tetra-def} is invertible, this equation is the same as to say that the torsion 2-form vanishes
\begin{equation}
\nabla\pradownup{4}{e}{AA'}{}=0.\label{trsn-lss}
\end{equation}
The (primary) reality conditions impose that the four-dimensional soldering forms $\praupdown{4}{e}{A\Ap }{\mu}$ are anti-Hermitian. These in turn imply secondary reality conditions, which impose constraints on the connection that follow from \eref{trsn-lss} provided $\pradownup{4}{e}{A\Ap }{}$ is invertible. If these secondary equations are satisfied, the reality conditions are  preserved in time, which can be immediately seen from
\begin{equation}
\mathcal{L}_t\pradownup{4}{e}{A\Ap }{}=\iota_t\nabla\pradownup{4}{e}{A\Ap }{}+\nabla t^{AA'}\approx \nabla t^{AA'},
\end{equation}
where $\mathcal{L}_t$ is the gauge covariant $\SL(2,\C)$ Lie derivative. Indeed, if the time-flow vector field is real, i.e.\ $\bar{t}^a=t^a\in TM$, such that $t^{AA'}=t^a \praupdown{4}{e}{A\Ap }{a}$ is anti-Hermitian, the reality conditions will be preserved under the time evolution, i.e.\ $\mathcal{L}_t\pradownup{4}{e}{AA'}{}+\mathcal{L}_t\pradownup{4}{e}{A'A}{\dagger}=0$.

Besides the reality conditions and torsionless equations, we also have the Einstein equations $\praupdown{4}{F}{A}{B}\wedge\praupdown{4}{e}{B\Ap }{}=0$. One may wonder if further tertiary reality conditions arise from those. This is not the case. No tertiary constraints arise from those, because the torsionless condition $\nabla\pradownup{4}{e}{AA'}{}=0$ implies the integrability condition $\nabla^2\pradownup{4}{e}{AA'}{}=\pradownup{4}{F}{A}{B}\wedge \pradownup{4}{e}{BA'}{}-\CC=0$, which weakly vanishes already if $\pradownup{4}{F}{B}{A}{}\wedge\praupdown{4}{e}{B\Ap }{}=0$ is satisfied.

To summarise, we are left with the following task: we need to identify the necessary conditions to be imposed on the Lagrange multipliers and phase space variables such that the reality conditions \eqref{realcond-1}, \eqref{tetra-def}  are compatible with the torsionless condition \eref{trsn-lss} under the implicit assumption that $\praupdown{4}{e}{A\Ap }{}$ is invertible.

\paragraph{Spinor frame fields} The reality conditions impose conditions on $e_{A\Ap  a}$ on $\Sigma$. Notice, however,  that by the very definition of our fundamental phase space variables $\updown{\psi}{A}{a}$ and $\updown{\phi}{A}{a}$ in terms of a spin dyad $(o^A, \iota^A)$, see \eqref{dyad}, the soldering forms $e_{A\Ap a}$ on $\Sigma$ have been washed away from our phase space. It is therefore not obvious at this stage how to impose the reality conditions on our variables.
To impose the reality conditions, we proceed as follows. First of all, we reconstruct a soldering form $e_{A\Ap a}$ from the phase space variables $\updown{\psi}{A}{a}$ and $\updown{\phi}{A}{a}$. This requires a choice of spinor frame. To choose such a spinor frame, notice first that the bispinors $(\updown{\psi}{A}{a},\updown{\phi}{A}{a})$ on $\Sigma$ define for every $x\in\Sigma$ a map from the complexified tangent bundle $[T_x\Sigma]_\C$, which is a three-dimensional complex vector space, into $\C^2\oplus \C^2$, which is complex four-dimensional. This map will have a kernel, which defines a spin dyad $(u^A,v^A)$, see \eref{uv-def} above. In what follows, we restrict ourselves to the case\footnote{The degenerate case in which $u_Av^A=0$ corresponds to the case in which the initial surface $\Sigma$ is light-like.} in which $u_Av^A\neq 0$. Without loss of generality, we may then also assume $u_Av^A=1$. Any such spin dyad defines a natural Hermitian metric 
\begin{align}
\delta_{A\Ap }:=u_A\bar{u}_{\Ap }+v_A\bar{v}_{\Ap }.\label{delta-def}
\end{align}
Given such a dyad, we can then also introduce the dressed Pauli matrices:
\begin{subequations}\begin{align}
\updown{\sigma}{A}{B1}&=u^Au_B-v^Av_B,\label{Pauli1}\\
\updown{\sigma}{A}{B2}&=-\I u
^Au_B-\I v^Av_B,\label{Pauli2}\\
\updown{\sigma}{A}{B3}&=-u^Av_B-v^Au_B,\label{Pauli3}
\end{align}\label{Pauli1-3}\end{subequations}
which form a basis in the space of traceless $2\times 2$ Hermitian matrices, i.e.\
\begin{equation}
\delta_{B\Ap }\updown{\sigma}{B}{Ai}-\delta_{A\Bp}\updown{\bar{\sigma}}{\Bp}{\Ap i}=0.
\end{equation}
\paragraph{Reality conditions on $^4e_{A\Ap }$}
We can now define the matrix-valued 1-form 
\begin{equation}
e_{A\Ap a}=\psi_{Aa}\bar{u}_{\Ap }+\phi_{Aa}\bar{v}_{\Ap }.
\end{equation}
Since $u,v$ are in the kernel, \eref{uv-def}, this matrix $e_{A\Ap a}$  has vanishing contractions with $\delta^{A\Ap }$. We can thus decompose it into the Pauli matrices (\ref{Pauli1-3}): 
\begin{equation}
e_{A\Ap a}=\frac{\I}{\sqrt{2}}{\sigma}_{A\Ap i}\updown{e}{i}{a},
\end{equation}
where we set
\begin{equation}
{\sigma}_{A\Ap i}=-\delta_{B\Ap }\updown{\sigma}{B}{Ai},\label{Pauli-matrx2}
\end{equation}
and we recall  that 
 $$\updown{e}{i}{a}=\frac{\I}{\sqrt{2}}\sigma^{A\Ap i}[\varphi^\ast_{t}\praupdown{4}{e}{}{A\Ap }]_a.
 $$
Note that at this stage $\updown{e}{i}{a}$ can be complex. To guarantee that ${}^4e_{A\Ap }$ is anti-Hermitian, we impose now the following reality conditions on the triads 

\begin{equation}
\updown{e}{i}{a}=(e^i{}_a)^\ast.\label{realcond-1}
\end{equation}
In addition, we need to impose corresponding reality conditions on the time components of the tetrad, specified in terms of the Lagrange multipliers $\lambda^A$ and $\mu^A$. Using the basis consisting of the (dressed) Pauli matrices \eref{Pauli-matrx2} and the Hermitian metric \eref{delta-def}, we can decompose the full coframe field as
\begin{equation}
\praupdown{4}{e}{}{A\Ap }=\frac{\I}{\sqrt{2}}\left[N^0\delta_{A\Ap }\d t+\sigma_{A\Ap i}\left(e^i+N^i\d t\right)\right],\label{tetra-def}
\end{equation}
where $N^i=(N^1,N^2,N^3)$ is the shift vector. 
The reality conditions on the lapse and shift are then naturally
\begin{equation}
{(N}^0)^*=N^0,\quad ({N}^i)^*=N^i.
\end{equation}
This means that the Lagrange multipliers $\lambda$ and  $\mu$, defined in \eqref{lambd-def1} and  \eqref{lambd-def2} respectively as the time component of the spinor fields $\psi$ and $\phi$, can be written as 
\begin{subequations}\begin{align}
\lambda^A&=-\frac{\I}{\sqrt{2}}(N^0-N^3)v^A-\I\sqrt{2}N^+u^A,\\
\mu^A&=\frac{\I}{\sqrt{2}}(N^0+N^3)u^A+\I\sqrt{2}N^-v^A,
\end{align}\end{subequations}
where
\begin{equation}
N^\pm=\frac{1}{2}(N^1\mp\I N^2).
\end{equation}

\smallskip

The torsionless condition \eqref{trsn-lss} can be easily split into its spatial and spacetime components.
In this way, we can infer the time derivatives of $e^i$ and infer secondary constraints and further conditions on the Lagrange multipliers  following from the stability of the primary constraints.  To do so, it is useful to introduce spin rotation coefficients $\updown{A}{i}{a}$ relative to the spinor dyad $(u^A,v^A)$. These spin connections define a dressed version of the self-dual Ashtekar connection.

\paragraph{Spatial covariant derivatives of the spin dyad: dressed Ashtekar connection}
In here, we introduce the \emph{dressed} Ashtekar connection  $\updown{A}{i}{a}$, relative to the $(u^A,v^A)$ spin dyad. We define $\updown{A}{i}{a}$ by setting
\begin{subequations}\begin{align}
D_au^A&=\frac{1}{2\I}\updown{\sigma}{A}{Bi}\updown{{A}}{i}{a}u^B,\label{D-u}\\
D_av^A&=\frac{1}{2\I}\updown{\sigma}{A}{Bi}\updown{A}{i}{a}v^B,\label{D-v}
\end{align}\end{subequations}
where $\updown{A}{i}{a}$ is a $\C^3$-valued 1-form on $\Sigma$. We introduce its real and imaginary parts,
\begin{equation}
\updown{A}{i}{a}=\updown{\Gamma}{i}{a}+\I\updown{K}{i}{a},
\end{equation}
where both $\updown{\Gamma}{i}{a}$ and $\updown{K}{i}{a}$ are $\R^3$-valued 1-forms on $\Sigma$. Since the Hermitian metric $\delta$ and the Pauli matrices $\sigma_i$ are expressed themselves in terms of the spin dyad alone, we can calculate
that 
\begin{subequations}\begin{align}
D_a\delta_{A\Ap }&=\sigma_{A\Ap i}\updown{K}{i}{a},\label{K-def}\\
D_a\sigma_{A\Ap i}&=\delta_{A\Ap }K_{ia}+\downup{\varepsilon}{li}{k}\sigma_{A\Ap k}\updown{\Gamma}{l}{a},
\end{align}\end{subequations}
where $\varepsilon_{ijk}$ is the three-dimensional internal Levi-Civita tensor. The first equation informs us about the geometric role of $\updown{K}{i}{a}$: Taking into account that the Hermitian metric $\delta_{A\Ap }$ defines the internal normal to the hypersurface (recall $\delta_{A\bar A}e^{A\Ap }=0$), we can identify the $\R^3$-valued 1-form $\updown{K}{i}{a}$ on $\Sigma$ with  the \textit{extrinsic curvature}.

\paragraph{Time derivative of the spin dyad: Lagrange multipliers}
In the same way, we can deal with the time derivatives of the spin dyad. By its very definition, the spin dyad is a functional of the fundamental phase space fields $\updown{\psi}{A}{a}$ and $\updown{\phi}{A}{a}$, see \eref{uv-def}. Hence, in general we will have $\{H,u^A\}\neq 0$ and $\{H,v^A\}\neq 0$. Since we deal with a normalized spin dyad, and since any two normalized spin dyads can be mapped into each other by an element of $\mathrm{SL}(2,\C)$, the time derivatives $\{H,u^A\}\neq 0$ and $\{H,v^A\}\neq 0$ determine an element of $\mathfrak{sl}(2,\C)$. This $\mathfrak{sl}(2,\C)$ Lie algebra element can  be then split into a rotation component $\varphi$ and a boost component $\chi$. Furthermore, to restore covariant time derivatives, it is useful to subtract the time component of the Lorentz convection, which is given by the Lagrange multiplier $\alpha\in \mathfrak{sl}(2,\C)$, see \eref{lambd-def} above. We thus write
\begin{subequations}\begin{align}
\frac{\di}{\di t}u^A&:=\big\{H_o,u^A\big\}=\frac{1}{2\I}\updown{\sigma}{A}{Bi}(\varphi^i+\I\chi^i-\alpha^i)u^B,\label{dot-u}\\
\frac{\di}{\di t}v^A&:=\big\{H_o,v^A\big\}=\frac{1}{2\I}\updown{\sigma}{A}{Bi}(\varphi^i+\I\chi^i-\alpha^i)v^B.\label{dot-v}
\end{align}\end{subequations}
The combination $\varphi^i+\I\chi^i$ is the analogue of the selfdual Ashtekar connection, but now in time rather than space. 
In what follows, the real and imaginary parts $\varphi^i$ and $\chi^i$ play the role of Lagrange multipliers. This is not too surprising, given their nature as the time components of a 1-form on $M$. In a similar way,
$\alpha^i$  parametrizes the $\mathfrak{sl}(2,\C)$-valued Lagrange multiplier $\updown{\alpha}{A}{B}$, see \eref{lambd-def}, via
\begin{equation}
\updown{\alpha}{A}{B}=\frac{1}{2\I}\updown{\sigma}{A}{Bi}\alpha^i\equiv \varphi^\ast_t[\iota_t\praupdown{4}{A}{A}{B}].
\end{equation}

\paragraph{Secondary constraints}
Using all these preparations, we can now split the torsionless condition \eref{trsn-lss} into its time and space components. With the reality conditions given in \eref{realcond-1}, the spacetime components of the torsionless condition impose
\begin{subequations}\begin{align}
\chi_ie^i&=\d N^0+K_iN^i,\label{bst-angl}\\
\dot{e}^i&=N^0K^i+\d N^i+\updown{\varepsilon}{i}{lk}\Gamma^lN^k+\updown{\varepsilon}{i}{lk}\varphi^le^k,\label{e-evolv}
\end{align}\end{subequations}
where $\dot{e}^i=\iota_t \d e^i$ is the time-derivative of the triad, see \eref{tetra-def}. Equation \eref{bst-angl} is a condition on the Lagrange multiplier $\chi^i$. Equation \eref{e-evolv}, on the other hand, is merely an evolution equation for the triads. 

On the other hand, we  also have the purely spatial components of the torsion 2-form which are real and given by
\begin{subequations} \begin{align}
K_i\wedge e^i&=0,\label{SU2-Gauss}\\
\d e^i+\updown{\varepsilon}{i}{jk}\Gamma^j\wedge e^k&=0.\label{3d-trsn}
\end{align}\end{subequations}
The first of these implies the familiar condition that the extrinsic curvature is symmetric, i.e.\ $K_{ab}=K_{ib}\updown{e}{i}{a}=K_{ba}$. The second equation implies an additional secondary constraint: $\updown{\Gamma}{i}{a}$ is constrained to be the three-dimensional spin connection associated to the triad $e^i$. Equation \eref{SU2-Gauss} represents three equations per point on the initial hypersurface, and equation \eref{3d-trsn} represents $3\times 3=9$ conditions. It is important to note, however, that there is a redundancy in the description. If $e^i=({e}^i)^\ast$, all of \eref{SU2-Gauss} and three out of the nine conditions of \eref{3d-trsn} follow already from
\begin{align}
&D(\psi_{(A}\wedge\phi_{B)})=0\Leftrightarrow K_i\wedge e^i=0\;\text{and}\; e^{[i}\wedge \d e^{j]}+e^{[i}\wedge \updown{\varepsilon}{j]}{kl}\Gamma^k\wedge e^l=0.
\end{align}
Thus, \eref{SU2-Gauss} is already contained in the Gauss law \eref{G-cons}, and so are three of the nine torsionless conditions given in \eref{3d-trsn}. The six missing components of the torsionless equations, which are independent of \eref{G-cons}, are given by the following secondary constraints:  
\begin{align}
e^{(i}\wedge\d e^{j)}+e^{(i}\wedge\updown{\varepsilon}{j)}{kl}\Gamma^k\wedge e^l&=0.\label{scndry-cons}
\end{align}
In summary,  
by imposing reality conditions on $e^i$ we found that the torsionless equation imposes an 
 an additional condition on the Lagrange multiplier $\chi^i$, see \eref{bst-angl}, and the  secondary constraint \eref{scndry-cons}. 
It is straightforward to check that the reality conditions \eref{realcond-1} and the resulting secondary constraints \eref{scndry-cons} are second-class.\footnote{The probably simplest way to see this is to impose these constraints by introducing auxiliary Lagrange multipliers and adding the constraints to the action. It is then easy to see that the resulting field equations are strong enough to imply that these Lagrange multipliers are constrained to vanish. From a Hamiltonian perspective, this can only happen if the constraints are second-class.}

\paragraph{Counting real degrees of freedom}
To conclude, let us count the dimensions of the resulting physical phase space. At the kinematical level, the phase space dimension is
\begin{equation}
\underbrace{2\times 2\times 2\times 3}_{\updown{\psi}{A}{a},\downup{\hat{\pi}}{A}{a}}+\underbrace{2\times 2\times 2\times 3}_{\updown{\phi}{A}{a},\downup{\hat{\rho}}{A}{a}}+\underbrace{2\times 2\times 3\times 3}_{\updown{A}{A}{Ba},\downup{\hat{\Pi}}{A}{Ba}}=84.
\end{equation}
First-class constraints, on the other hand, are in one-to-one correspondence with the undetermined Lagrange multipliers, these are given by
\begin{equation}
\underbrace{1+3}_{N^0,N^i}+\underbrace{3+3}_{\updown{\alpha}{A}{B}}+\underbrace{3}_{\varphi^i}=13.
\end{equation}
Finally, we have the second-class constraints. We infer them by considering first the conditions on the Lagrange multipliers. First of all, we recall that $\updown{\gamma}{A}{Ba}$ is fully constrained, see \eref{mu-cons} above. Next, we have the Lagrange multipliers $\updown{\eta}{A}{a}$ and $\updown{\zeta}{A}{a}$. Reinstating a covariant notation, we have
\begin{subequations}\begin{align}
\updown{\eta}{A}{a}&=-\varphi^\ast_t[\iota_t(\pradownup{4}{e}{A\Ap }{}\wedge\nabla \bar{u}^{A'})]_a,\label{u-def-4d}\\
\updown{\zeta}{A}{a}&=-\varphi^\ast_t[\iota_t(\pradownup{4}{e}{A\Ap }{}\wedge\nabla \bar{v}^{ A'})]_a.\label{v-def-4d}
\end{align}\end{subequations}
Going back to \eref{dot-u}, \eref{dot-v} and \eref{D-u}, \eref{D-v}, \eref{tetra-def}, we can see that the right-hand sides of these two equations are merely linear combinations of the Lagrange multipliers $N^0, N^i$ and $\varphi^i$ and $\chi^i$. Now, $\chi^i$ is determined by \eref{bst-angl} as a linear combination of $\di N^0$ and $N^i$ as well. The terms that are linear in $N^0$, $\di N^0$ and $N^i$ contribute to the diffeomorphism generators.  As in above, see \eref{u-hom}, \eref{v-hom}, the general solution for the $2\times 2\times 2\times 3=24$ Lagrange multipliers $\updown{\eta}{A}{a}$ and $\updown{\zeta}{A}{a}$ can be written as a particular solution, which is linear in $N^0$, $\d N^0$ and $N^i$, and a homogenous solution, which is linear in $\varphi^i$. The homogenous solution that is linear in $\varphi^i$ generates an additional internal $\SU(2)$ symmetry. This additional $\SU(2)$ symmetry generates rotations preserving the Hermitian metric $\delta_{A\Ap }$, see \eref{delta-def}. The $2\times 2\times 2\times 3=24$ fake-flatness constraints \eref{R-cons} and \eref{P-cons} split, therefore, into three first class constraints (corresponding to the Lagrange multiplier $\varphi^i$, which appears in no other constraint) and $24-3=21$ second-class constraints, whose multipliers are fully constrained by $N^0$ and $N^i$.  In summary, there are $24-3+6\times 3=39$ primary second-class constraints. In addition, there are the $3\times 3=9$ reality conditions \eref{realcond-1} on $e^i$. These in turn imply secondary constraints \eref{scndry-cons}, which are six further second-class constraints. The total number of second class constraints is,
\begin{equation}
\underbrace{24-3}_{(\updown{\eta}{A}{a},\updown{\zeta}{A}{a})/\varphi^i}+\underbrace{6\times 3}_{\updown{\gamma}{A}{Ba}}+\underbrace{3\times 3}_{e^i=({e}^i)^\ast}+\underbrace{6}_{\Gamma_{(ij)}=0}=54.
\end{equation} 
We conclude that the physical phase space has
\begin{equation}
\boxed{84-2\times 13-54=2\times 2}
\end{equation}
phase space dimensions, representing \textit{two physical degrees of freedom} per point on $\Sigma$.

\section{Particle coupling} \label{sec:part}

In this section, we demonstrate that a spinless  particle can be incorporated in exactly the same manner in both the topological and gravitational sectors of our framework.

This highlights a key advantage of our approach compared to the Pleba\'{n}ski formulation, which is conceptually the closest. In the Pleba\'{n}ski model, one begins with a purely topological 
BF-theory, in which no explicit frame field is present. Point-like defects can then be introduced as sources for the 2-curvature (the covariant derivative of the $B$-field) or, equivalently, for the Gauss constraint. However, once the simplicity constraints are imposed, this equation becomes related to torsion. As a result, such topological defects are naturally interpreted as sourcing spin degrees of freedom, but not mass degrees of freedom that would directly couple to a frame field.
To introduce genuine massive degrees of freedom, one must reconstruct the frame-field, or metric, from the 
$B$-field—via the Ur\-ba\-nt\-ke  \cite{HUrbantke:1984} construction, for instance. This reconstruction involves highly non-linear functions of the $B$-field, making the coupling to matter and its quantization technically cumbersome.

An alternative strategy is provided by the McDowell–Mansouri approach \cite{Freidel:2006hv, fairbain-2008}, where both the frame field as well as the spin connection are already present at the topological level. In that setting, the full set of particle charges (momentum and spin) can be introduced directly through topological defects, and the transition from the topogical action to the gravitational regime does not alter their structure. This feature makes the framework particularly appealing for spinfoam models, where the addition of standard matter fields is notoriously difficult.

In the spinorial formulation we adopt here, the frame field is likewise present already at the topological level. Consequently, a massive spinless particle can be coupled in the same way to both the topological and gravitational theories.
By contrast, internal spin degrees of freedom couple directly to the spin connection, i.e.\ the coupling is mediated by the local Lorentz symmetry. Since the purely topological sector does not have such a symmetry, spin does not couple in the same universal manner at that level. We stress that spin can still be introduced at the topological level as well, but it is not so enlightening.   For this reason, we restrict the present discussion to the spinless case.

\medskip 

Let the worldline $W=\{p\in M:\exists \tau\in\R: x(p)=z(\tau)\}$ be a one-dimensional submanifold of $M$ parametrized by $\tau\in\R$,  with 
\begin{align*}
    z:&\,\R\rightarrow \R^4,\\ &\tau\rightarrow z(\tau)
\end{align*}
 its coordinate embedding. 
Consider then the following worldline action:
\begin{align}
    S_W&=\int_W \varphi^*_W(p_{A}\phi^{ A}-q_A\psi^A)\equiv \int_W(p_{A}\phi_{\mu}^A-q_{A}\psi^A_\mu)\dot{z}^\mu \d\tau,\label{mass-defect-in-spinors}
\end{align}
where $\dot f(\tau)\equiv\frac{d}{d\tau}f(\tau)$ and $\varphi^*_W$ denotes the pullback to the worldline $W$. The contributions $p_{A}$ and $q_A$ will characterize the massive degree of freedom with the mass-shell constraint
\begin{equation}
     p_{A}\,q^{A}=-\frac{m^2}{2},\label{mass-shell}
\end{equation}
which can be introduced through a Lagrange-multiplier term in the initial action:
\begin{equation}\label{Skappa}
S_{\varkappa}=\int_W\varkappa\left(p_{A}\,q^{A}+\frac{m^2}{2}\right)\d\tau.
\end{equation}
Such a defect can be added to either the topological \eqref{action-topo} or the gravitational \eqref{action-Robinson} action without change. Its dynamics and coupling to the field theory is determined by the action
\begin{equation}
    S_m=S_W+S_\kappa.\label{full-def-action}
\end{equation}
Note that, due to the condition \eqref{mass-shell}, a pair of spinors $p_A$ and $q_A$ can be used to define a spin dyad $(o_A,\iota_A)$ on a worldline, which then selects a preferred reference frame in $\C^2$.

\paragraph{Equations of motion in the topological case}

The  variation of the action \eqref{action-topo} with the defect action $S_m$  with respect to the spinors $\psi$, $\phi$ modifies the 2-flatness equations (since these defects are point-like):
\begin{subequations}\begin{align}
    \d\pis_A&=\I q^W_{A},\\
    \d\pif_A&=-\I p^W_{A},
\end{align}\end{subequations}
 where we introduced the 3-form source $p^W_A(x)$, which describes an embedding of the world-line momentum into the manifold:
\begin{equation}
    p^W_A(x)=\frac{1}{6}\int_W\d\tau\,{\epsilon}_{\mu\nu\lambda\rho}{\delta}^{(4)}(x-z(\tau))\Dot{z}^\mu(\tau)p_A(\tau) \d x^\nu\wedge \d x^\lambda \wedge \d x^\rho.
\end{equation} 
The variation with respect to the embedding coordinate $z$ leads to 
\begin{equation}
    \psi_{\mu }^A\partial_\tau q_A-\phi_{\mu}^A\partial_\tau p_A+\left(p_A(\partial_\mu\phi_{\nu}^A-\partial_\nu\phi_{\mu}^A)-q_A(\partial_\mu\psi_{\nu }^A-\partial_\nu\psi_{\mu }^A) \right)\dot{z}^\nu=0.\label{MP-topo}
\end{equation}
The term proportional to $\dot z^\mu$ is similar to  a torsion contribution (as derivative of frame field components). 

\paragraph{Equations of motion in the gravitational case}
We now introduce  local Lorentz invariance. The action $S_W$ is not affected. Let ${\nabla}$ denote the self-dual covariant derivative: ${\nabla}=\d +A$, as before.

Performing variation of the action \eqref{action-Robinson}   with respect to the spinor fields $\phi,\psi$, we recover now the analogue of Einstein equation with a source massive particle:  
\begin{subequations}\begin{align}
    \nabla\pis_A&=\I q^W_{A},\\
    \nabla\pif_A&=-\I p^W_{A}.
\end{align}\end{subequations}
Since the matter content we introduce did not explicitly depend on the spin connection or the momentum variables, the rest of the equations of motion (\ref{fake-flatness-pi-psi}--\ref{fake-flatness-pi-phi}) are not modified.

The variation with respect to the embedding coordinate $z$ leads to the  equivalent of the  geodesic equation  in the spinorial variables, which takes the same form as in the topological case \eqref{MP-topo}.

\paragraph{Variations over momentum}

Variation of \eqref{full-def-action} with respect to the defect momenta $p$ and $q$ gives the same equations in the topological and gravitational cases:
\begin{subequations}
    \begin{align}
        &\psi^A_\tau+\varkappa p^A=0,\label{p-k-eqn}\\
        &\phi^A_\tau+\varkappa q^A=0,\label{q-k-eqn}
    \end{align}\label{kappa-eom}
\end{subequations}
where $\psi_\tau\equiv \psi_\mu\dot{z}^{\mu}$. Each of the latter equations can be easily solved for $\varkappa$ on the mass shell \eqref{mass-shell}. Contraction of the equations \eqref{p-k-eqn} and \eqref{q-k-eqn} with $q_A$ and $p_A$ respectively results in
\begin{equation}
    \varkappa=\frac{2}{m^2}p_{A}\phi^A_\tau=-\frac{2}{m^2}q_{A}\psi^A_\tau.
\end{equation}
This implies a constraint: 
\begin{align}
    p_{A}\phi^A_\tau=-q_{A}\psi^A_\tau,\label{matter-eom-2}
\end{align}
the meaning of which will be discussed below.

\paragraph{Equivalence with the standard formulation}
Let us now show that the action \eqref{full-def-action} actually describes a non-spinning massive particle. For this, we use the definition of a particle as a Poincar\'e preferred reference frame \cite{Balachandran:1983oit,Freidel:2006hv}.  

Contribution to the action due to a point defect with a momentum $P^I$ is given by
\begin{equation}
    S_m=\int_W P_{I}e^I_\mu \dot{z}^\mu \d\tau, 
    \label{balach-mass-action}
\end{equation}
supplemented by a mass-shell condition to be imposed as a constraint. 

We apply the transformation \eqref{spinor-from-lor-vector} to convert the Lorentz indices into the spinor ones:
\begin{align}
&P_{AA'}=\frac{\I}{\sqrt{2}}\sigma^I_{AA'}P_{I}.
\label{Lambda-into-spinors}
\end{align}
We introduce the spinor components
\begin{subequations}\label{momentum-split}
\begin{align}
    &p_{A}=P_{AA'}\iota^{A'},\\
    &q_{A}=-P_{AA'}o^{A'},
\end{align}
\end{subequations}
so that 
\begin{equation}
    P_{AA'}:=p_{A}o_\Ap +q_{A}\iota_\Ap .
\end{equation}
Then,
\begin{equation}
    P^IP_I=-m^2 \quad\iff\quad P_{AA'}P^{AA'}=-m^2 \quad \iff \quad  p_{A}\,q^{A}=-\frac{m^2}{2}
\end{equation} Substitution of \eqref{Lambda-into-spinors}, \eqref{momentum-split} and \eqref{lor-vector-from-spinor} into the action \eqref{balach-mass-action} gives exactly \eqref{mass-defect-in-spinors}. We then add the Lagrange-multiplier term $S_\kappa$ to restore the mass-shell condition.

The equation of motion \eqref{MP-topo} is easily recognized to be a geodesic equation \cite{Balachandran:1983oit}:
\begin{equation}
    \dot{P}_{\mu} = \dot{z}^{\nu} P_I\partial_{\mu} e^{I}_\nu.
\end{equation}
In order to understand equation \eqref{matter-eom-2}, let us rewrite it as follows:
\begin{subequations}\begin{align}
    0&=p_{A}\phi^A_\tau+q_{A}\psi^A_\tau=-P_{AA'}\iota^\Ap  e_\tau^A{}_\Bp  o^\Bp-P_{AA'} o^\Ap  e_\tau^A{}_\Bp \iota^\Bp=\\&=-P_{AA'}e_\tau^A{}_\Bp(\iota^\Ap  o^\Bp+o^\Ap \iota^\Bp)=-2P^\mu e_{\mu AA'}e_\nu^A{}_\Bp \dot{z}^\nu\iota^{(\Ap } o^{\Bp)}=\\&=-2P^\mu e_{\mu A(A'}e_\nu^A{}_{\Bp)} \dot{z}^\nu \iota^{\Ap } o^{\Bp}=-2P^\mu e_{[\mu AA'}e_{\nu]}^A{}_{\Bp} \dot{z}^\nu \iota^{\Ap } o^{\Bp}=\\
    &=-2P^{[\mu} \dot{z}^{\nu]} e_{\mu AA'}e_{\nu}^A{}_{\Bp} \iota^{\Ap } o^{\Bp}\sim P^{[\mu} \dot{z}^{\nu]}\Sigma_{\mu\nu \Ap \Bp},\label{new-eom-expansion}
\end{align}\end{subequations}
where $\Sigma_{\mu\nu \Ap \Bp}$ is a basis of anti-self-dual 2-forms on spacetime. To bring this to the form that appears in the literature \cite{Balachandran:1983oit}, we use tetrads and rewrite:
\begin{equation}
    P^{[\mu} \dot{z}^{\nu]}=P^Ie_I^{[\mu} \delta_\lambda^{\nu]}\dot{z}^\lambda=P^Ie_I^{[\mu} e_J^{\nu]}e^J_\lambda\dot{z}^\lambda=P^Ie_{[I}^{\mu} e_{J]}^{\nu}e^J_\lambda\dot{z}^\lambda=P^{[I}e^{J]}_\lambda\dot{z}^\lambda e_I^{\mu} e_J^{\nu}.
\end{equation}
Substituting this back into \eqref{new-eom-expansion}, we get
\begin{equation}
    0=P^{[I}e^{J]}_\lambda\dot{z}^\lambda \Sigma_{IJ \Ap \Bp},
\end{equation}
where $\Sigma_{IJ \Ap \Bp}$ is now a basis of anti-self-dual 2-forms in the internal Minkowski space. Hence, \eqref{matter-eom-2} turns out to be the angular momentum conservation equation (see e.g. \cite{Balachandran:1983oit}):
\begin{equation}
    P^{[I}e^{J]}_\lambda\dot{z}^\lambda\equiv J^{[IJ]}=0.
\end{equation}
This concludes the proof that \eqref{full-def-action} defines a  particle in the usual sense. Our formulation allows its identical expression in both topological and gravitational cases, as well as a convenient massless limit implemented at $S_\kappa$. We also note that Weyl spinors can be coupled to the dual sector analogously.

\section*{Outlook and Discussion} \label{sec:outlook}
The geometric formulation of four-dimensional gravity admits many equivalent descriptions: the metric formulation, the teleparallel approach, the Pleba\'{n}ski formulation, the Einstein–Cartan theory, and the McDowell–Mansouri construction, to name a few. Each of these frameworks relies on different fundamental variables---in particular, different notions of connection. While the solution space is classically equivalent,   technical advantages differ significantly when it comes to quantization.

The metric formulation is historically the most widely used. However, its Hamiltonian description involves constraints that are non-polynomial in nature---such as square roots and inverse powers of the metric. This complicates canonical quantization. The Einstein–Cartan formulation, by treating the frame field and spin connection as independent variables, alleviates some of these difficulties and {has} naturally led to the canonical framework of loop quantum gravity. The Pleba\'{n}ski formulation, recasting gravity as a constrained topological BF-theory, provides a powerful setting for path integral quantization: the topological theory serves as a backbone, and gravity is recovered by appropriately implementing the simplicity constraints. This strategy underlies the four-dimensional spinfoam approach. The McDowell-Mansouri formulation, although comparatively less explored, offers an intriguing perturbative perspective, in which gravity emerges from symmetry breaking of a topological theory with a larger gauge group.

In this work, we investigate yet another formulation, based on spinor-valued $p$-forms. In doing so, we have rediscovered Robinson’s action \cite{Robinson_1996} and carried out a careful and systematic analysis of its gauge symmetries, charges, and Hamiltonian structure. The metric data is encoded in a pair of spinor-valued 1-forms. However, one can first define a purely topological theory formulated  in terms of these spinorial variables {alone}. This topological theory carries, in addition to its (higher) gauge symmetries, a global $\mathrm{SL}(2,\mathbb{C})$ symmetry. Gravity arises precisely when this global symmetry is promoted to a local one. Since gravity is obtained by gauging a global symmetry, many structural properties of the topological theory survive this procedure. In particular, the kinematical phase space remains the same, and symmetry charges are  matched between the two theories.    
This mechanism is reminiscent of the Pleba\'{n}ski construction, although the role of constraints differs: instead of imposing simplicity constraints, we introduce a new constraint which is at the same time the generator of a gauge symmetry. This is markedly different {from} the Pleba\'{n}ski construction, in which the simplicity constraints do not correspond to gauge symmetries. They contain second-class constraints that do not close under the Poisson bracket. The {simplicity} constraints can only be imposed weakly at the quantum level, which can be understood very clearly already at the level of certain effective models \cite{Asante:2020qpa}.

Within our formulation, we have also identified a novel type of a (non-topological) theory emerging in the limit $G \rightarrow 0 $. Previous studies of this regime \cite{Smolin:1992wj,Varadarajan:2018tei,Zarate:2025erv,Bakhoda:2024mth,Bakhoda:2020ril,Bakhoda:2020fiy,Thiemann:2022all,Sahlmann:2024pba} typically relied on rescaling the {gauge} connection, so that in the limit the ($\SLC$) gauge symmetry effectively became abelian.
In the spinorial framework, however, one can instead rescale the frame field. This leads to a qualitatively different realization of the $G \rightarrow 0 $ limit, probing a new sector of the theory, which is not captured by the usual contraction of the gauge algebra nor the standard perturbation theory. The resulting theory is not purely topological and retains non-trivial dynamics, although its precise physical interpretation remains to be clarified.

Although not emphasized extensively, the underlying four-dimensional topological theory is naturally formulated as a higher gauge theory \cite{baez, Borsten_2025}.  It is 
interesting to note the similarities with the interpretation of teleparallel gravity as a 2-gauge
theory \cite{Baez_2014}: in our work, we treated some components of torsion as a 2-connection. However, unlike in the earlier realization of this idea \cite{Baez_2014}, our
1-connection is the (spinorial) coframe field, rather than the Lorentz connection. A closer
study of the teleparallel theory from this perspective could be interesting.

\medskip 

\noindent Our results indicate that the spinorial formulation possesses several promising features for quantization. Let us therefore highlight several specific  properties,   which are currently under investigation.

From the path integral perspective, one may again use the topological theory as the starting point and implement the constraint that promotes the Lorentz symmetry to a local one. Unlike in the Pleba\'{n}ski case, the construction would not rely on simplicity constraints, which are  second-class. Here,  first-class constraints break the topological shift symmetries. The strategy is therefore fundamentally different from standard spinfoam models.

As we alluded previously, 2-gauge structures characterize the topological theory.
The canonical variables consist of two pairs of 1-connections and 2-connections,
$(\rho,\phi)$, $(\pi,\psi)$, 
providing a concrete realization of the structure of higher gauge symmetries \cite{baez, Borsten_2025}. Consequently, we expect the quantum theory of the backbone topological theory to involve 2-representations and 2-intertwiners \cite{douglas2018}. 

Since the theory is built on $\mathbb{C}^2$, both the discrete and quantum formulations should be naturally expressed in terms of complex variables decorating faces and edges. In the Hamiltonian analysis, we examined the reality conditions and demonstrated their compatibility with the dynamics. At the quantum level, one may expect these conditions to emerge from holomorphicity requirements on physical states, suggesting a coherent complex polarization.

An important feature of the spinorial formulation---already emphasized by Nester, Tung, and Jacobson, among others---is that the derivation of the gravitational energy as a boundary term is straightforward in this formalism. Positivity of energy can be established along the lines of Witten’s argument \cite{Witten:1981mf, Nester:1981bjx}. This suggests that, in the quantum theory, the Hamiltonian is bounded from below, a highly desirable property for a consistent quantum theory of gravity (unitarity, stability of the vacuum state).

The structure of the constraint algebra---and in particular the associated structure functions---plays a central role in any quantization scheme. {As already mentioned, } in the metric formulation, these structure functions typically depend on non-polynomial expressions such as square roots or inverse powers of the metric, which makes the quantization procedure technically challenging. The Einstein--Cartan framework, and especially the introduction of Ashtekar variables, significantly simplifies the structure of the constraint algebra. In the spinorial formulation, the situation improves even further: the constraint algebra takes an even simpler form, with structure functions that are more tractable and potentially better suited for a canonical quantization. 

We have also looked into a coupling of matter defects.
A point particle couples to geometry through the frame field. In the standard $\SO(1,3)$ BF theory, which underlies the Plebański framework, there is no frame field, and no natural coupling, therefore, to point particles. In our framework, the frame field is already present at the level of the underlying topological theory. As a result, particle degrees of freedom can be incorporated as defects in a uniform manner, whether one works in the purely topological model or in the gravitational model obtained after gauging the symmetry. In this sense, matter can be introduced without altering the structural backbone of the theory. This will be very useful in the spinfoam approach when coupling matter (a particle defect) to gravity. We can introduce the particle degree of freedom in the quantum topological model, and proceed to gauging the Lorentz symmetry to recover a quantum gravity model with defects.

All these exciting directions are currently under investigation.

\bigskip

\noindent \emph{Acknowledgments.} Work on this project was supported in part by  Deutsche Forschungsgemeinschaft  and Perimeter Institute for Theoretical Physics. Research at Perimeter Institute is supported in part by the Government of Canada through the Department of Innovation, Science and Economic Development and by the Province of Ontario through the Ministry of Colleges, Universities, Research Excellence and Security. O.H.\ is supported by a NSERC Discovery grant
awarded to M.D. The contribution of W.W.\ to this research was funded by the Heisenberg programme of the Deutsche Forschungsgemeinschaft (DFG, German
Research Foundation)--–543301681.

\appendix

\section*{Appendix: Constraints on Lagrange multipliers}\label{app:LM-constr}
\addcontentsline{toc}{section}{Appendix: Constraints on Lagrange multipliers}
\renewcommand{\theequation}{A.\arabic{equation}}
\renewcommand{\thesubsection}{A.\arabic{subsection}}\label{appdxA1}
\subsection{Constraints on $\updown{\gamma}{A}{B}$}
We show that the solution \eref{mu-cons} is the unique solution that sets \eref{H0-P} and \eref{H0-R} to zero given a non-degenerate (regular) configuration of $(\psi_{Aa},\phi_{Aa})$, i.e\ $d^3v\neq0$ in (\ref{three-vol1}, \ref{three-vol2}). Given such a regular configuration, we can use the spinor-valued 1-forms $(\psi_{Aa},\phi_{Aa})$ as a complexified basis in co-tangent space. Taking into account the definition of the spinor dyad $(u_A,v_A)$, see \eref{uv-def}, we can thus write the $\mathfrak{sl}(2,\C)$-valued 1-form $\updown{\gamma}{A}{Ba}$ as
\begin{equation}
\updown{\gamma}{A}{Ba}=\updown{\mu}{A}{BC}\updown{\psi}{C}{a}+\updown{\nu}{A}{B}u_C\updown{\phi}{C}{a},\label{mu-comp}
\end{equation}
where $\mu_{ABC}=\mu_{BAC}$ and $\nu_{AB}=\nu_{BA}$. We split these spinors into irreducible parts, obtaining
\begin{subequations}\begin{align}
\mu_{ABC}&=-\frac{2}{3}\epsilon_{C(A}\mu_{B)}+\mu_{(ABC)},\label{mu-comp-1}\\
\nu_{AB}u_C&=-\frac{2}{3}\epsilon_{C(A}\nu_{B)}+\nu_{(AB}u_{C)}\label{mu-comp-2},
\end{align}\end{subequations}
where $\mu_A=\downup{\mu}{EA}{E}$ and $\nu_A={\nu}_{AB}u^B$. The homogenous part of the conditions on the Lagrange multiplier $\updown{\gamma}{A}{Ba}$, see \eref{H0-P} and \eref{H0-R}, is
\begin{subequations}\begin{align}
\updown{\gamma}{A}{B}\wedge\psi^B=0,\label{homgen-mu-cons1}\\
\updown{\gamma}{A}{B}\wedge\phi^B=0.\label{homgen-mu-cons2}
\end{align}\end{subequations}
Inserting the decompositions (\ref{mu-comp}, \ref{mu-comp-1}, \ref{mu-comp-2}) back into \eref{homgen-mu-cons1}, we obtain
\begin{equation}
\updown{\gamma}{A}{B}\wedge\psi^B=-\mu_A E_2+\nu_{AB}v^B E_0+2\nu_{AB}u^B E_1\stackrel{!}{=}0,\label{mu-comp-3}
\end{equation}
where
\begin{equation}
E_{2}=v^Av^B\psi_A\wedge\phi_B,\quad E_1=u^Av^B\psi_{(A}\wedge\phi_{B)},\quad E_0=u^Au^B\psi_A\wedge\phi_B.
\end{equation}
If the configuration of $(\psi_{Aa},\phi_{Aa})$ is regular, the triple $(E_0,E_1,E_2)$ is linearly independent. Going back to \eref{mu-comp-3}, we thus see
\begin{equation}
\mu_A=0,\quad\nu_{AB}=0.
\end{equation}
We are now left with \eref{homgen-mu-cons2}. Taking into account \eref{mu-comp-1} and  \eref{mu-comp-2}, we obtain
\begin{equation}
\mu_{(ABC)}\psi^B\wedge\phi^C=0.
\end{equation}
Since the three complexified bivectors $E_{AB}=\psi_{(A}\wedge\phi_{B)}$ are linearly independent, this implies
\begin{equation}
\mu_{(ABC)}=0.
\end{equation}
In short, the only solution for $\updown{\gamma}{A}{Ba}$ to the homogenous equations \eref{homgen-mu-cons1} and \eref{homgen-mu-cons2} is $\updown{\gamma}{A}{Ba}=0$. Therefore, equation \eref{mu-cons} is the unique solution for $\updown{\gamma}{A}{Ba}$ that satisfies, for given Lagrange multipliers $\lambda^A$ and $\mu^A$, the stability conditions for the fake-flatness constraints under the Hamiltonian evolution, see \eref{H0-P} and \eref{H0-R}. 
\subsection{Constraints on $\updown{\eta}{A}{a}$ and $\updown{\zeta}{A}{a}$}\label{appdxA2}

In Section \ref{sec3.4}, we saw that the Hamiltonian vector field preserves the void-momentum constraint \eref{C-cons} provided the Lagrange multipliers $\updown{\eta}{A}{a}$ and $\updown{\zeta}{A}{a}$ satisfy
\begin{equation}
\eta_{(A}\wedge\phi_{B)}-\zeta_{(A}\wedge\psi_{B)}=g^\ii\Xi_{AB},\label{uv-condXi}
\end{equation}
see \eref{uv-cond}. The $\mathfrak{sl}(2,\C)$-valued 2-form $\updown{\Xi}{A}{Bab}$ is given by
\begin{equation}
\Xi_{AB}=\lambda_{(A}D\phi_{B)}-\mu_{(A}D\psi_{B)}.
\end{equation}
To solve equation \eref{uv-condXi} for given $\Xi_{ABab}$, we spilt $\updown{\eta}{A}{a}$ and $\updown{\zeta}{A}{a}$ into a solution of the homogenous equation (for $\Xi_{ABab}=0$) and a particular solution $\praupdown{o}{\eta}{A}{a}$ and $\praupdown{o}{\zeta}{A}{a}$:
\begin{subequations}\begin{align}
g\updown{\eta}{A}{a}&=a\updown{\psi}{A}{a}+b\updown{\phi}{A}{a}+\praupdown{o}{\eta}{A}{a},\\
g\updown{\zeta}{A}{a}&=c\updown{\psi}{A}{a}-a\updown{\phi}{A}{a}+\praupdown{o}{\zeta}{A}{a}.
\end{align}\end{subequations}
Notice that $(a,b,c)\in\C$ are completely arbitrary. They parametrise the three-complex-dimensional kernel of equation \eref{uv-condXi} for $\Xi_{ABab}=0$.

To find an inhomogenous solution $\praupdown{o}{\eta}{A}{a}$ and $\praupdown{o}{\zeta}{A}{a}$ of \eref{uv-condXi} for given $\Xi_{ABab}$, we split the 1-forms $\psi_{Aa}$ and $\phi_{Aa}$ into components with respect to the spin dyad $(u^A,v^A)$. Going back to \eref{uv-def}, we obtain:
\begin{subequations}\begin{align}
\updown{\psi}{A}{a}&=\frac{\I}{\sqrt{2}}v^A\updown{e}{3}{a}-\I\sqrt{2}u^A\updown{e}{+}{a},\\
\updown{\phi}{A}{a}&=\frac{\I}{\sqrt{2}}u^A\updown{e}{3}{a}+\I\sqrt{2}v^A\updown{e}{-}{a}.
\end{align}\end{subequations}
For a regular configuration of $\updown{\psi}{A}{a}$ and $\updown{\phi}{A}{a}$, the triple $(\updown{e}{3}{a},\updown{e}{+}{a},\updown{e}{-}{a})$ are linearly independent.\footnote{Such that the three-volume (\ref{three-vol1}, \ref{three-vol2}) is non-vanishing. If the reality conditions are satisfied as discussed in  Section \ref{sec3.6},  $\updown{e}{3}{a}$ is real and $\updown{e}{-}{a}$ is the complex conjugate of $\updown{e}{+}{a}$.} We can thus split the Lagrange multipliers $\updown{\eta}{A}{a}$ and $\updown{\zeta}{A}{a}$ with respect to the co-triad $(\updown{e}{3}{a},\updown{e}{+}{a},\updown{e}{-}{a})$. The dual $\mathfrak{sl}(2,\C)$ transformations allow us to redefine $\beta=\left(\begin{smallmatrix}a&b\\c&-a\end{smallmatrix}\right)$ in such a way that we can choose, without loss of generality, that
\begin{subequations}\begin{align}
g\, \praupdown{o}{\eta}{A}{a}&=u^A_+\updown{e}{+}{a}+u^A_-\updown{e}{-}{a},\\
g\, \praupdown{o}{\zeta}{A}{a}&=vu^A\updown{e}{3}{a}+v^A_+\updown{e}{+}{a}+v^A_-\updown{e}{-}{a}.
\end{align}\end{subequations}
Inserting the decompositon back into \eref{uv-condXi}, we obtain
\begin{subequations}\begin{align}\nonumber
+\frac{\I}{\sqrt{2}}\left(u_+^{(A}u^{B)}_{\phantom{+}}-2vu^Au^B-v_+^{(A}v^{B)}_{\phantom{+}}\right)e^+\wedge e^3+\\
+\I\sqrt{2}\left(u_+^{(A}v^{B)}_{\phantom{+}}-v^{(A}_-u^{B)}_{\phantom{-}}\right)+\frac{\I}{\sqrt{2}}\left(u_-^{(A}u^{B)}_{\phantom{-}}-v_-^{(A}v^{B)}_{\phantom{-}}\right)e^-\wedge e^3=\Xi^{AB}.
\end{align}\end{subequations}
Next, we split the spinors $u_\pm^A$ and $v_\pm^A$ into components with respect to the spinor dyad $(u^A,v^A)$, introduced in \eref{uv-def}. We write
\begin{subequations}\begin{align}\nonumber
u^A_\pm&=u^\downarrow_\pm u^A + u^\uparrow_\pm v^A,\\
v^A_\pm&=v^\downarrow_\pm u^A + v^\uparrow_\pm v^A.\label{uv-condXi2}
\end{align}\end{subequations}
Contracting $\Xi_{AB}$ with $u^Au^B$ and $v^Av^B$ and taking into account that we restrict ourselves here to configurations in which $e^+$, $e^-$ and $e^3$ are linearly independent, we obtain from \eref{uv-condXi2} immediately $v_+^\uparrow$, $u_+^\downarrow-2v$, $u_+^\uparrow$, $v_-^\downarrow$, $u_-^\downarrow$ and $v_-^\uparrow$. The mix terms $\Xi_{AB}u^Av^B$ determine the remaining components $v^\downarrow_+$, $u^\downarrow_+$ and $u_-^\uparrow$.

\bibliography{main}
\bibliographystyle{biblio}

\end{document}